\pdfoutput=1
\documentclass[aps,prd,amsmath,floats,floatfix, twocolumn,
superscriptaddress,nofootinbib,showpacs,longbibliography]{revtex4-1}

\usepackage[T1]{fontenc}
\usepackage[utf8]{inputenc}
\usepackage{lmodern}

\usepackage{verbatim}

\usepackage[dvipsnames, usenames]{xcolor}
\definecolor{linkcolor}{rgb}{0.0,0.3,0.5}
\usepackage[hypertexnames=false, unicode, colorlinks=true, linkcolor=linkcolor,
citecolor=linkcolor, filecolor=linkcolor,urlcolor=linkcolor,
pdfusetitle]{hyperref}

\usepackage[all]{hypcap}
\usepackage{graphicx}
\usepackage{xspace}
\usepackage{amssymb}

\usepackage{microtype}

\usepackage[english]{babel}
\usepackage{blindtext}

\usepackage[normalem]{ulem}
\usepackage{bm}

\graphicspath{%
	{figs/}%
}

\DeclareMathAlphabet{\mathpzc}{OT1}{pzc}{m}{it}

\newcommand{\h}{\mathpzc{h}}
\newcommand{\hlm}{\mathpzc{h}_{\ell m}}

\newcommand{\bchi}{\bm{\chi}}

\newcommand{\ecc}{e_{\rm ref}}
\newcommand{\meanano}{l_{\rm ref}}

\usepackage[nomessages]{fp}
\newcommand{\MySXSID}[1]{\FPeval{\result}{clip(2265+#1)}\result\xspace}

\begin{document}
\title{Eccentric binary black hole surrogate models for the gravitational waveform and remnant properties: comparable mass, nonspinning case}
\newcommand\caltech{\affiliation{TAPIR 350-17, California Institute of
    Technology, 1200 E California Boulevard, Pasadena, CA 91125, USA}}
\newcommand{\cornell}{\affiliation{Cornell Center for Astrophysics
    and Planetary Science, Cornell University, Ithaca, New York 14853, USA}} 
\newcommand\cornellPhys{\affiliation{Department of Physics, Cornell
    University, Ithaca, New York 14853, USA}}
\newcommand{\UMassDMath}{\affiliation{Department of Mathematics,
		University of Massachusetts, Dartmouth, MA 02747, USA}}
\newcommand{\UMassDPhy}{\affiliation{Department of Physics,
		University of Massachusetts, Dartmouth, MA 02747, USA}}
\newcommand{\CSCVR}{\affiliation{Center for Scientific Computing and Visualization 	Research, University of Massachusetts, Dartmouth, MA 02747, USA}}
\newcommand{\URI}{\affiliation{Department of Physics, 
    University of Rhode Island, Kingston, RI 02881, USA}}    
\newcommand{\AEI}{\affiliation{Max Planck Institute for Gravitational
    Physics (Albert Einstein Institute), Am M\"uhlenberg 1, Potsdam 14476,
    Germany}} %
\newcommand{\bham}{\affiliation{School of Physics and Astronomy \& Institute for Gravitational Wave Astronomy, \\ University of Birmingham, Birmingham, B15 2TT, United Kingdom}}

\author{Tousif Islam}
\email{tislam@umassd.edu}
\UMassDPhy
\UMassDMath
\CSCVR

\author{Vijay Varma}
\thanks{Klarman fellow}
\cornellPhys
\cornell 
\caltech

\author{Jackie Lodman}
\caltech

\author{Scott E. Field}
\UMassDMath
\CSCVR

\author{\\Gaurav Khanna}
\UMassDPhy
\CSCVR
\URI

\author{Mark A. Scheel}
\caltech

\author{Harald P. Pfeiffer}
\AEI

\author{Davide Gerosa}
\bham

\author{Lawrence E. Kidder}
\cornell 

\hypersetup{pdfauthor={Islam et al.}}

\date{\today}

\begin{abstract}
We develop new strategies to build numerical relativity surrogate models for
eccentric binary black hole systems, which are expected to play an increasingly
important role in current and future gravitational-wave detectors.  We
introduce a new surrogate waveform model, \texttt{NRSur2dq1Ecc}, using 47
nonspinning, equal-mass waveforms with eccentricities up to $0.2$ when measured
at a reference time of $5500M$ before merger. This is the first waveform model
that is directly trained on eccentric numerical relativity simulations and does
not require that the binary circularizes before merger.  The model includes the
$(2,2)$, $(3,2)$, and $(4,4)$ spin-weighted spherical harmonic modes. We also
build a final black hole model, \texttt{NRSur2dq1EccRemnant}, which models the
mass, and spin of the remnant black hole.  We show that our waveform model can
accurately predict numerical relativity waveforms with mismatches $\approx
10^{-3}$, while the remnant model can recover the final mass and dimensionless
spin with absolute errors smaller than $\approx 5 \times 10^{-4}M$ and
$\approx 2 \times10^{-3}$ respectively. We demonstrate that the waveform model
can also recover subtle effects like mode-mixing in the ringdown signal without
any special ad-hoc modeling steps. Finally, we show that despite being trained
only on equal-mass binaries, \texttt{NRSur2dq1Ecc} can be reasonably extended
up to mass ratio $q\approx3$ with mismatches $\simeq 10^{-2}$ for
eccentricities smaller than $\sim 0.05$ as measured at a reference time of
$2000M$ before merger.  The methods developed here should prove useful in the
building of future eccentric surrogate models over larger regions of the
parameter space.
\end{abstract}

\maketitle
\section{Introduction}
\label{Sec:Introduction}
Detection of gravitational waves (GWs)~\cite{LIGOScientific:2018mvr,
Abbott:2020niy} by the LIGO~\cite{TheLIGOScientific:2014jea} and
Virgo~\cite{TheVirgo:2014hva} detectors has opened a new window in astrophysics
to probe binary compact objects -- binary black holes (BBHs) being the most
abundant source for these detectors.  Both detection and extraction of source
properties from the GW signal relies on the availability of accurate
inspiral-merger-ringdown (IMR) waveform models for BBHs.  While numerical
relativity (NR) provides the most accurate gravitational waveforms for BBHs,
they are computationally expensive, taking weeks to months to generate a single
waveform. Data-driven surrogate modeling strategies~\cite{Field:2013cfa,
    Purrer:2014fza, Blackman:2015pia, Blackman:2017pcm, Blackman:2017dfb,
Varma:2018mmi, Chua:2018woh, Lackey:2018zvw, Varma:2019csw, Williams:2019vub,
Khan:2020fso, Haegel:2019uop} have been shown to be capable of producing
waveforms that are nearly indistinguishable from NR with evaluation times of
less than $0.1$ seconds.  While NR surrogate waveform models for
nonspinning~\cite{Blackman:2015pia}, aligned-spin~\cite{Varma:2018mmi}, and
precessing BBHs~\cite{Blackman:2017dfb, Varma:2019csw} are well developed, NR
surrogate modeling of eccentric systems is completely unexplored.

So far, all GW detections of BBHs are consistent with signals emitted from
quasicircular binaries~\cite{Salemi:2019owp, Romero-Shaw:2019itr,
Lenon:2020oza, Yun:2020aow, Wu:2020zwr, Nitz:2019spj, Ramos-Buades:2020eju}.
In fact, eccentricity has been traditionally ignored in most GW data analyses
(for e.g.  Refs.~\cite{Abbott:2020niy, LIGOScientific:2018mvr}).  This is
motivated by the expectation that even if a binary is formed with a non-zero
eccentricity, it should circularize before reaching the frequency band of
ground based detectors, as eccentricity gets radiated away via GWs during the
long inspiral~\cite{peters1964gravitational}.  However, this assumption may not
always hold, especially for binaries formed in dense environments like globular
clusters or galactic nuclei~\cite{Giesler:2017uyu, Rodriguez:2018pss,
OLeary:2005vqo, Samsing:2017xmd, Fragione:2019hqt, Kumamoto:2018gdg,
OLeary:2008myb,Gondan:2020svr}. Indeed, recent follow-up analysis of GW190521
\cite{Abbott:2020tfl} claim this event to be consistent with a BBH source with
eccentricity ranging from  $\sim 0.1$ \cite{Romero-Shaw:2020thy} up to $\sim
0.7$ \cite{Gayathri:2020coq} (see also \cite{CalderonBustillo:2020odh,
CalderonBustillo:2020srq}).

Eccentricity, if present in GW signals, carries precious astrophysical
information about the environment in which the binary was formed. The detection
of an eccentric merger would not only be a smoking-gun signature of sources
formed via dynamical encounters, but would point towards specific type of
interactions, namely GW captures~\cite{Zevin:2018kzq}, taking place in those
environments. Catching eccentric sources in the mHz regime targeted by the
LISA space mission is also a promising avenue to distinguish astrophysical
formation channels~\cite{Nishizawa:2016eza, Nishizawa:2016jji, Breivik:2016ddj,
Fang:2019dnh, Rodriguez:2017pec, Gondan:2017wzd, Tagawa:2020jnc}.

Furthermore, ignoring eccentricity in our models can lead to systematic biases
if the actual signal corresponds to an eccentric
system~\cite{Ramos-Buades:2019uvh}.  Such biases can also lead to eccentric
systems being misidentified as a violation of general relativity (GR).  Even if
all binaries are found to be circular, eccentric models are necessary to place
bounds on the eccentricity.  Therefore, including eccentricity in our GW models
is important, especially as the detectors become more sensitive.

In the last few years, a handful of eccentric
inspiral-only~\cite{Klein:2018ybm, Tiwari:2020hsu, Moore:2018kvz,
Moore:2019xkm, Liu:2019jpg, Tanay:2019knc} and IMR
models~\cite{Hinderer:2017jcs, Hinder:2017sxy, Huerta:2017kez, Chen:2020lzc,
Chiaramello:2020ehz, Cao:2017ndf} have become available. We highlight some
recent eccentric IMR models in the following.
\texttt{ENIGMA}~\cite{Huerta:2017kez, Chen:2020lzc} is a nonspinning eccentric
BBH model that attaches an eccentric post-Newtonian (PN) inspiral to a
quasicircular merger based on an NR surrogate model~\cite{Blackman:2015pia}.
\texttt{SEOBNRE}~\cite{Cao:2017ndf} modifies an aligned-spin quasicircular EOB
waveform model~\cite{Taracchini:2012ig} to include some effects of
eccentricity. Similarly, Ref.~\cite{Chiaramello:2020ehz} modifies a different
aligned-spin EOB multipolar waveform model for quasicircular
BBHs~\cite{Nagar:2018zoe, Nagar:2020pcj} to include some effects of
eccentricity. The model is then further improved by replacing the carrier
quasicircular model with a generic eccentric one~\cite{Nagar:2021gss}.
In addition to these models, Ref.~\cite{Setyawati:2021gom} recently
developed a method to add eccentric modulations to existing quasicircular
BBH models.

Notably, all of these models rely on the assumption that the binary
circularizes by the merger time. While this is approximately true for many
expected sources~\cite{Huerta:2017kez, Habib:2019cui}, this necessarily places
a limit on the range of validity of these models. In addition, none of these
models are calibrated on eccentric NR simulations, even though their accuracy
is tested by comparing against eccentric simulations.

Apart from the waveform prediction, BBH remnant modeling from eccentric sources
is also of crucial astrophysical importance~\cite{Sperhake:2007gu,
Hinder:2007qu, Sopuerta:2006et, Sperhake:2019wwo}. For example, recoils from
eccentric mergers can be up to $25\%$ higher than the circular
case~\cite{Sopuerta:2006et, Sperhake:2019wwo}, which result in a higher
likelihood of ejections from astrophysical hosts like star clusters and
galaxies.

It is, therefore, timely to invest in building faithful eccentric BBHs waveform
and remnant models that address some of these limitations. In this paper, we
develop a detailed framework for constructing a surrogate model with eccentric
NR data.  We then build a two-dimensional surrogate model,
\texttt{NRSur2dq1Ecc}, over parameters that describe eccentricity for
equal-mass, nonspinning systems to demonstrate the efficacy of the proposed
methods. This is the first eccentric waveform that is directly trained on
eccentric NR simulations and does not need to assume that the binary
circularizes before merger. The model can produce waveforms that are of
comparable accuracy to the NR simulations used to train it.  Furthermore,
despite being trained only on equal-mass eccentric BBHs, we find that the model
can be reasonably evaluated beyond its training range upto mass ratio
$q\approx3$ provided the eccentricities are small.

In addition to the waveform model, we build a surrogate model for the remnant
mass and spin, \texttt{NRSur2dq1EccRemnant}, which can provide accurate
predictions for the final state of eccentric binary mergers. This work paves
the way forward for building future eccentric surrogate models: we expect that
the methods developed here can be applied straightforwardly to aligned-spin
eccentric BBHs, while the precessing case requires significantly more work.

The rest of the paper is organized as follows. Sec.~\ref{Sec:NR_Runs} describes
the NR simulations. Sec.~\ref{Sec:buildSurrogate} describes data decomposition,
parameterization and construction of the surrogate model. In
Sec.~\ref{sec:results}, we test the surrogate model by comparing against NR
waveforms. We end with some concluding remarks in Sec.~\ref{Sec:Conclusion}.

\section{Numerical Relativity Data}
\label{Sec:NR_Runs}

NR simulations for this work are performed using the Spectral Einstein
Code (SpEC)~\cite{SpECwebsite} developed by the Simulating eXterme Spacetimes (SXS)
collaboration~\cite{SXSWebsite}. We follow the procedure outlined in
Ref.~\cite{Chatziioannou:ecccontrol_inprep} to construct initial orbital
parameters that result in a desired eccentricity. The constraint equations are
solved employing the extended conformal thin sandwich
formalism~\cite{York:1998hy, Pfeiffer:2002iy} with superposed harmonic Kerr
free data~\cite{Varma:2018sqd}. The evolution equations are solved employing
the generalized harmonic formulation~\cite{Lindblom:2005qh,Rinne:2008vn}.
The time steps during the simulations are
chosen nonuniformly using an adaptive time-stepper~\cite{Boyle:2019kee}. Further details
can be found in Ref.~\cite{Boyle:2019kee} and references within. We perform 47
new eccentric NR simulations that have been assigned the identifiers
SXS:BBH:\MySXSID{1} - SXS:BBH:\MySXSID{47}, and will be made available
through the SXS public catalog~\cite{SXSCatalog}.

The component BH masses, $m_{1}$ and $m_{2}$, and dimensionless spins, $\bchi_{1}$
and $\bchi_{2}$, are measured on the apparent horizons~\cite{Boyle:2019kee} of the
BHs, where index 1 (2) corresponds to the heavier (lighter) BH. The component masses at
the relaxation time~\cite{Boyle:2019kee} are used to define the mass ratio
$q=m_1/m_2\geq1$ and total mass $M=m_{1}+m_{2}$. 
Unless otherwise specified, all masses in this paper are
given in units of the total mass.
When training the surrogate model, we restrict
ourselves to $q=1$, $\bchi_{1}, \bchi_{2} = 0$ in this work.

The waveform is extracted at several extraction spheres at varying finite radii
from the origin and then extrapolated to future null
infinity~\cite{Boyle:2019kee, Boyle:2009vi}. These extrapolated waveforms are then 
 corrected to account for the initial drift of the center of
mass~\cite{Boyle:2015nqa, scri}. The
spin-weighted spherical harmonic modes
at future null infinity, scaled to unit mass and unit distance, are denoted as
$\hlm(t)$ in this paper.

The complex strain $\h = h_{+} -i h_{\times}$ is given by:
\begin{equation}
    \h(t, \iota, \varphi_0) = \sum^{\infty}_{\ell=2} \sum_{m=-l}^{l}
        \hlm(t) ~_{-2}Y_{\ell m}(\iota, \varphi_0),
\label{eq:spherical_harm}
\end{equation}
where $h_+$ ($h_{\times}$) is the plus (cross) polarization of the waveform,
$_{-2}Y_{\ell m}$ are the spin$\,=\!\!-2$ weighted spherical harmonics, and
$\iota$ and $\varphi_0$ are the polar and azimuthal angles on the sky in the
source frame. We model modes with $(\ell,m)={(2,2),(3,2),(4,4)}$.  Because of
the symmetries of equal-mass, nonspinning BBHs, all odd-$m$ modes are
identically zero, and the $m<0$ modes can be obtained from the $m>0$ modes.
Therefore, we model all non-zero $\ell\leq3$ and $(4, \pm 4)$ modes, except the
$m=0$ modes.  We exclude $m = 0$ memory modes because (non-oscillatory)
Christodoulou memory is not accumulated sufficiently in our NR
simulations~\cite{Favata:2008yd}; this defect was recently addressed in both
Cauchy characteristic extraction (CCE)~\cite{Mitman:2020pbt,Barkett:2019uae,
Moxon:2020gha} and extrapolation~\cite{Mitman:2020bjf} approaches.  The $(4,2)$
mode, on the other hand, was found to have significant numerical error in the
extrapolation procedure~\cite{Boyle:2019kee, Boyle:2009vi}. We expect this
issue to be resolved with CCE as well. Therefore, in future models, we should
be able to include the $m=0$ modes as well as modes like the (4,2) mode.

The remnant mass $m_{f}$ and spin $\bchi_{f}$ are determined from the common
apparent horizon long after the ringdown, as described in
Ref.~\cite{Boyle:2019kee}.  For nonprecessing systems like the ones considered
here,  the final spin is directed along the direction of the orbital angular
momentum. Unlike previous surrogate models~\cite{Varma:2019csw, Varma:2018aht,
Varma:2020nbm}, we do not model the recoil kick in this work, as the symmetries
of equal-mass, nonspinning BBHs restrict the kick to be zero.

\section{Surrogate methodology for eccentric waveforms}
\label{Sec:buildSurrogate}

In this section, we describe our new framework to build NR surrogate
models for eccentric BBHs. We begin by applying the following post processing
steps that simplify the modeling procedure.

\subsection{Processing the training data}
\label{Sec:PostProcessing}
In order to construct parametric fits (cf. Sec.~\ref{Sec:model_building}) for
the surrogate model, it is necessary to align all the waveforms such that their
peaks occur at the same time. We define the peak of each waveform,
$\tau_{peak}$, to be the time when the quadrature sum,
\begin{equation}
A_{\rm tot}(\tau)=\sqrt{\sum_{l,m} |\hlm(\tau)|^2} \,,
\end{equation}
reaches its maximum. Here the summation is taken over all the modes being modeled.
We then choose a new time coordinate,
\begin{equation}
t=\tau - \tau_{\rm peak} \,,
\end{equation}
such that $A_{\rm tot}(t)$ for each waveform peaks at $t=0$.

Next, we use cubic splines to interpolate the real and imaginary parts of the
waveform modes onto a common time grid of [$-5500M$, $75M$] with a uniform time
spacing of $dt=0.1 M$; this is dense enough to capture all frequencies of
interest, including near merger. The initial time of $-5500M$ is chosen so
that we can safely eliminate spurious initial transients in the waveform, also
known as junk radiation~\cite{Boyle:2019kee}, for each waveform in our dataset.

Once all the waveforms are interpolated onto a common time grid, we
perform a frame rotation of the waveform modes about the z-axis such
that the orbital phase is zero at $t=-5500M$. The orbital phase is obtained
from the $(2,2)$ mode [cf. Eq.~(\ref{Eq:orb_phase})]. Because of the symmetry
of the equal-mass, equal-spin systems considered here, the odd-$m$ modes are
identically zero and so we need not worry about remaining $\phi_{\rm orb}
\rightarrow \phi_{\rm orb} +\pi$ rotational freedom as was necessary in
Refs.~\cite{Blackman:2015pia, Blackman:2017pcm, Blackman:2017dfb,
Varma:2018mmi, Varma:2019csw}. This preprocessing of time and phase ensures
that the waveform varies smoothly across the parameter space, which in turn
makes modeling easier.

\subsection{Measuring eccentricity and mean anomaly}
Departure of NR orbits from circularity is measured by a time-dependent
eccentricity and mean anomaly. Eccentricity takes values between
$[0,1]$ where the boundary values correspond to a quasicircular binary
and an unbound orbit~\cite{Chandrasekhar:1983mtbh}, respectively.
Mean anomaly, on the other hand, is
bounded by $[0,2\pi)$.  While it may seem most natural to estimate orbital
parameters from the BH trajectories, this task is complicated by the fact that
any such measurement will be impacted by the gauge conditions chosen by the NR
simulation. We instead choose to estimate eccentricity and anomaly parameters
directly from the waveform data at future null infinity.

\subsubsection{Measuring eccentricity}
\label{Sec:measuring_ecc}
Various methods to extract the eccentricity from NR simulations have been
proposed in the literature~\cite{Habib:2019cui, Healy:2017zqj, Mroue:2010re,
Purrer:2012wy}. As the eccentricity evolves during the binary's
orbit~\cite{peters1964gravitational}, these methods use dynamical quantities
such as some combination of the $(2,2)$ mode's amplitude, phase, or frequency.
All of these methods reduce to the eccentricity
parameter in the Newtonian limit.  The estimated value of the eccentricity  may
differ slightly depending on the method used and the noise in the numerical
data. However, as long as they provide a consistent measurement of eccentricity
that decays monotonically with time, one can use any of the eccentricity
estimators for constructing a surrogate
waveform model. For this work, we use the following definition
of eccentricity based on orbital frequency~\cite{Mora:2002gf}:
\begin{equation} e(t)=
    \frac{\sqrt{\omega_p(t)} - \sqrt{\omega_a(t)}}{\sqrt{\omega_p(t)} + \sqrt{\omega_a(t)}},
\label{Eq:ecc_estimator}
\end{equation}
where $\omega_a$ and $\omega_p$ are the orbital frequencies at apocenter (i.e.
point of furthest approach) and pericenter (i.e. point of closest approach), respectively.
Unlike several other eccentricity estimators proposed in literature
\cite{Habib:2019cui, Healy:2017zqj, Mroue:2010re, Purrer:2012wy}, the one
defined in Eq.~\eqref{Eq:ecc_estimator} is normalized and reduces to
the eccentricity parameter in the Newtonian limit at both low and high
eccentricities~\cite{Ramos-Buades:2019uvh}.

We first compute the orbital frequency,
\begin{equation}
\omega_{\rm orb} = \frac{d\phi_\mathrm{\rm orb}}{dt} \,,
\end{equation}
where $\phi_\mathrm{\rm orb}$ is the orbital phase
inferred from the (2,2) mode (cf.
Eq.~\eqref{Eq:orb_phase}), and the derivative is approximated using
second-order finite differences.  We then find the times where $\omega_{\rm
orb}$ passes through a local maxima (minima) and associate those to pericenter
(apocenter) passages, to obtain $\omega_p$ ($\omega_a$). We find that using the
local maxima/minima of the amplitude of the $(2,2)$ mode to identify the
pericenter/apocenter times leads to a consistent value for the eccentricity.
We then interpolate $\omega_p$ and $\omega_a$ onto the full time grid using
cubic splines. This gives us $\omega_p(t)$ and $\omega_a(t)$, which are used in
Eq.~\eqref{Eq:ecc_estimator}.

\begin{figure}[t]
\includegraphics[width=\columnwidth]{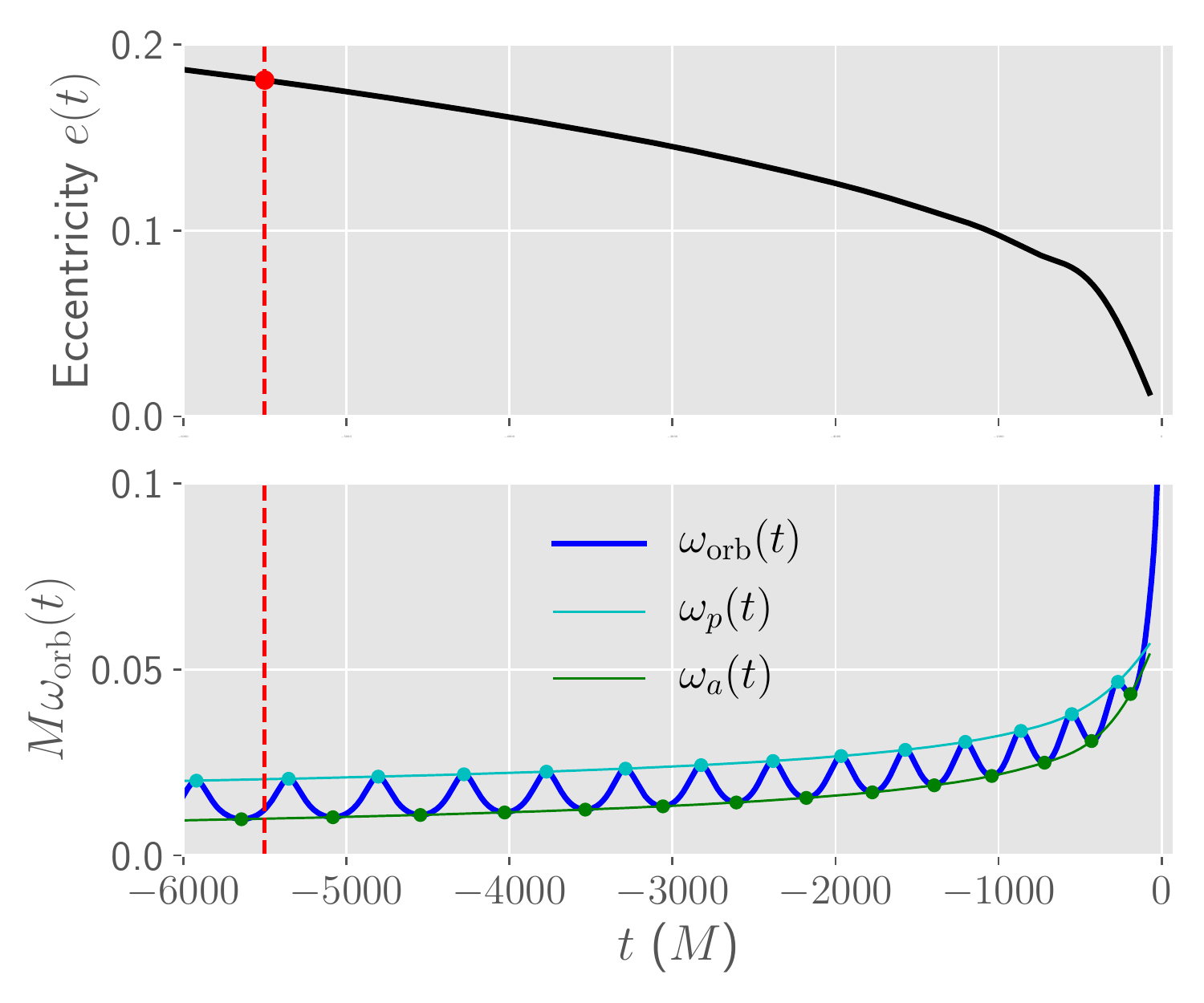}
\caption{Time evolution of the eccentricity $e(t)$ ({upper panel}) and the
orbital frequency $\omega_{\rm orb}(t)$ ({lower panel}) for NR Simulation
SXS:BBH:2304.  $\omega_p$ and $\omega_a$ denote, respectively, the the orbital
frequency at pericenter (local maxima, cyan circles) and apocenter passages
(local minima, green circles). From this data we construct spline interpolants
to obtain $\omega_p(t)$ (cyan curve) and $\omega_a(t)$ (green curve). The
eccentricity is then estimated using Eq.~\eqref{Eq:ecc_estimator}. The red
dashed vertical line corresponds to the reference time $t_{\rm ref}=-5500M$ at
which the surrogate model is parameterized.
\label{Fig:EccID39}
}
\end{figure}

Figure~\ref{Fig:EccID39} shows an example of the measured eccentricity for the
NR simulation SXS:BBH:\MySXSID{39}. We see that our method provides a smooth,
monotonically decreasing $e(t)$. The estimate become unreliable near
merger where finding local maxima/minima in $\omega_{\rm orb}$
becomes problematic as the orbit transitions from inspiral to
plunge. The estimate also becomes problematic whenever the eccentricity 
is extremely small, thereby preventing the appearance of
an identifiable local maxima/minima.
This does not affect our modeling, however, as we only require an
eccentricity value at a reference time while the binary is still in the
inspiral phase. We select a reference time of $t_{\rm ref}=-5500M$ and
parameterize our waveform model by
\begin{equation}
	\ecc = e(t_{\rm ref}) \,.
\end{equation}
While estimating $\ecc$, we include the data segment slightly before $t_{\rm ref}$
as this allows us to interpolate, rather than extrapolate, when constructing
$e(t)$ in Eq.~\eqref{Eq:ecc_estimator}.

\subsubsection{Measuring mean anomaly}
In the Newtonian context, the mean anomaly $l$ of an eccentric orbit is defined
as
\begin{eqnarray} \label{eq:ma}
l &\equiv& 2\pi\frac{t-t_0}{P},
\end{eqnarray}
where $t_0$ is a time corresponding to the previous pericenter passage and $P$
is the radial period, which is defined to be the time between two successive
pericenter passages.  In the Newtonian case $P$ is a constant, but in GR it
changes as the binary inspirals.
However, one can continue to use Eq.~\eqref{eq:ma} as a meaningful
measurement of the radial oscillation's phase for the purpose
of constructing a waveform model~\cite{Hinder:2017sxy}.

\begin{figure}[t]
\includegraphics[width=\columnwidth]{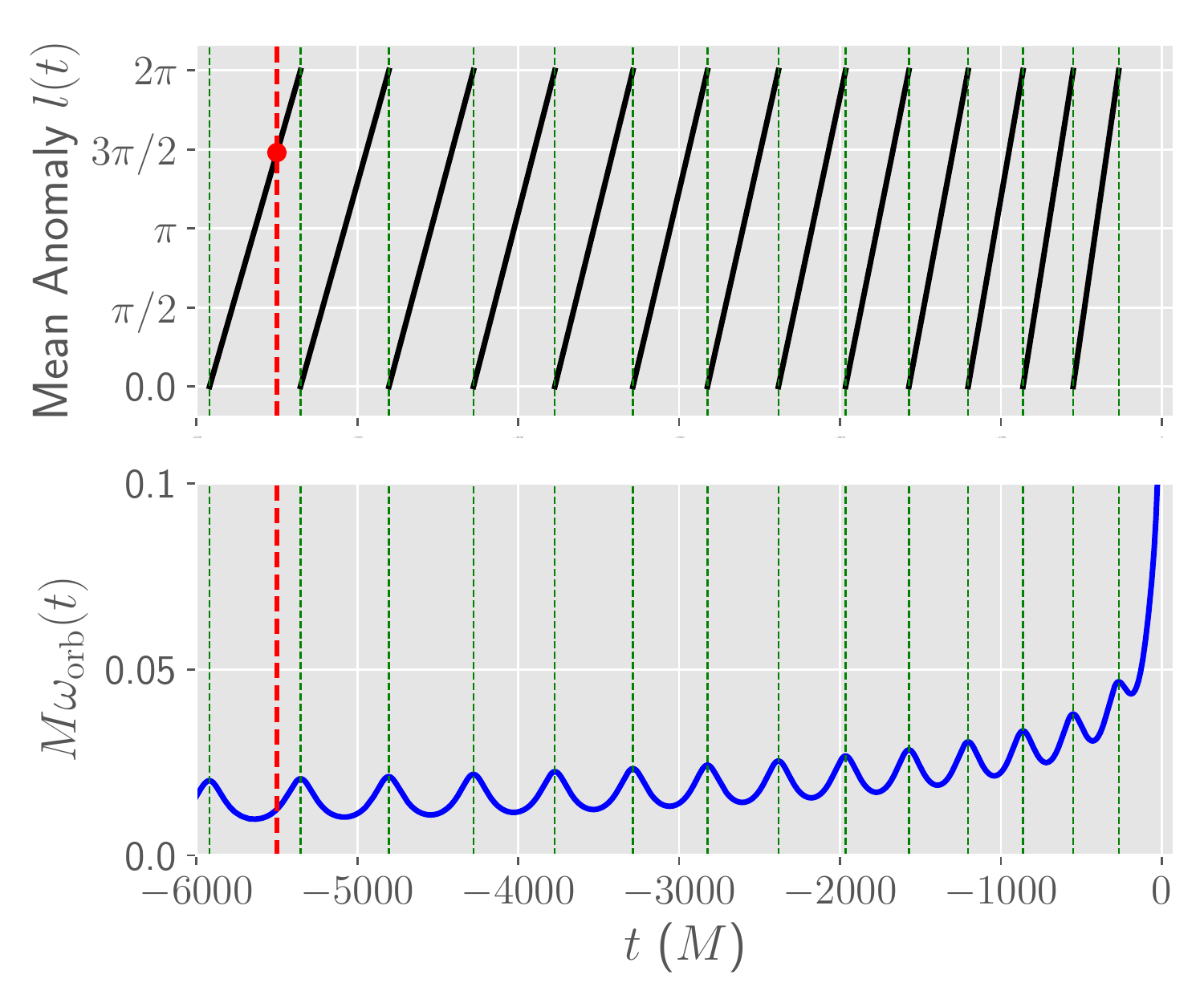}
\caption{
Time evolution of the mean anomaly $l(t)$ ({upper panel}) and the orbital
frequency $\omega_{\rm orb}(t)$ ({lower panel}) for the NR Simulation
SXS:BBH:2304. Green dashed vertical lines indicate the times for pericenter
passages. The anomaly $l(t)$ grows linearly with time over $[0,2\pi)$ in
between two successive pericenters. The red dashed vertical line corresponds to
the reference time $t_{\rm ref}=-5500M$ at which the surrogate model is
parametrized.
}
\label{Fig:AnomalyID39}
\end{figure}

For each NR waveform, we compute the times for all pericenter passages using
the same procedure as in Sec.~\ref{Sec:measuring_ecc}. We divide the time array
into different orbital windows defined as $[t_{i}^{\rm peri},t_{i+1}^{\rm
peri})$, where $t_{i}^{\rm peri}$ is the time for $i^{th}$ pericenter passage.
The orbital period in each window is given by $P_i=t_{i+1}^{\rm
peri}-t_{i}^{\rm peri}$, and the mean anomaly by
\begin{eqnarray}
    l_{i}(t) = 2\pi\frac{t - t_{i}^{\rm peri}}{P_i} \,.
\label{Eq:anomaly}
\end{eqnarray}
Note that each $l_{i}(t)$ grows linearly with time over $[0, 2\pi)$ for the
window $[t_{i}^{\rm peri},t_{i+1}^{\rm peri})$. To obtain the full $l(t)$, we
simply join each $l_{i}(t)$ for consecutive orbits. Finally, the value for mean
anomaly parameterizing our waveform model is then simply the evaluation of the
mean anomaly at $t_{\rm ref}=-5500M$.
\begin{eqnarray}
    \meanano = l(t_{\rm ref}) \,.
\label{Eq:l_compute}
\end{eqnarray}
Figure~\ref{Fig:AnomalyID39} shows an example application of our method to
estimate the mean anomaly of the NR simulation
SXS:BBH:\MySXSID{39}.

\subsubsection{Targeted parameter space} \label{sec:ps}
\begin{figure}[thb]
\includegraphics[width=\columnwidth]{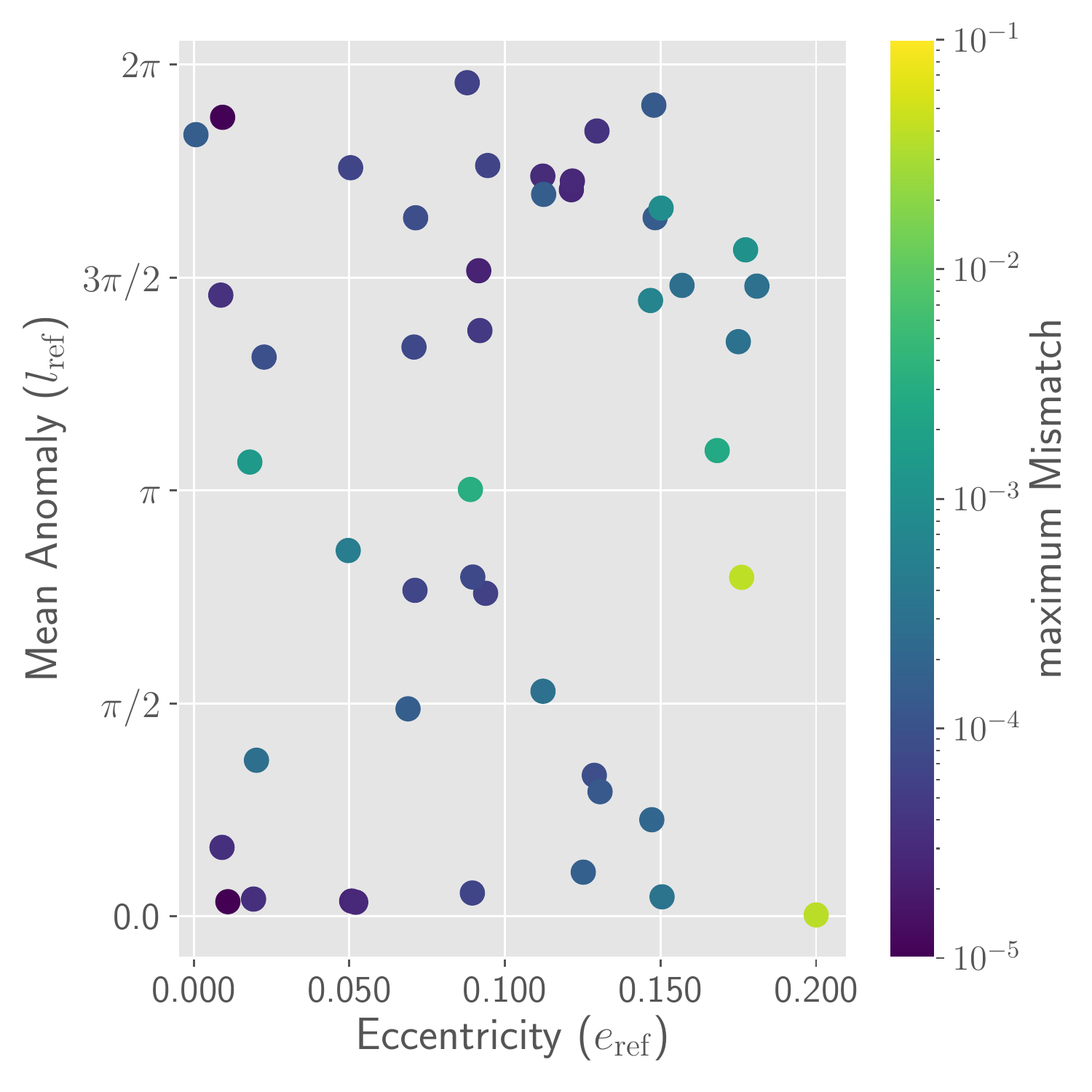}
\caption{The parameter space covered by the 47 NR waveforms (circle markers)
    used in the construction of our surrogate model. The axes show the
    eccentricity and mean anomaly values at $t_{\rm ref}$. We also show the
dependence of the maximum (over the sky of the source frame) flat-noise
mismatches on the parameters eccentricity and mean
anomaly (cf.~Sec.~\ref{Sec:Freqdom_mismatch}). The colors
indicate the maximum mismatch, which systematically increases near the high
eccentricity boundary where few training data points are available.
\label{Fig:MM_heat_map}
}
\end{figure}

In Fig.~\ref{Fig:MM_heat_map}, we show the measured values for eccentricity and mean
anomaly at $t_{\rm ref}$ for all 47 NR waveforms, which leads to the
following 2d parameter space for our model:
\begin{itemize}
	\item eccentricity: $\ecc \in [0, 0.2]$;
        \item mean anomaly: $\meanano \in [0, 2\pi)$.
\end{itemize}
Fig.~\ref{Fig:MM_heat_map} shows a large gap in the parameter space, which
reflects an inherent limitation in our current approach to achieve target
eccentricity parameters from the initial data. The method we use to construct
initial orbital parameters~\cite{Chatziioannou:ecccontrol_inprep} seeks to
achieve target values of $(\ecc, \meanano)$ at a time $500M$ after the start of
the simulation. The initial orbital frequency is chosen such that time to
merger is $6000M$, as predicted by a leading-order PN calculation.
Unfortunately, this is only approximate, leading to different merger times for
different simulations. Consequently, when we estimate the eccentricity
parameters at $t_{\rm ref}=-5500M$, this is no longer a fixed time from the
start of the simulation. The eccentricity parameters evolve differently for
different simulations during this time, leading to the clustering in
Fig.~\ref{Fig:MM_heat_map}. In the future, we plan to resolve this using a
higher order PN expression, or an eccentric waveform
model~\cite{Huerta:2017kez, Chen:2020lzc, Chiaramello:2020ehz} to predict the
time to merger.

\subsection{Waveform data decomposition}
\label{Sec:DataDecomposition}
Building a surrogate model becomes more challenging for oscillatory and
complicated waveform data. One solution is to transform or decompose the
waveform data into several simpler ``{waveform data pieces}'' that also vary
smoothly over the parameter space. These simpler data pieces can then be
modeled more easily and recombined to get back the original waveform.
Successful decomposition strategies have been developed for
quasi-circular NR 
surrogates~\cite{Blackman:2015pia,Varma:2018mmi,Varma:2019csw, Blackman:2017pcm,Blackman:2017dfb}.
In order to develop similar strategies for eccentric
waveform data, we have pursued a variety of options.  We now summarize the most
successful decomposition technique we have tried, while relegating some
alternatives to Appendix~\ref{app:decomp}.

\subsubsection{Decomposing the quadrupolar mode $\h_{22}$}
The complex $(2,2)$ waveform mode,
\begin{gather}
	\label{Eq:AmpPhase_22}
	\h_{22} = A_{22} ~e^{-\mathrm{i} \phi_{22}} \,,
\end{gather}
can be decomposed into an amplitude, $A_{22}$, and phase, $\phi_{22}$.
For non-precessing systems in quasicircular orbit, $A_{22}$ and $\phi_{22}$ are
slowly varying functions of time, and have therefore been used as waveform data
pieces for many modeling efforts. For eccentric
waveforms, however, both amplitude and phase show highly oscillatory
modulations on the orbital time scale (cf. Figs.~\ref{Fig:EccID39} and
\ref{Fig:AnomalyID39} for the frequency, which is a time-derivative of the
phase). This demands further decomposition of the waveforms into even simpler
data pieces. One natural solution could have been to build interpolated
functions of the local maxima and minima of $A_{22}$ and $\phi_{22}$.  The
secular trend of these functions can then be subtracted out from the original
amplitude and phase. The resulting residual amplitude and phase data may be
easier to model.  Unfortunately, as mentioned in Sec.~\ref{Sec:measuring_ecc},
finding the local maxima/minima becomes problematic near the merger.

\begin{figure*}[thb]
\includegraphics[width=0.49\textwidth]{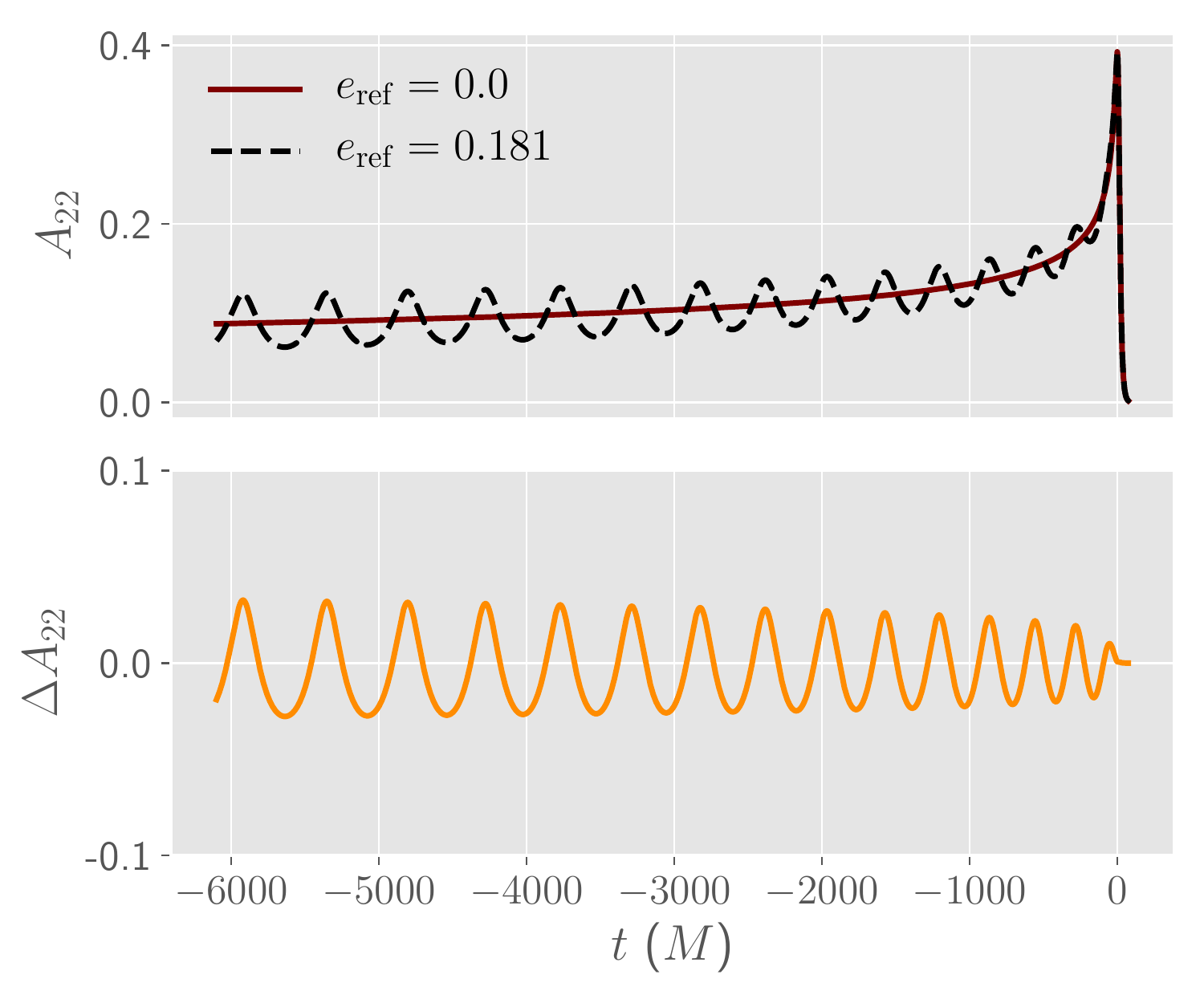}
\includegraphics[width=0.49\textwidth]{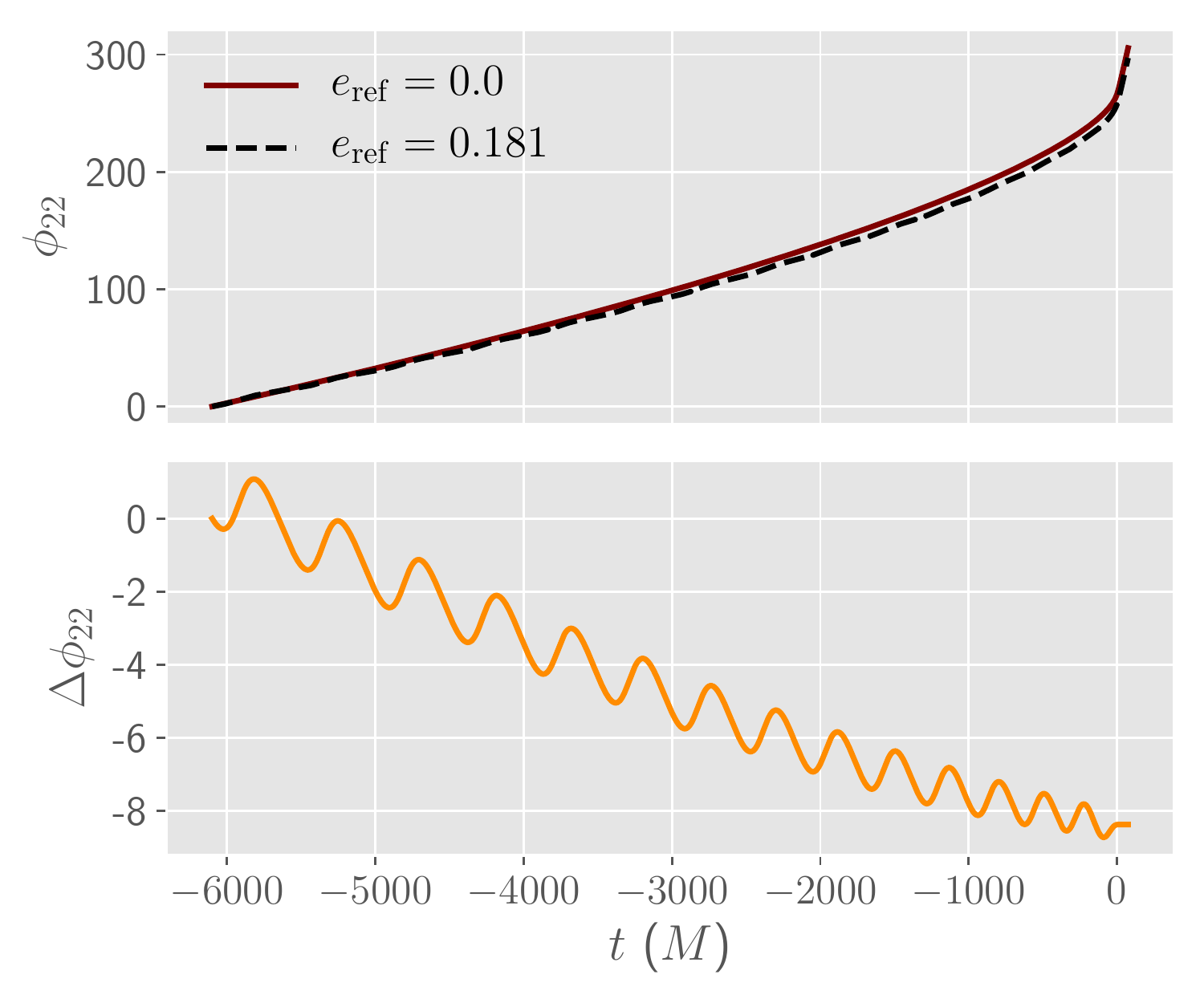}
\caption{
Example decomposition of the amplitude and phase of the $(2,2)$ mode.
\textit{Upper left:} Amplitude $A_{22}$ of the eccentric waveform SXS:BBH:2304
(with eccentricity $\ecc=0.181$) along with the amplitude $A_{22}^{0}$ of the
noneccentric waveform SXS:BBH:1155. \textit{Lower left}: The residual
amplitude $\Delta A_{22}=A_{22} - A_{22}^{0}$. \textit{Upper right:} Phase $\phi_{22}$ of the
eccentric waveform SXS:BBH:2304 and the phase $\phi_{22}^{0}$ of the
noneccentric waveform SXS:BBH:1155. \textit{Lower right}: The residual phase
$\Delta \phi_{22}=\phi_{22} - \phi_{22}^{0}$. In this work we model $\Delta A_{22}$ and
$\Delta \phi_{22}$
}
\label{Fig:ResAbsPhaseID39}
\end{figure*}

We instead follow a simpler approach whereby the amplitude and phase of a
quasicircular $q=1$, nonspinning NR waveform
(SXS:BBH:1155) is used as a proxy for the secular trend of the
amplitude and phase. We then compute the residual amplitude and phase,
\begin{gather}
	\label{Eq:AmpRes}
	\Delta A_{22} =  A_{22} - A_{22}^{0}, \\
	\label{Eq:PhaseRes}
	\Delta \phi_{22} = \phi_{22} - \phi_{22}^{0} ,
\end{gather}
where $A_{22}^{0}$ and $\phi_{22}^{0}$ are the amplitude and phase of the
noneccentric waveform, respectively, which have been aligned according the
same procedure outlined in Sec.~\ref{Sec:PostProcessing}.  In the upper-left
panel of Fig.~\ref{Fig:ResAbsPhaseID39}, we show the amplitude of an eccentric
waveform (SXS:BBH:\MySXSID{39}) along with the amplitude of its noneccentric
counterpart (SXS:BBH:1155) which traces the secular trend of the
nonmonotonically increasing eccentric amplitude.
The difference of these two amplitudes, $\Delta A_{22}$, is then plotted in the
lower-left panel. $\Delta A_{22}$ is simpler to model than $A_{22}$, as it
isolates the oscillatory component\footnote{In fact, the relatively simple
    oscillatory behavior of $\Delta A_{22}$ suggests the use of a Hilbert
    transform for further simplification.  However, we found that this does not
    improve the accuracy of our model. Such further simplifications, may become
    necessary for larger eccentricities than considered in this work, as the
modulations will be more pronounced.} of $A_{22}$. Similarly, in the right
panels of Fig.~\ref{Fig:ResAbsPhaseID39}, we show the phase evolution of the
same eccentric waveform (SXS:BBH:\MySXSID{39}), its noneccentric counterpart
(SXS:BBH:1155), and their difference $\Delta \phi_{22}$ which isolates the
oscillatory component of $\phi_{22}$. Note that noneccentric waveform data is
plentiful~\cite{Boyle:2019kee} and accurate surrogate models have been built
for noneccentric NR waveforms~\cite{Varma:2018mmi, Varma:2019csw}. So
extending the residual amplitude and phase computation to spinning,
unequal-mass systems is straightforward. For instance the surrogate model of
Ref.~\cite{Varma:2018mmi} can be used to generate $A_{22}^{0}$ and
$\phi_{22}^{0}$ for generic aligned-spin systems.

\subsubsection{Decomposing the higher order modes}
In this paper, we model the quadrupolar mode and the higher-order modes
differently. For $\h_{22}$, we model data pieces closely associated with the
amplitude and phase as described above. On the other hand,
for higher order modes, we first transform the waveform into a
co-orbital frame in which the waveform is described by a much simpler and
slowly varying function. This is done by applying a time-dependent rotation
given by the instantaneous orbital phase:
\begin{gather}
	\label{Eq:coorb_frame}
	\hlm^C = \hlm ~e^{\mathrm{i} m \phi_\mathrm{orb}}, \\
	\phi_\mathrm{orb} = \frac{\phi_{22}}{2},
	\label{Eq:orb_phase}
\end{gather}
where $\phi_{22}$ is the  phase of the $(2,2)$ mode (cf.
Eq.~\eqref{Eq:AmpPhase_22}), $\phi_\mathrm{orb}$ is the orbital phase, and
$\hlm^C$ represents the complex modes in the co-orbital frame.

\begin{figure}[thb]
\includegraphics[width=\columnwidth]{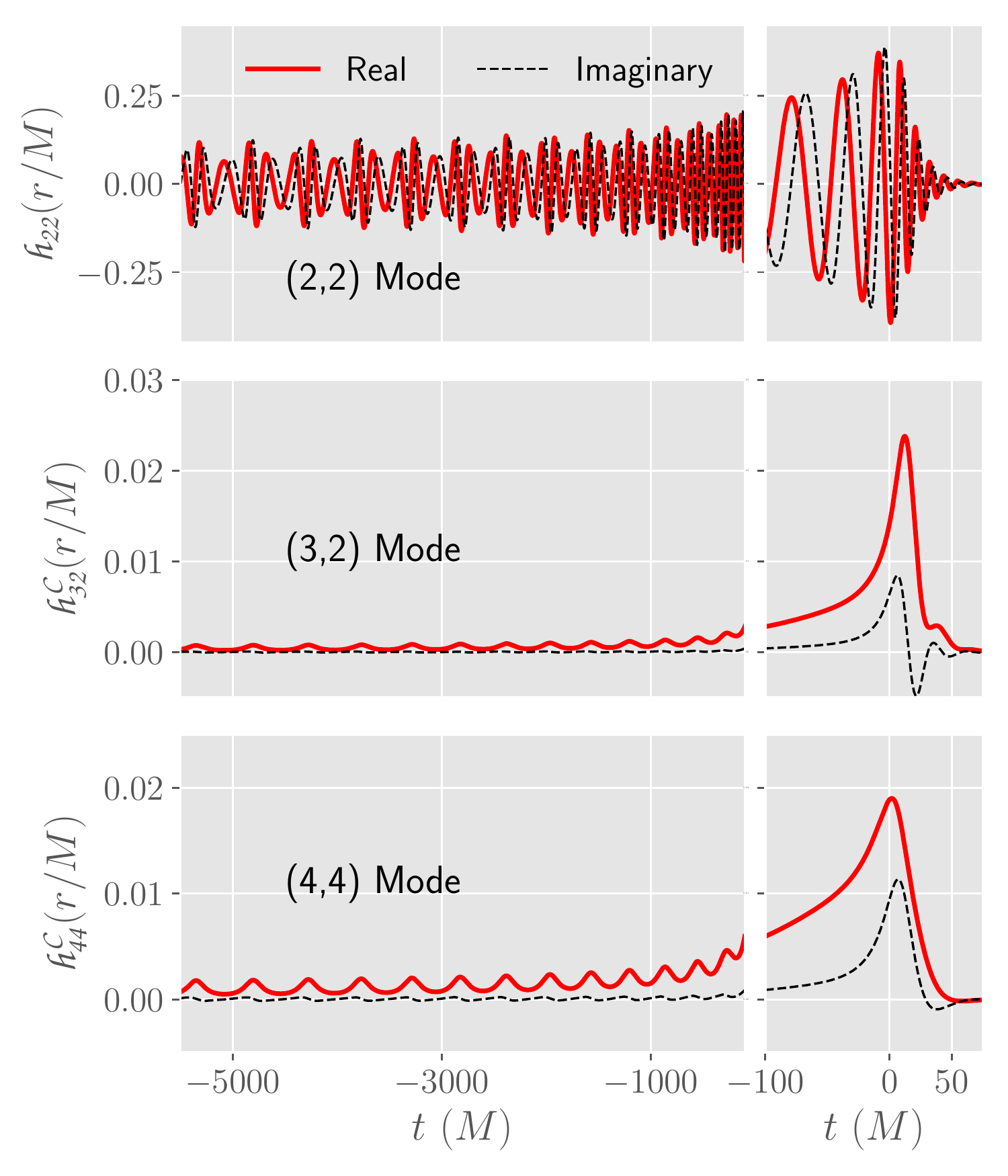}
\caption{
The waveform modes for NR Simulation SXS:BBH:2304 ($\ecc=0.181$) are
shown. The top panel shows the dominant $(2,2)$ mode in the inertial frame.
Two higher-order modes $(3,2)$ and $(4,4)$ in the co-orbital frame are shown in
the middle and lower panels respectively. The waveform is aligned such that the
peak of the amplitude occurs at $t=0$ and the orbital phase is zero at $t_{\rm
ref}=-5500M$.}
\label{Fig:EccID39waveform}
\end{figure}

We use the real and imaginary parts of $\hlm^C$ as our waveform data
pieces for the nonquadrupole modes. As shown in Fig.~\ref{Fig:EccID39waveform},
the $\hlm^C$ data have less structure, making them easier to model. We find
that using quasicircular $\hlm^C$ to subtract off the secular trend does not
provide any modeling advantage. We, therefore, model the real and imaginary
parts of $\hlm^C$ without any further data decomposition.

\subsubsection{Summary of waveform data pieces}
\label{Sec:DataPieceSummary}
To summarize, the full set of waveform data pieces we model is as follows:  $\Delta A_{22}$,
$\Delta \phi_{22}$ for the (2,2) mode, and real and imaginary parts of $\hlm^C$
for the (3,2) and (4,4) modes.

\subsection{Building the waveform model}
\label{Sec:model_building}
We decompose the inertial frame waveform data into many
waveform data pieces as summarized in Sec.~\ref{Sec:DataPieceSummary}. For each
of these data pieces, we now build a surrogate model using reduced basis,
empirical interpolation, and parametric fits
across the parameter space.  The detailed procedure is outlined in
Refs.~\cite{Blackman:2017dfb,Field:2013cfa}, which we only briefly describe
here.

For each waveform data piece, we employ a greedy algorithm to construct a
reduced basis \cite{Field:2011mf} such that the projection errors (cf. Eq. (5)
of Ref. \cite{Blackman:2017dfb}) for the entire data set onto this basis are
below a given tolerance. We use a basis tolerance of $10^{-2}$ radians for
$\Delta \phi_{22}$ , $1.5\times10^{-3}$ for
$\Delta A_{22}$ and $2\times10^{-5}$ for the real part of $\h_{32}^C$.
For all other data pieces, basis tolerance is set to $5\times10^{-5}$.

These choices are made so that we include sufficient number of basis functions
for each data piece [9 for $\Delta A_{22}$, 12 for $\Delta \phi_{22}$, 7 (5)
for the real (imaginary) part of $\h_{32}^C$ and 10 (6) for the real
(imaginary) part of $\h_{44}^C$] to capture the underlying physical features in
the simulations while avoiding over fitting. We perform additional visual
inspection of the basis functions to ensure that they are not noisy in which
case modeling accuracy can become comprised (cf. Appendix B of
Ref.~\cite{Blackman:2017dfb}).

The next step is to construct an empirical interpolant in time using a greedy
algorithm which picks the most representative time nodes~\cite{Maday:2009,
chaturantabut2010nonlinear, Field:2013cfa, Canizares:2014fya}.  The number of
the time nodes for each data piece is equal to the number of basis functions
used. The final surrogate-building step is to construct
parametric fits for each data piece at each of the empirical time nodes across
the two-dimensional parameter space $\{ \ecc, \meanano \}$. We do this using
the Gaussian process regression (GPR) fitting method as described in
Refs.~\cite{Taylor:2020bmj, Varma:2018aht}.

\begin{figure}[thb]
	\includegraphics[width=\columnwidth]{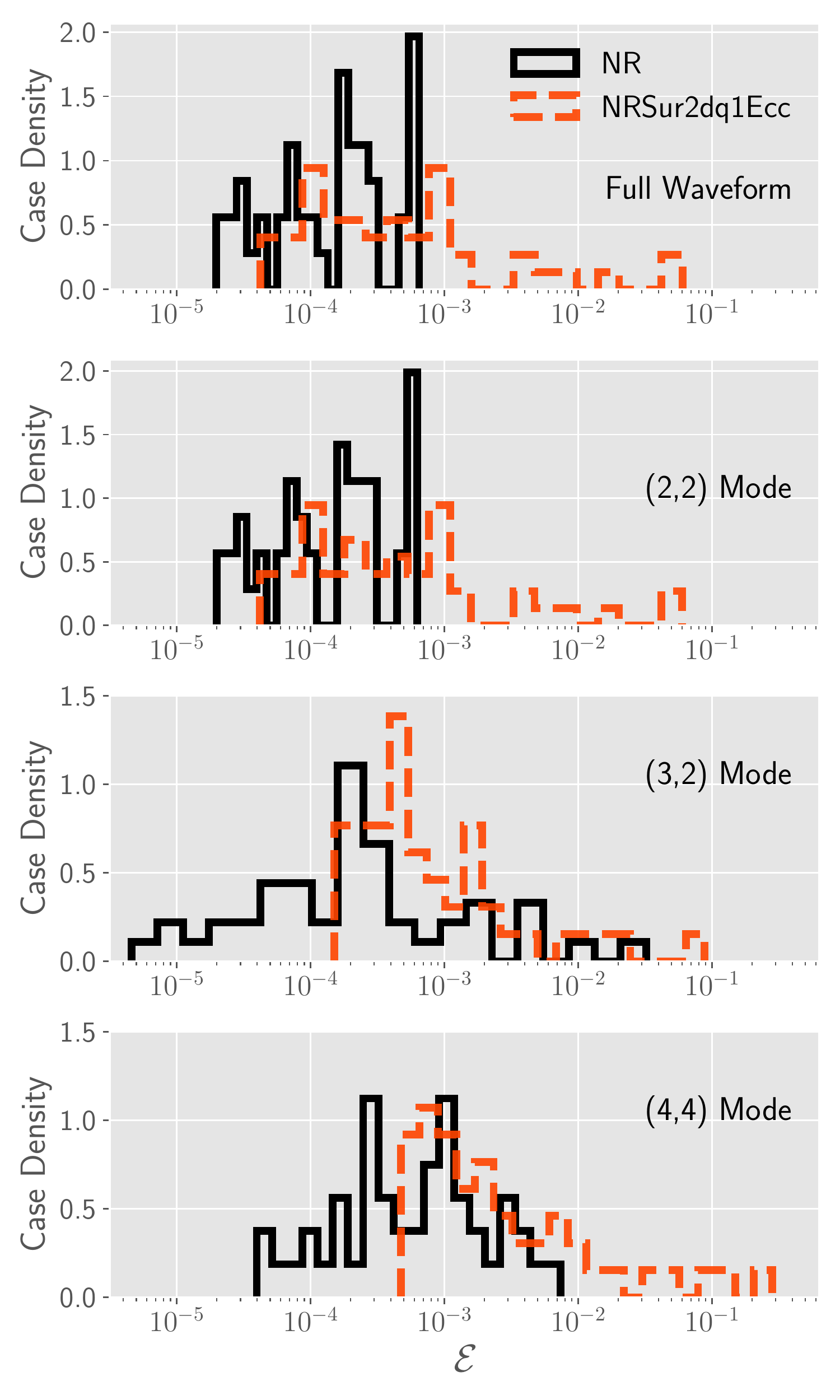}
	\caption{
		Time-domain leave-one-out errors $\mathcal{E}$, defined in Eq.~\eqref{Eq:cost_function}, for
		the full waveform as well as the individual modes considered in the
		model. For comparison, we also show the NR error between the
		two highest resolutions. The largest errors are found near the 
        parameter domain's boundary where the trial surrogate, built as part of
        the cross-validation study, is extrapolating.
	}
	\label{Fig:TD_L2modebymode}
\end{figure}

\subsection{Evaluating the waveform surrogate}
\label{Sec:Surrogate_evaluation}
To evaluate the \texttt{NRSur2dq1Ecc} surrogate model, we provide the
eccentricity $\ecc$ and mean anomaly $\meanano$ as inputs.  We then evaluate
the parametric fits for each waveform
data pieces at each time node. Next, the empirical
interpolant is used to reconstruct the full waveform data pieces (cf.
Sec.~\ref{Sec:DataPieceSummary}).

We compute the amplitude and phase of the $(2,2)$ mode,
\begin{gather}
	A_{22}^S = \Delta A_{22}^S + A_{22}^{0}, \\
	\phi_{22}^S = \Delta \phi_{22}^S + \phi_{22}^{0} ,
\end{gather}
where $\Delta A_{22}^S \approx \Delta A_{22}$  and $\Delta \phi_{22}^S \approx
\Delta \phi_{22}$ are the surrogate models for $\Delta A_{22}$ and $\Delta
\phi_{22}$ respectively while $A_{22}^{0}$ and $\phi_{22}^{0}$ are the
amplitude and phase of the quasicircular NR waveform used in the
decompositions~[cf. Eqs.~(\ref{Eq:AmpRes}-\ref{Eq:PhaseRes})]. We
obtain the (2,2) mode complex strain as $\h^S_{22} = A_{22}^S ~ e^{-\mathrm{i}
\phi_{22}^S}$.

For the nonquadrupole modes, we similarly evaluate the
surrogate models for the real and imaginary parts of the co-orbital frame
waveform data pieces $\hlm^{C,S} \approx \hlm^C$ and treat it as $\hlm^C$.
Finally, we use Eqs.~(\ref{Eq:AmpPhase_22}), (\ref{Eq:coorb_frame}), and
(\ref{Eq:orb_phase}) to obtain the surrogate prediction for the inertial frame
strain $\hlm^S$ for these modes.

\begin{figure*}[thb]
\includegraphics[width=0.49\textwidth]{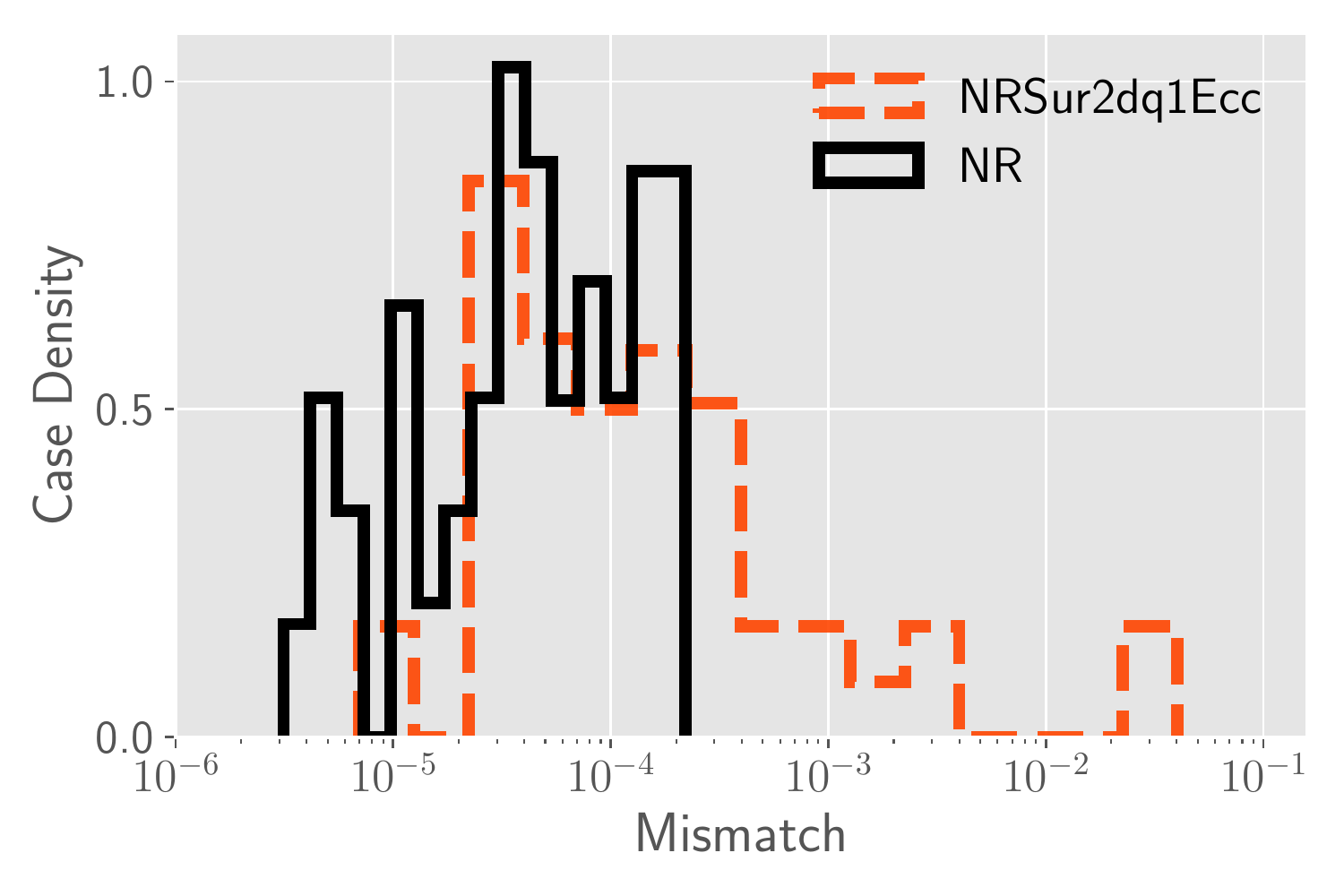}
\includegraphics[width=0.49\textwidth]{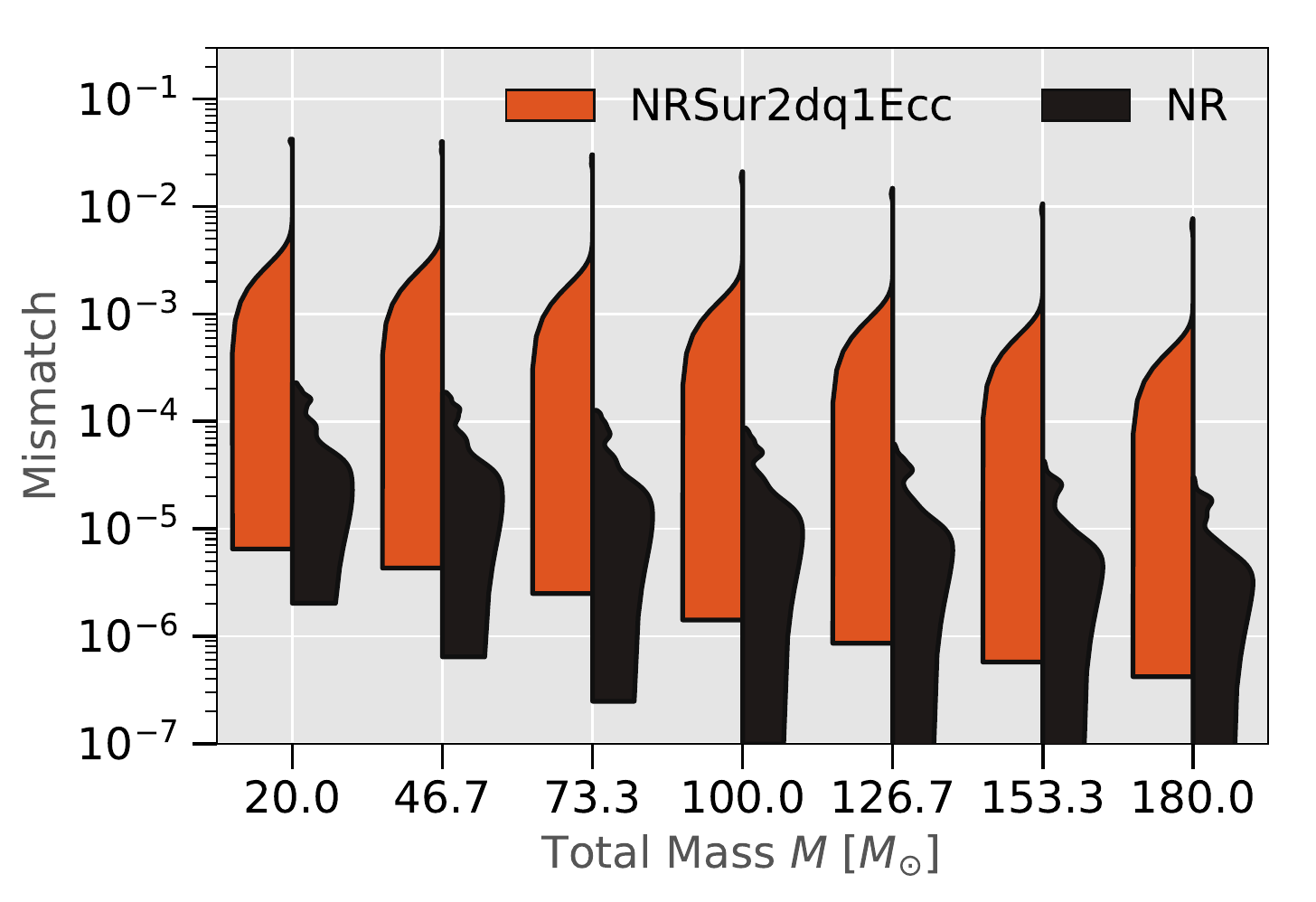}
\caption{
\textit{Left panel:} Flat noise mismatch between the \texttt{NRSur2dq1Ecc}
model (following the leave-one-out validation procedure) and the
highest-resolution NR waveform data.  For comparison, we also show the NR
resolution error, obtained by comparing the two highest available resolutions.
\textit{Right panel:} \texttt{NRSur2dq1Ecc} (validation) mismatches computed
using the advanced LIGO design sensitivity noise curve, as a function of the
total mass of the binary.  For comparison, we also show the NR mismatches. For
each mass, the distribution of mismatches are shown as a smoothed vertical
histogram (or a violin). The histograms are normalized so that all violins have
equal width.  The largest errors are found near the parameter domain's boundary
where the trial surrogate, built as part of the cross-validation study, is
extrapolating. 
}
\label{Fig:mismatch}
\end{figure*}

\subsection{Building the remnant surrogate}
\label{Sec:remnant_building}
In addition to the waveform model, we also construct the \textit{first}
model for the remnant quantities of eccentric BBHs. The new remnant model,
\texttt{NRSur2dq1EccRemnant}, predicts the final mass $m_f$ and the component
of the final spin, $\chi_{fz}$, along the orbital angular momentum direction.
The remnant model takes eccentricity $\ecc$ and mean anomaly $\meanano$ as its
inputs and maps to the final state of the binary. The final mass and spin fits
are also constructed using the GPR fitting method as described in
Refs.~\cite{Taylor:2020bmj, Varma:2018aht}.

\section{Results}
\label{sec:results}

In this section we demonstrate the accuracy of \texttt{NRSur2dq1Ecc} and
\texttt{NRSur2dq1EccRemnant} by comparing against the eccentric NR simulations
described in Sec.~\ref{Sec:NR_Runs}. We do this by performing a
leave-one-out cross-validation study. In this study, we
hold out one NR waveform from the training set and build a 
trial surrogate from the remaining 46 eccentric NR waveforms. 
We then evaluate the trial surrogate at the parameter value corresponding
to the held out data, and compare its prediction with the highest-resolution NR waveform.
We refer to the errors obtained by comparing against the left-out NR waveforms
as cross-validation errors. These represent conservative error estimates for
the surrogate models against NR. Since we have 47 eccentric NR waveforms, we
build 47 trial surrogates for each error study.
We compare these errors to the NR resolution
error, estimated by comparing the two highest available NR simulations.

\subsection{NRSur2dq1Ecc errors}
\subsubsection{Time domain error without time/phase optimization}
In order to quantify the accuracy of \texttt{NRSur2dq1Ecc}, we first compute
the normalized $L_{2}$-norm between the NR data and surrogate approximation
\begin{gather}
	\label{Eq:cost_function}
	\mathcal{E} [\mathpzc{h} , \mathpzc{\tilde{h}}] =
	\frac{1}{2} \frac{\sum_{\ell,m}\int_{t_1}^{t_2}|
		\mathpzc{h}_{\ell m}(t) - \mathpzc{\tilde{h}}_{\ell m}(t)|^2 dt}
	{\sum_{\ell, m} \int_{t_1}^{t_2} |\mathpzc{h}_{lm}(t)|^2 dt},
\end{gather}
where $\h(t)$ and $\tilde{\h}(t)$ correspond to the complex strain for NR and
\texttt{NRSur2dq1Ecc} waveforms, respectively. Here,
$t_1$ and $t_2$ denote the start and end
of the waveform data.  As the NR waveforms are already aligned in time and
phase, the surrogate reproduces this alignment.  Therefore, we compute the
time-domain error $\mathcal{E}$ without any further time/phase shifts.

In Fig.~\ref{Fig:TD_L2modebymode}, we report both the full waveform and
individual mode errors for \texttt{NRSur2dq1Ecc}. For comparison, we also show
the NR resolution errors. When computing the full waveform error we use all
modes included in the surrogate model $(\ell,m)=(2,2),(3,2),(4,4)$ in
Eq.~\eqref{Eq:cost_function}.  To compute errors for individual modes, we
restrict the sum in Eq.~\eqref{Eq:cost_function} to only the mode of interest.
The \texttt{NRSur2dq1Ecc} errors are comparable to the NR errors in
Fig.~\ref{Fig:TD_L2modebymode}.

However, we find that the surrogate errors have an extended tail 
around two orders of magnitude larger than the largest NR mismatch.
While this could imply over-fitting, we find that highest mismatches correspond
to the parameter space adjacent to the higher eccentricity $\ecc$ boundary
where only few (to none) training waveforms are used.  As will be discussed in
Sec.~\ref{Sec:Freqdom_mismatch}, the sparsely sampled region of the training
domain around $(\ecc=0.2, \meanano\lesssim2)$ leads to this extended high-error
tail in Fig.~\ref{Fig:TD_L2modebymode}.

We further note in Fig.~\ref{Fig:TD_L2modebymode} that the highest error in
each mode corresponds to the same point in the parameter space indicating
consistency in our modeling.  Furthermore, as we only deal with mass ratio
$q=1$ waveforms, the contribution of the higher modes are expected to be
negligible compared to the dominant $(2,2)$ mode (see for example,
Ref.~\cite{Varma:2016dnf}). Therefore, even though the $(3,3)$ and $(4,4)$
modes have larger relative errors compared to the $(2,2)$ mode, their
contribution to the total error is much smaller. This can be verified by
comparing the full waveform errors to the (2,2) mode errors in
Fig.~\ref{Fig:TD_L2modebymode}.

\begin{figure*}[t]
\includegraphics[width=1.0\textwidth]{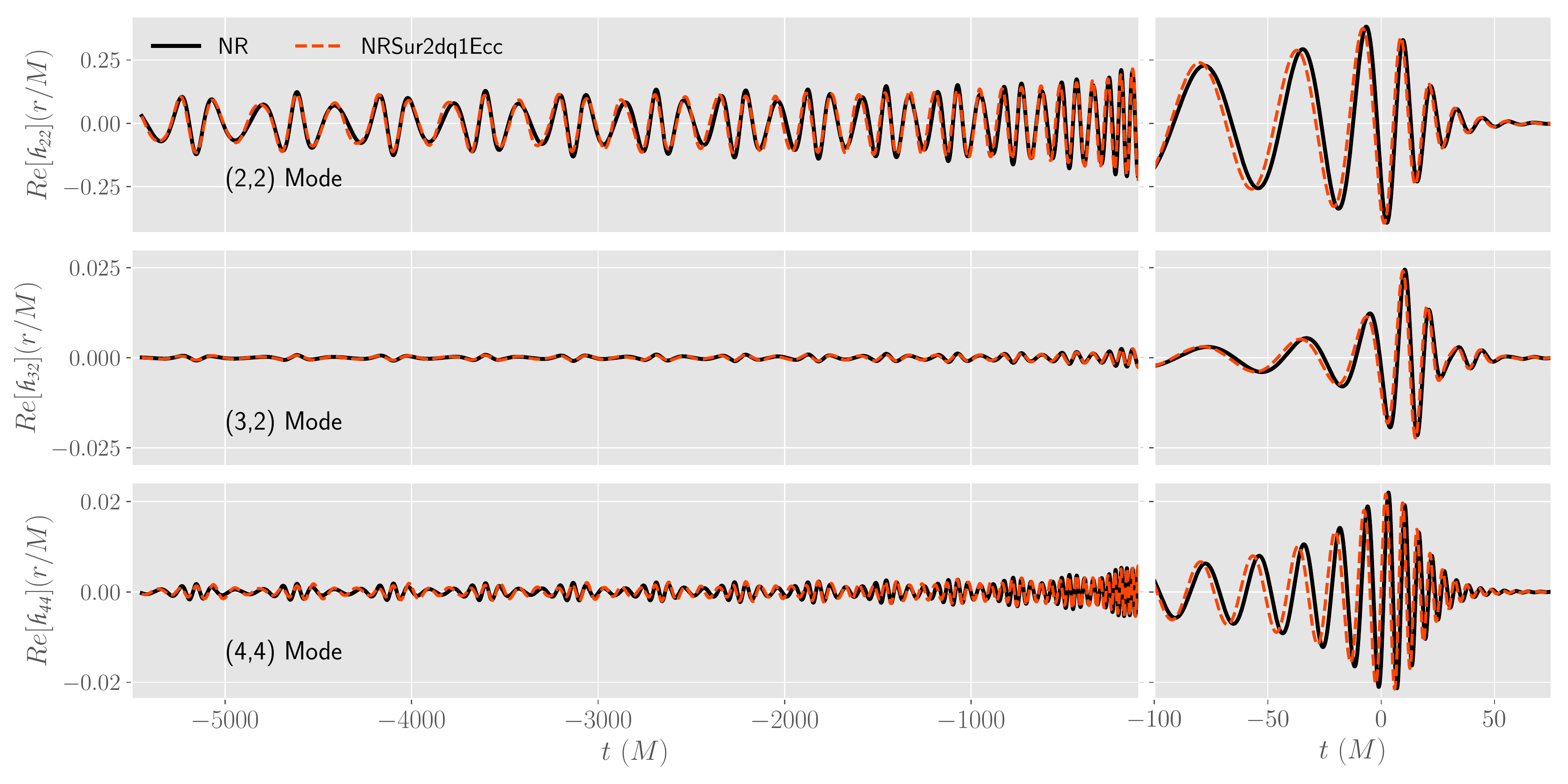}
\caption{
Real part of the waveform modes for the case that results in the largest flat
noise mismatch ($\sim0.04$) for \texttt{NRSur2dq1Ecc} (red dashed line)
in the left panel of Fig~\ref{Fig:mismatch}. We also show the corresponding NR
waveform, SXS:BBH:2308 (black solid line).  The parameter values for this
waveform are: $\ecc=0.176$ and $\meanano=2.51$.  Note that this plot is
generated using a trial surrogate that was not trained using this NR waveform
data.
}
\label{Fig:secondworsterror}
\end{figure*}

\subsubsection{Frequency domain mismatch with time/phase optimization}
\label{Sec:Freqdom_mismatch}
In this section, we estimate leave-one-out cross-validation errors
by computing mismatches between the NR
waveform and the trial surrogate waveform in the frequency domain.
The frequency domain mismatch between two waveforms, $\h_1$ and $\h_2$ is
defined as:
\begin{gather}
	\left<\h_1, \h_2\right> = 4 \mathrm{Re}
	\int_{f_{\mathrm{min}}}^{f_{\mathrm{max}}}
	\frac{\tilde{\h}_1 (f) \tilde{\h}_2^* (f) }{S_n (f)} df,
	\label{Eq:freq_domain_Mismatch}
\end{gather}
where $\tilde{\h}(f)$ indicates the Fourier transform of the complex strain
$\h(t)$, $^*$ indicates a complex conjugation, $\mathrm{Re}$ indicates the real
part, and $S_n(f)$ is the one-sided power spectral density of a GW detector.

Before transforming the time domain waveform to the frequency domain, we first
taper the time domain waveform using a Planck window~\cite{McKechan:2010kp},
and then zero-pad to the nearest power of two. The tapering at the start of the
waveform is done over $1.5$ cycles of the $(2,2)$ mode. The tapering at the end
is done over the last $20M$. 
Once we obtain the frequency domain waveforms, we
compute mismatches following the procedure described in Appendix D of
Ref.~\cite{Blackman:2017dfb}. The mismatches are optimized over shifts in time,
polarization angle, and initial orbital phase. We compute the mismatches at 37
points uniformly distributed on the sky of the source frame, and use all
available modes for the surrogate model. 

We consider a flat noise curve $S_n(f)=1$ as well as the Advanced-LIGO design
sensitivity Zero-Detuned-HighP noise curve from
Ref.~\cite{aLIGODesignNoiseCurve}.  We take $f_{\mathrm{min}}$ to be the
frequency of the $(2,2)$ mode at the end of the initial tapering window while
$f_{\mathrm{max}}$ is set at $4f^{\rm peak}_{22}$, where $f^{\rm peak}_{22}$ is
the frequency of the $(2,2)$ mode at its peak.  This ensures that the peak
frequencies of all modes considered in our model are captured well, and we have
confirmed that our mismatch values do not change for larger values of
$f_{\mathrm{max}}$.  Note that when computing mismatches using Advanced LIGO
noise curve, for masses below $\sim70M_{\odot}$, $f_{\mathrm{min}}$ is greater
than $20$Hz, meaning that the signal starts within the detector sensitivity
band.

The mismatches computed using the flat noise curve are shown in the left panel
of Fig.~\ref{Fig:mismatch}. The histograms include mismatches for all 47 NR
waveforms and source-frame sky locations. We find that the typical surrogate
mismatches are $10^{-5}-10^{-3}$, which are comparable to but larger than the
NR errors. As an example, Fig.~\ref{Fig:secondworsterror} shows the surrogate
and NR waveforms for the case that leads to the largest mismatch in the left
panel of Fig.~\ref{Fig:mismatch}.

In Fig.~\ref{Fig:MM_heat_map}, we show the dependence of the mismatches on the
parameter space. It can be easily recognized that the surrogate yields largest
errors at and around $(\ecc=0.2, \meanano\lesssim2)$ where the training grid
becomes sparse. Further, when these sparse data points themselves are left out
when computing the cross-validation errors, the surrogate is effectively
extrapolating in parameter space. This indicates that the surrogate accuracy
could be improved by adding new NR simulations in this high-eccentricity
region.  However, achieving target values of $\ecc$ and $\meanano$ has proven
difficult. We return to this issue in the conclusions.

The right panel of Fig.~\ref{Fig:mismatch} shows the mismatches computed using
advanced LIGO design sensitivity noise curve~\cite{aLIGODesignNoiseCurve} for
different total masses $M$ of the binary. For each $M$, we compute the
mismatches for all 47 NR waveforms and source-frame sky locations and show the
distribution of mismatches using vertical histograms known as violin plots.
Over the mass range $20-180 M_{\odot}$, the surrogate mismatches are at the
level of $\sim 10^{-4}-10^{-3}$ but with an extending tail as before. However,
we note that these errors are typically smaller than the mismatches for other
eccentric waveform models~\cite{Huerta:2017kez, Chen:2020lzc,
Chiaramello:2020ehz}.

\begin{figure}[htb]
\includegraphics[width=\columnwidth]{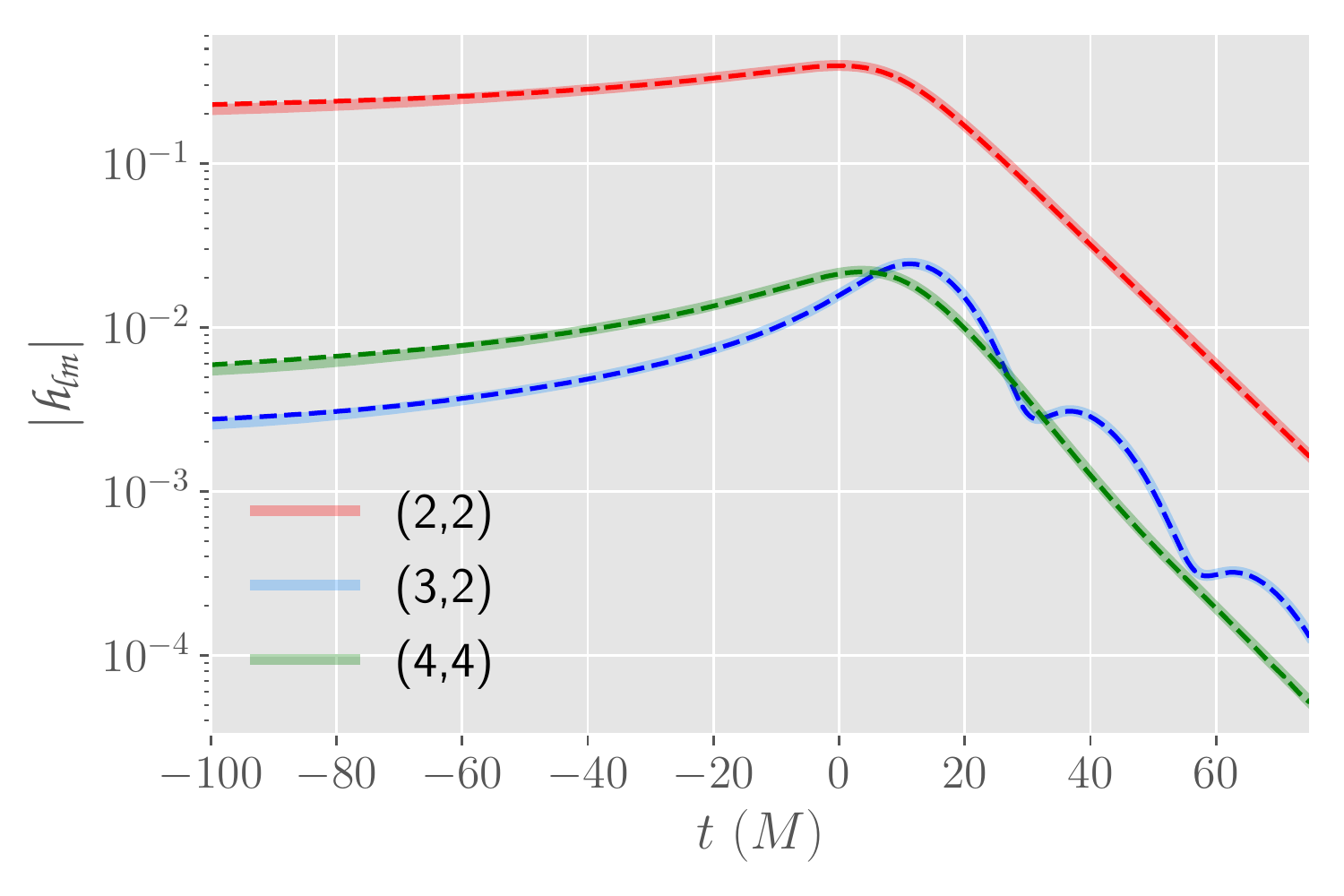}
\caption{
The absolute values of different spherical-harmonic modes are shown as dashed
(solid) curves for the surrogate (NR) for SXS:BBH:2308, for which the
surrogate produces largest flat noise mismatch ($\sim0.04$). The
parameter values for this waveform are: $\ecc=0.176$ and $\meanano=2.51$. Mode
mixing for the $(3,2)$ mode is clearly seen in the ringdown signal of the NR
waveform and is accurately reproduced by the surrogate.
}
\label{Fig:modemixing}
\end{figure}

\begin{figure*}[htb]
\includegraphics[width=1.0\textwidth]{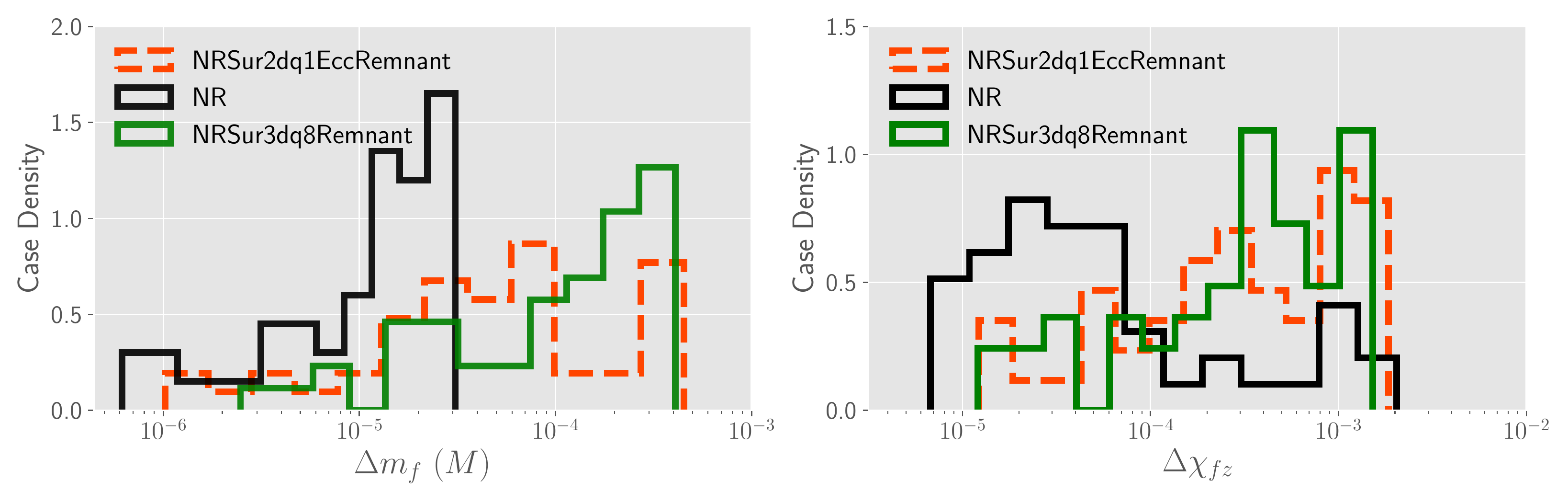}
\caption{
Leave-one-out error histograms of \texttt{NRSur2dq1EccRemnant} (red) for the
remnant mass $m_f$ (left) and remnant spin $\chi_{fz}$ (right). 
For comparison we plot the NR errors (black), estimated by comparing the two
highest resolution NR simulations, and errors for the noneccentric model
\texttt{NRSur3dq8cRemnant} (green).
}
\label{Fig:remnant}
\end{figure*}

\subsection{Mode mixing}

NR waveforms are extracted as spin-weighted spherical harmonic modes
\cite{Newman:1966ub,Goldberg:1966uu}.  However, during the ringdown, the system
can be considered a single Kerr black hole perturbed by quasinormal modes;
perturbation theory tells us that the angular eigenfunctions for these modes
are the spin-weighted \emph{spheroidal}
harmonics~\cite{Teukolsky:1973ha,Teukolsky:1972my}.  A spherical harmonic mode
$\hlm$ can be written as a linear combination of all spheroidal harmonic modes
with the same $m$ index. During the ringdown, each (spheroidal-harmonic)
quasinormal mode decays exponentially in time, but each spherical-harmonic mode
has a more complicated behavior because it is a superposition of multiple
spheroidal-harmonic modes (of the same $m$ index) with different decay rates.
This more complicated behavior is referred to as mode mixing, since power flows
between different spherical-harmonic modes~\cite{Berti:2014fga}.  This mixing
is particularly evident in the $(3,2)$ mode as significant power of the
dominant $(2,2)$ spherical-harmonic mode can leak into the $(3,2)$
spherical-harmonic mode.
As the surrogate accurately reproduces the spherical harmonic modes from the NR
simulations, it is also expected to capture the effect of mode mixing without
any additional effort~\cite{Varma:2018mmi}. We demonstrate this for an example
case in Fig.~\ref{Fig:modemixing} where we plot the amplitude of individual
modes of the waveform during the ringdown.  We show that the mode mixing in the
$(3,2)$ mode is effectively recovered by the surrogate model.

\subsection{\texttt{NRSur2dq1EccRemnant} errors}
In addition to the waveform surrogate, we also build a remnant surrogate model,
\texttt{NRSur2dq1EccRemnant}, that predicts the mass and spin of the final BH
left behind after the merger. This is the first such model for eccentric BBHs
(but see e.g. Refs.~\cite{Sopuerta:2006et, Sperhake:2019wwo}).
Figure~\ref{Fig:remnant} shows the cross-validation errors of
\texttt{NRSur2dq1EccRemnant} in predicting the remnant mass and spin. We find
that \texttt{NRSur2dq1EccRemnant} can predict the final mass and spin with an
accuracy of $\lesssim 5\times10^{-4} M$ and $\lesssim 2\times 10^{-3}$
respectively. 
We further compute the errors for a noneccentric remnant model,
\texttt{NRSur3dq8Remnant}~\cite{Varma:2018aht}, when compared against the same
eccentric NR simulations, finding that errors in \texttt{NRSur3dq8Remnant} are
comparable with \texttt{NRSur2dq1EccRemnant} errors. This suggests that
noneccentric remnant models may be sufficient for equal-mass nonspinning
binaries with eccentricities $\ecc\leq0.2$. However, we expect such models
to disagree with eccentric simulations in the more general case of
unequal-mass, spinning binaries (see for e.g.
Ref.~\cite{Ramos-Buades:2019uvh}).

\begin{figure}[thb]
\includegraphics[width=\columnwidth]{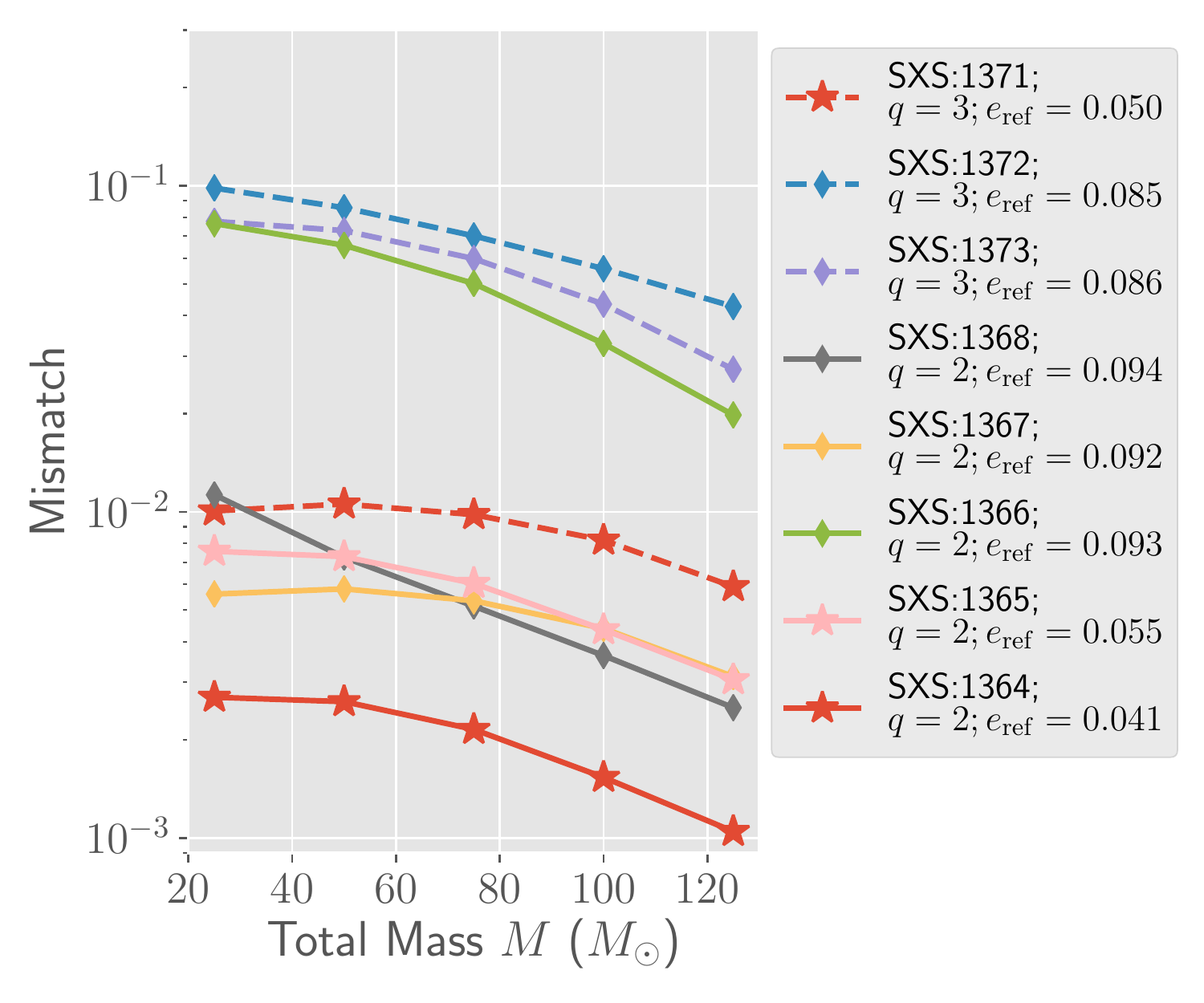}
\caption{
Mismatches against NR for the \texttt{NRSur2dq1Ecc+} model (a simple extension
of \texttt{NRSur2dq1Ecc}) when the surrogate is evaluated beyond its training
parameter range ($q=1$). The mismatches are shown as a function of the binary
total mass $M$ (at $\iota=\pi/3$, $\varphi_0=0.0$), and are computed using
the advanced LIGO design sensitivity noise curve. We show mismatches for $q=2$
($q=3$) as solid lines (dashed lines). We use star markers to denote waveforms
with $\ecc$ smaller than $\sim 0.05$ and diamond markers for the rest. All
eccentricity values are computed at a reference time of $t_{\rm ref}=-2000 M$.
}
\label{Fig:extrapolation}
\end{figure}

\begin{figure*}[t]
\includegraphics[width=1.0\textwidth]{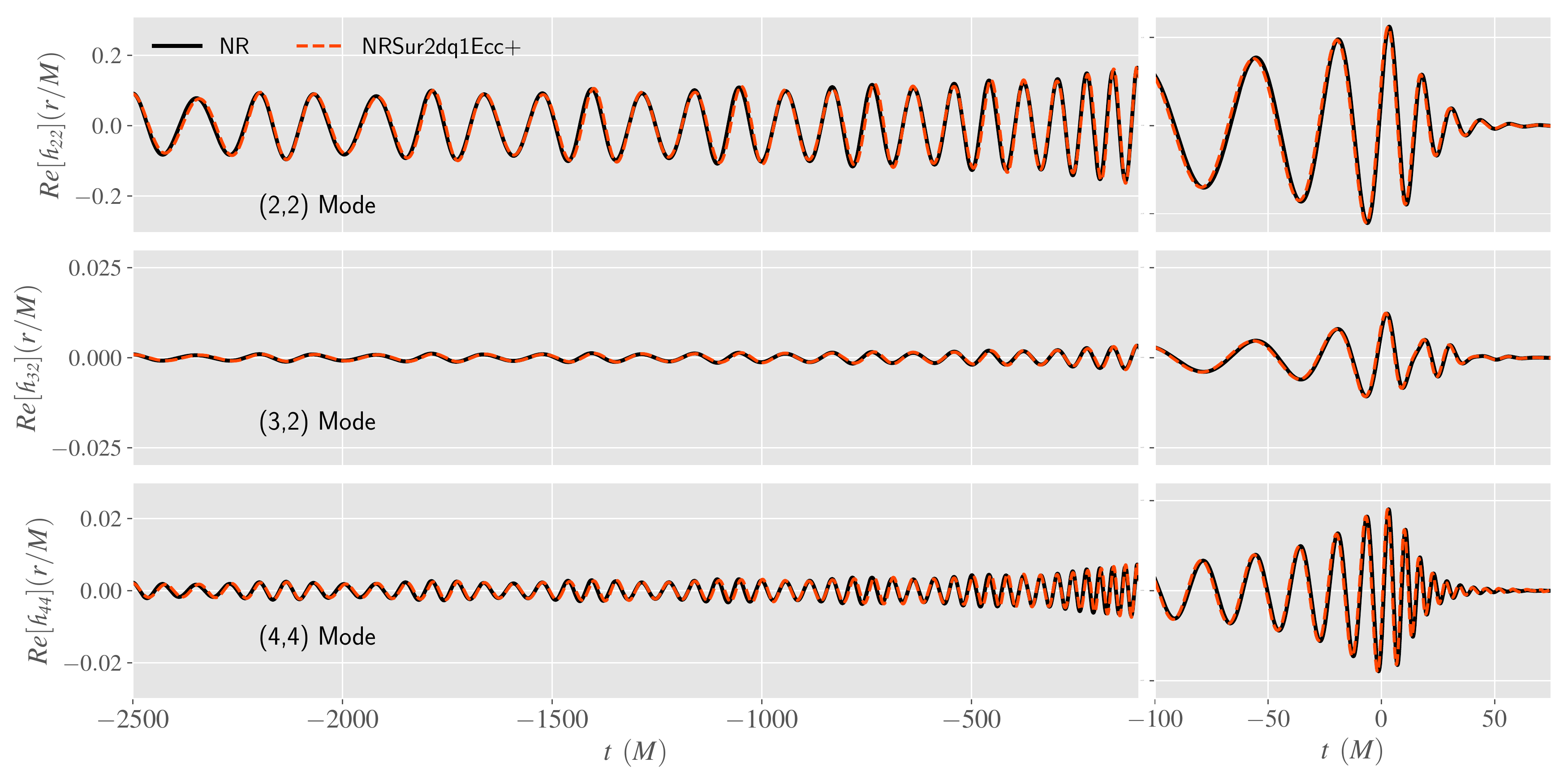}
\caption{
We show the \texttt{NRSur2dq1Ecc+} prediction (red dashed line) beyond training
range ($q=1$) of the surrogate for the case that results in the largest
mismatch (Fig.~\ref{Fig:extrapolation}) in the region defined by $\ecc$ (at
$t_{\rm ref}=-2000M$) smaller than $\sim 0.05$.  We also show the corresponding
NR waveform SXS:BBH:1371 (black solid line). The parameters for this waveform
are: $q=3$, $\ecc=0.050$ and $\meanano=2.45$ (at $t_{\rm ref}=-2000M$).  
}
\label{Fig:sxs1371}
\end{figure*}

\subsection{Extending \texttt{NRSur2dq1Ecc} to comparable mass systems}
We now assess the performance of \texttt{NRSur2dq1Ecc} when evaluated beyond
its training parameter range ($q=1$).  To generate surrogate predictions at a
given $(q, \ecc, \meanano)$, we first evaluate \texttt{NRSur2dq1Ecc} at
$(q\!=\!1, \ecc, \meanano)$ and refer to the output as $\h^S_{lm}(q\!=\!1,
\ecc, \meanano)$. We then evaluate the noneccentric surrogate model
\texttt{NRHybSur3dq8}~\cite{Varma:2018mmi} at the given mass ratio $q$ and mass
ratio $q=1$, and refer to the output as $\hlm^0(q)$ and $\hlm^0(q=1)$. We then
compute the difference in amplitude and phase between $\h^S_{lm}(q\!=\!1, \ecc,
\meanano)$ and $\hlm^0(q=1)$:
\begin{align}
& \Delta A^S_{lm}(q\!=\!1, \ecc, \meanano) \nonumber \\
& ~~~~~~~ = A^S_{lm}(q\!=\!1, \ecc, \meanano) - A^0_{lm}(q\!=\!1) \,,
\end{align}
\begin{align}
    & \Delta \phi^S_{lm}(q\!=\!1, \ecc, \meanano) \nonumber \\
    & ~~~~~~~ = \phi^S_{lm}(q\!=\!1, \ecc, \meanano) - \phi^0_{lm}(q\!=\!1).
\end{align}
Even though these amplitude and phase differences are computed at $q=1$, we
treat them as a proxy for the modulations due to eccentricity at any $q$. We
then add these modulations to the amplitude and phase of $\hlm^0(q)$, the
noneccentric surrogate model evaluated at the given $q$, to get the full
amplitude and phase:
\begin{align}
& A^S_{lm}(q, \ecc, \meanano)     \nonumber \\
& ~~~~~~~~~ = \Delta A^S_{lm}(q\!=\!1, \ecc, \meanano) + A^0_{lm}(q) \,,
\end{align}
\begin{align}
& \phi^S_{lm}(q, \ecc, \meanano)     \nonumber \\
& ~~~~~~~~~ = \Delta \phi^S_{lm}(q\!=\!1, \ecc, \meanano) + \phi^0_{lm}(q).
\end{align}
The final surrogate prediction, which we view as a new, simple model
\texttt{NRSur2dq1Ecc+}, is then:
\begin{equation}
\hlm^S(q, \ecc, \meanano) = A^S_{lm}(q, \ecc, \meanano)e^{-\mathrm{i}\phi^S_{lm}(q, \ecc, \meanano)}.
\end{equation}

\begin{figure*}[t]
\includegraphics[width=1.0\textwidth]{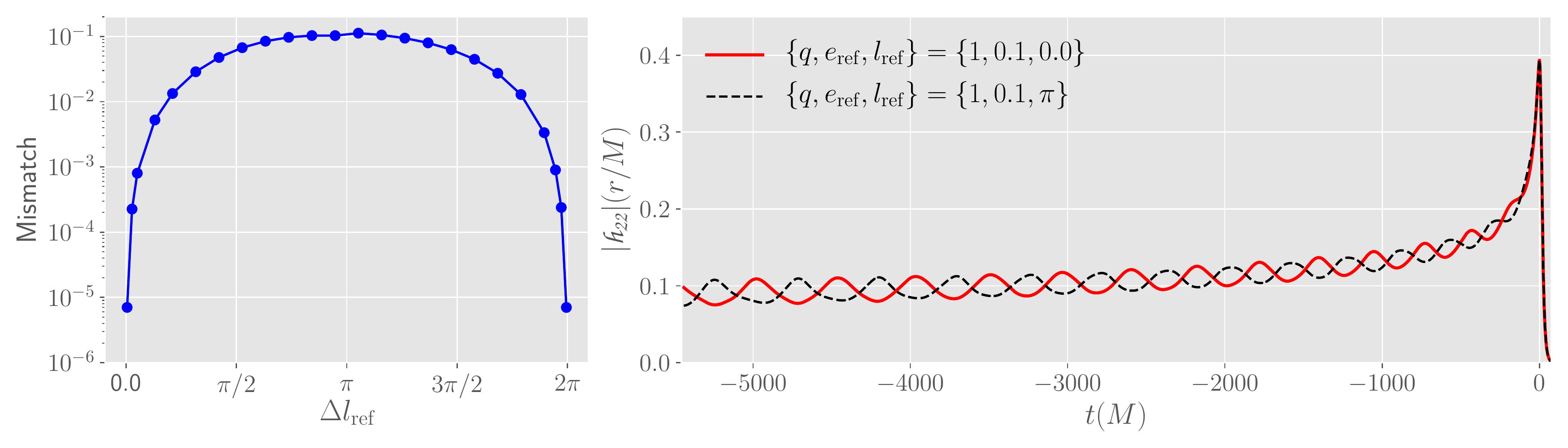}
\caption{
Importance of mean anomaly for waveform modeling and data analysis.
\textit{Left panel:} Flat noise mismatch (optimized over time, phase and
polarization angle shifts) between \texttt{NRSur2dq1Ecc} predictions with
$\meanano=0.0$ and $\meanano=\Delta\meanano$, at fixed $q=1$ and $\ecc=0.1$.
While the mismatch, as expected, is $\sim0$ for $\Delta\meanano=0.0$ and
$\Delta\meanano=2\pi$, it reaches values $\sim0.1$ near $\Delta\meanano=\pi$.
\textit{Right panel:} The $(2,2)$ amplitude of the waveforms leading to the
maximum mismatch, i.e.  $\meanano=0$ and $\meanano=\pi$. These differences
cannot be accounted for by a time or phase shift, therefore, mean anomaly is an
important parameter to include for waveform modeling and data analysis of
eccentric binaries.
}
\label{Fig:anomaly_mismatch}
\end{figure*}

To assess the accuracy of \texttt{NRSur2dq1Ecc+}
we compare against eight publicly available eccentric NR simulations with
$q=2$ and $q=3$ \cite{Hinder:2017sxy, Boyle:2019kee}.  These NR waveforms are
shorter in length than the ones used to train our surrogate model.  To ensure
fair comparison between surrogate predictions and NR waveforms, we build a test
surrogate\footnote{While building the test surrogate, we exclude
    SXS:BBH:\MySXSID{29} ($\ecc=7\times10^{-4}$, $\meanano=5.766$ at $t_{\rm
    ref}=-5500M$) from the training set as the binary circularizes enough by
    $t=-2000M$ such that our eccentricity estimator defined in
Eq.(\ref{Eq:ecc_estimator}) becomes unreliable.} which is parameterized by
$\ecc$ and $\meanano$ at $t_{\rm ref}=-2000M$.

In Fig.~\ref{Fig:extrapolation}, we show mismatches computed using the advanced
LIGO design sensitivity noise curve, between the \texttt{NRSur2dq1Ecc+} model
and eccentric NR data at $q=2,3$. We include all modes available in the model
while computing the mismatch. For simplicity, we only consider a single point
in the source-frame sky, with an inclination angle of $\pi/3$.  For $\ecc$ (at
$t_{\rm ref}=-2000M$) smaller than $\sim 0.05$, mismatches are always smaller
than $10^{-2}$. As we increase $\ecc$ (at $t_{\rm ref}=-2000M$) to
$0.09$, the mismatches become significantly worse, especially for $q=3$,
reaching values $\sim 10^{-1}$. As an example, Fig.~\ref{Fig:sxs1371} shows the
surrogate prediction (and NR waveform) for the case that leads to the largest
mismatch in Fig.~\ref{Fig:extrapolation} with $\ecc$ (at $t_{\rm ref}=-2000M$)
smaller than $\sim 0.05$.

This suggests that our scheme to extend the surrogate model to comparable mass
systems produces reasonable waveforms for small eccentricities. However, we
advise caution with extrapolation-type procedures in general. 

\subsection{Importance of mean anomaly for data analysis}
Many existing waveform models~\cite{Huerta:2017kez,
Chen:2020lzc,Chiaramello:2020ehz, Cao:2017ndf} for eccentric binaries
parameterize eccentric characteristics of the waveform by only one parameter
$\ecc$ while keeping $\meanano$ fixed. We, however, use both $\ecc$ and
$\meanano$ as parameters in our model. We find that not allowing $\meanano$ as
an independent parameter results in large modeling error, indicating that the
mean anomaly is important to consider when modeling the GW signal from
eccentric binaries.

To demonstrate the importance of mean anomaly also in data analysis, we present
a simple study. We generate \texttt{NRSur2dq1Ecc} predictions $\hlm^S(q=1,
\ecc=0.1, \meanano)$ with $\meanano\in[0.0,2\pi]$.  The left panel of
Fig.~\ref{Fig:anomaly_mismatch} shows mismatches between the waveform at
$\meanano=0$ and various $\meanano$, parametrized by $\Delta\meanano =
\meanano-0$.  For simplicity, we only consider a single point in the
source-frame sky, at $\iota=\pi/3$, $\varphi_0=0.0$. As expected, we find that
$\Delta\meanano=0.0$ and $\Delta\meanano=2\pi$ produce identical waveforms.
However, the mismatch reaches a value of $\sim0.1$ at $\Delta\meanano=\pi$. As
we already account for allowed time and frame shifts when computing the
mismatch, ignoring this difference can lead to modeling errors or biased
parameter estimation. In the right panel of Fig.~\ref{Fig:anomaly_mismatch}, we
show the waveform amplitude for the cases with $\meanano=0$ and $\meanano=\pi$.
The clear differences in the amplitude reinforce our assertion that this
mismatch cannot be accounted for by a time or frame shift.

\section{Conclusion}
\label{Sec:Conclusion}
We present \texttt{NRSur2dq1Ecc}, the first eccentric NR surrogate waveform
model. This model is trained on 47 NR waveforms of equal-mass nonspinning BBH
systems with eccentricity $\ecc\leq0.2$, defined at a reference time $t_{\rm
ref}=-5500M$ before the waveform peak. The model includes the $(2,2)$, $(3,2)$
and $(4,4)$ spin-weighted spherical harmonic modes.  Due to the symmetries of
the equal-mass, nonspinning systems considered here, this is equivalent to
including all $\ell\leq3$ and $(4, \pm 4)$ modes, except the $m=0$ modes. This
is the first eccentric BBH model that is directly trained on eccentric NR
simulations and does not require that the binary circularizes before merger. We
also present \texttt{NRSur2dq1EccRemnant}, the first NR surrogate model for the
final BH properties of eccentric BBH mergers. This model is also trained on the
same set of simulations. We use Gaussian process regression to construct the
parametric fits for both models. Both \texttt{NRSur2dq1Ecc} and
\texttt{NRSur2dq1EccRemnant} will be made publicly available in the near
future.

Through a leave-one-out cross-validation study, we show that
\texttt{NRSur2dq1Ecc} accurately reproduces NR waveforms with a typical
mismatch of $\sim 10^{-3}$.  We further demonstrate that our remnant model,
\texttt{NRSur2dq1EccRemnant}, can accurately predict the final mass and spin of
the merger remnant with errors $\lesssim 5\times10^{-4} M$ and $\lesssim
2\times 10^{-3}$ respectively.  We showed that despite being trained
on equal-mass binaries, \texttt{NRSur2dq1Ecc} can be reasonably extended up to
mass ratio $q\approx3$ with mismatches $\simeq 10^{-2}$ for eccentricities
$\ecc\lesssim0.05$ at $t_{\rm ref}=-2000M$.  Finally, we demonstrate that the
mean anomaly, which is often ignored in waveform modeling and parameter
estimation of eccentric binaries, is an important parameter to include.
Exclusion of mean anomaly can result in poor modeling accuracy and/or biased
parameter inference.

The NR simulations used for this work were performed using the Spectral
Einstein Code (SpEC)~\cite{SpECwebsite}. SpEC's development efforts have been
primarily focused on evolutions of binary black hole systems in quasi-circular
orbits~\cite{Boyle:2019kee}. To efficiently generate accurate training data for
high eccentricity systems, it may be necessary to improve certain algorithmic
subroutines. For example, as noted in Sec.~\ref{sec:ps}, we found it difficult
to achieve target values of $(\ecc, \meanano)$ at a reference time before
merger. We also noticed that the waveform's numerical error was noticeably
larger near pericenters, suggesting better adaptive mesh refinement
algorithms~\cite{Szilagyi:2009qz} may be necessary for highly eccentric
simulations.

We have also explored several data decomposition techniques and
parametrizations for building eccentric NR surrogate models, which can guide
strategies for future models. 
Our final framework for building eccentric NR surrogates is quite general, and
we expect that it can be applied straightforwardly to higher dimensional
parameter spaces including unequal masses and aligned-spins. We leave these
explorations to future work.

\begin{acknowledgments}
We thank Geraint Pratten for comments on the manuscript.
We thank Nur Rifat and Feroz Shaik for helpful discussions.
We thank Katerina Chatziioannou for the implementation of an improved
eccentricity control system used in many of our simulations.
T.I.\ is supported by NSF grant PHY-1806665 and a doctoral fellowship provided
by UMassD Graduate Studies.
V.V.\ is supported by a Klarman Fellowship at Cornell, the Sherman
Fairchild Foundation, and NSF grants PHY–170212 and PHY–1708213 at Caltech. 
J.L.\ is supported by the Caltech Summer Undergraduate Research Fellowship
Program and the Rose Hills Foundation.
S.F.\ is supported by NSF grants No. PHY-1806665 and No. DMS-1912716.
G.K.\ acknowledges research support from NSF Grants No. PHY-2106755 and No.
DMS-1912716.
M.S. is supported by Sherman Fairchild Foundation and by NSF Grants
PHY-2011961, PHY-2011968, and OAC-1931266 at Caltech.
D.G. is supported by European Union H2020 ERC Starting Grant No. 945155--GWmining, Leverhulme Trust Grant No. RPG-2019-350, and
Royal Society Grant No. RGS-R2-202004.
L.K. is supported by the Sherman Fairchild Foundation, and NSF Grants
PHY-1912081 and OAC-1931280 at Cornell.
A portion of this work was carried out while a subset of the authors were in
residence at the Institute for Computational and Experimental Research in
Mathematics (ICERM) in Providence, RI, during the Advances in Computational
Relativity program.  ICERM is supported by the National Science Foundation
under Grant No. DMS-1439786.  
Simulations were performed on the Wheeler cluster at Caltech, which is
supported by the Sherman Fairchild Foundation and by Caltech; and on CARNiE at
the Center for Scientific Computing and Visualization Research (CSCVR) of
UMassD, which is supported by the ONR/DURIP Grant No.\ N00014181255.
Computations for building the model were performed on both CARNiE and Wheeler.
\end{acknowledgments}

\appendix
\section*{Appendix}
\label{appendix}

In this Appendix, we describe various alternate modeling strategies we pursued
before deciding on the formalism presented in the main text.

\section{Choice of data decomposition}
\label{app:decomp}

In this work, we have modeled the amplitude ($A_{22}$) and phase ($\phi_{22}$)
of the $(2,2)$ mode by modeling the residual ($\Delta A_{22}$, $\Delta
\phi_{22}$) of these quantities with respect to a quasicircular NR waveform
(cf. Sec.~\ref{Sec:DataDecomposition}). Alternatively, one could instead model
the amplitude and frequency (or their residuals), and then integrate the
frequency to obtain the phase. The frequency of the (2,2) mode is given by
\begin{gather}
    \label{Eq:Freq22}
    \omega_{22} = \frac{d\phi_{22}}{dt},
\end{gather}
where $\phi_{22}$ is defined in Eq.~\eqref{Eq:AmpPhase_22}. The corresponding residual is given by:
\begin{gather}
\label{Eq:FreqRes}
    \Delta \omega_{22} = \omega_{22} - \omega_{22}^{0} ,
\end{gather}
where $\omega_{22}^{0}$ is the frequency of $(2,2)$ mode for the quasicircular
NR waveform.

We, therefore, explore four different data decomposition strategies for the
$(2,2)$ mode, summarized below:
\begin{itemize}
    \item Model $\{A_{22}, \phi_{22}\}$ directly.
    \item Model $\{\Delta A_{22}, \Delta \phi_{22}\}$ and then
        add them to the amplitude and phase of the
        quasicircular NR waveform to obtain $\{A_{22},\phi_{22}\}$.
    \item Model $\{A_{22}, \omega_{22}\}$ and integrate the
        frequency data to get $\{A_{22},\phi_{22}\}$.
    \item Model $\{\Delta A_{22}, \Delta \omega_{22}\}$;
         add them to the amplitude and
        frequency of the quasicircular NR waveforms, and finally
        integrate the frequency data to obtain $\{A_{22},
        \phi_{22}\}$.
\end{itemize}

\begin{figure}[t]
\includegraphics[width=\columnwidth]{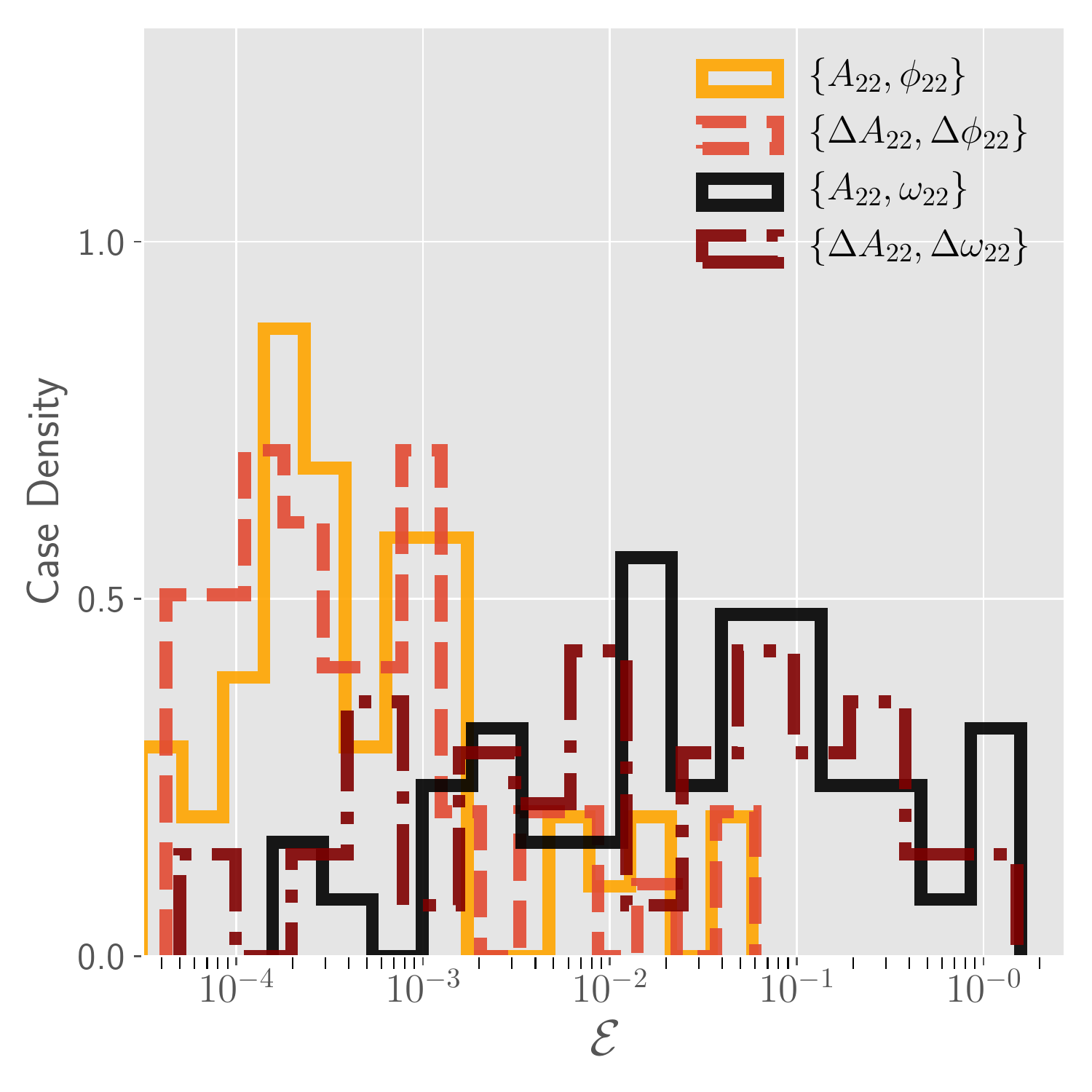}
\caption{Histograms of surrogate errors (defined in
Eqs.~\eqref{Eq:cost_function}) for the four different decomposition strategies
we consider. We find that modeling the residual amplitude $\Delta A_{22}$ and
residual phase $\Delta \phi_{22}$ yields the least errors.
}
\label{Fig:DecompositionError}
\end{figure}

In order to explore the effectiveness of these strategies, we build a separate
surrogate model using each strategy. When building the frequency surrogates
($\omega_{22}$ or $\Delta \omega_{22}$) we use a basis tolerance of $10^{-3}$
rad/$M$. For $A_{22}$ ($\phi_{22}$) we use the same tolerance as used for
$\Delta A_{22}$ ($\Delta \phi_{22}$) in Section \ref{Sec:model_building}. We
compute the normalized $L_{2}$-norm between the NR data and each surrogate
approximation using Eq.~(\ref{Eq:cost_function}). In
Fig.~\ref{Fig:DecompositionError}, we show the surrogate errors for all four
different strategies. We find that modeling the frequency $\omega_{22}$ or
residual frequency $\Delta \omega_{22}$ yields at least two-to-three orders of
magnitude larger $\mathcal{E}$ than when we model the phase $\phi_{22}$ or
$\Delta \phi_{22}$.  Furthermore, modeling the residual amplitude $\Delta
A_{22}$ proves to be slightly more accurate than the case where we model the
amplitude $A_{22}$ directly.  Therefore, in the main text, we build surrogate
models of the residual $\{\Delta A_{22}, \Delta \phi_{22}\}$ (cf.
Sec.~\ref{Sec:DataDecomposition}).

\section{Choice of fit parameterization}
\label{Sec:Parameterization}
When building the surrogate models in main text, fits across parameter
space are required for the waveform model as well as the remnant model (cf.
Sec.~\ref{Sec:model_building}). These fits are parameterized by the
eccentricity ($\ecc$) and mean anomaly $(\meanano)$ at the reference time
$t_{\rm ref}$. While $\{\ecc,\meanano\}$ is a natural choice, we also
explore the following choices of parameterizations:
\begin{itemize}
	\item $\{\ecc, \meanano\}$,
	\item $\{\ecc,\sin(\meanano/2)\}$,
	\item $\{\log_{10}(1-\ecc),\meanano\}$,
	\item $\{\log_{10}(1-\ecc),\sin(\meanano/2)\}$,
\end{itemize}
Here $\sin(\meanano/2)$ is considered because it maps the periodic parameter
$\meanano\in[0, 2\pi)$ uniquely to the range $[0, 1]$, 
while still mapping the physically equivalent points $\meanano=0$ and
$\meanano=2\pi$ to the same point ($\sin(\meanano/2)=0$). The same is not true
for other possible parameterizations such as $\sin(\meanano)$,
$\cos(\meanano)$, or $\cos(\meanano/2)$.
$\log_{10}(1 -
\ecc)$ is considered because it flattens the spread in eccentricity, which can
be useful if the eccentricity varies over several orders of magnitude in the NR
dataset.

\begin{figure}[thb]
\includegraphics[width=\columnwidth]{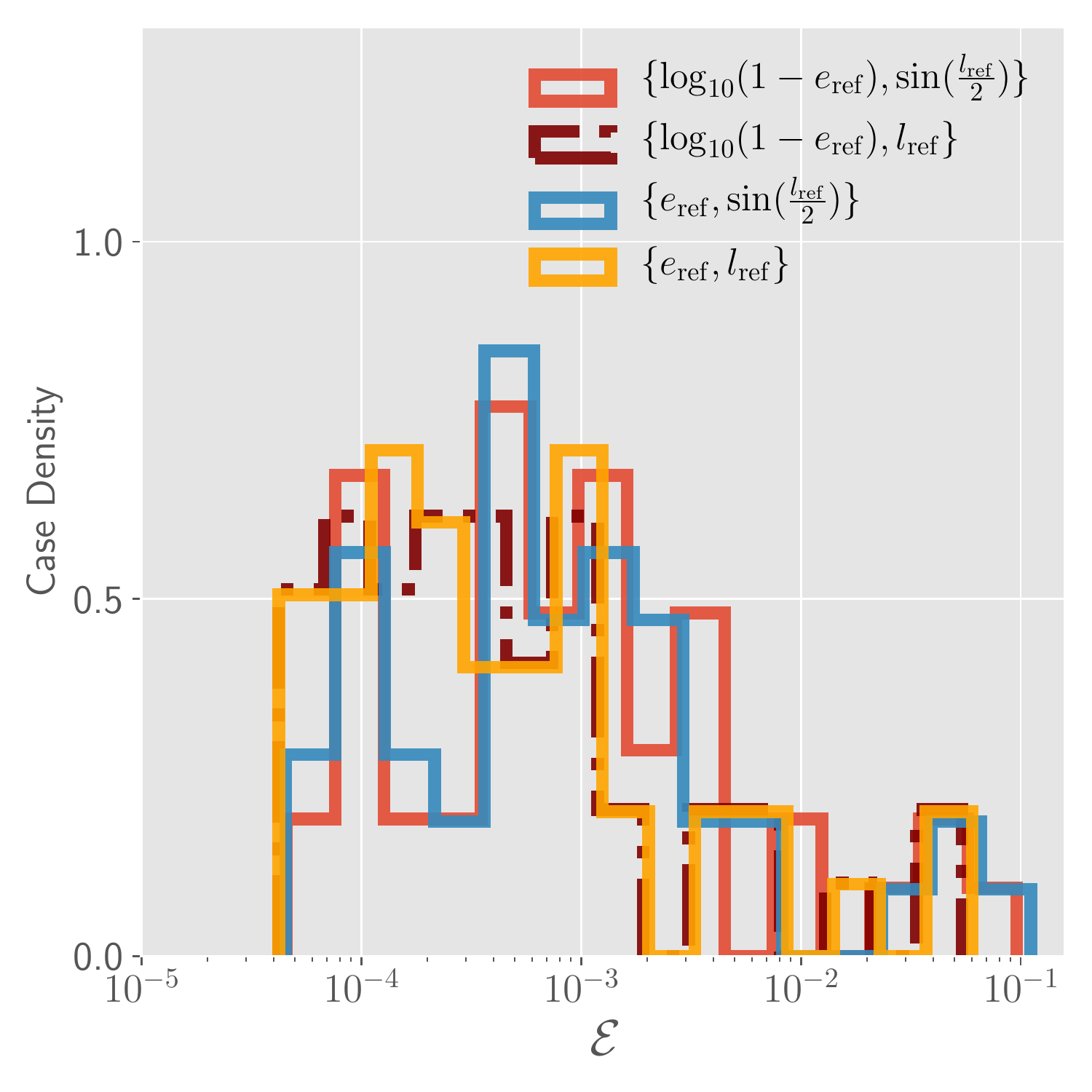}
\caption{Histograms of the error for the full waveform, for the six different fit parameterizations we
consider.
}
\label{Fig:ParameterizationError}
\end{figure}

Similarly to the previous section, to explore the effectiveness of these
strategies, we build a separate surrogate model using each strategy. Here,
however, we consider all modes included $(\ell,m)=(2,2),(3,2),(4,4)$ and
evaluate $\mathcal{E}$ errors [cf. Eq.~\eqref{Eq:cost_function}].  In
Fig.~\ref{Fig:ParameterizationError}, we show $\mathcal{E}$ errors for each
parameterization strategy. We find that while the alternative strategies using
either $\log_{10}(1-\ecc)$, or $\sin(\meanano/2)$, or both, may be comparable,
none of the them result in errors smaller than the original choice
$\{\ecc,\meanano\}$.  As we do not achieve a noticeable improvement with these
alternative parameterizations, we stick to the original choice
$\{\ecc,\meanano\}$ in the main text.

\bibliography{References}

\begin{thebibliography}{107}%
\makeatletter
\providecommand \@ifxundefined [1]{%
 \@ifx{#1\undefined}
}%
\providecommand \@ifnum [1]{%
 \ifnum #1\expandafter \@firstoftwo
 \else \expandafter \@secondoftwo
 \fi
}%
\providecommand \@ifx [1]{%
 \ifx #1\expandafter \@firstoftwo
 \else \expandafter \@secondoftwo
 \fi
}%
\providecommand \natexlab [1]{#1}%
\providecommand \enquote  [1]{``#1''}%
\providecommand \bibnamefont  [1]{#1}%
\providecommand \bibfnamefont [1]{#1}%
\providecommand \citenamefont [1]{#1}%
\providecommand \href@noop [0]{\@secondoftwo}%
\providecommand \href [0]{\begingroup \@sanitize@url \@href}%
\providecommand \@href[1]{\@@startlink{#1}\@@href}%
\providecommand \@@href[1]{\endgroup#1\@@endlink}%
\providecommand \@sanitize@url [0]{\catcode `\\12\catcode `\$12\catcode
  `\&12\catcode `\#12\catcode `\^12\catcode `\_12\catcode `\%12\relax}%
\providecommand \@@startlink[1]{}%
\providecommand \@@endlink[0]{}%
\providecommand \url  [0]{\begingroup\@sanitize@url \@url }%
\providecommand \@url [1]{\endgroup\@href {#1}{\urlprefix }}%
\providecommand \urlprefix  [0]{URL }%
\providecommand \Eprint [0]{\href }%
\providecommand \doibase [0]{http://dx.doi.org/}%
\providecommand \selectlanguage [0]{\@gobble}%
\providecommand \bibinfo  [0]{\@secondoftwo}%
\providecommand \bibfield  [0]{\@secondoftwo}%
\providecommand \translation [1]{[#1]}%
\providecommand \BibitemOpen [0]{}%
\providecommand \bibitemStop [0]{}%
\providecommand \bibitemNoStop [0]{.\EOS\space}%
\providecommand \EOS [0]{\spacefactor3000\relax}%
\providecommand \BibitemShut  [1]{\csname bibitem#1\endcsname}%
\let\auto@bib@innerbib\@empty
\bibitem [{\citenamefont {Abbott}\ \emph
  {et~al.}(2019{\natexlab{a}})\citenamefont {Abbott} \emph
  {et~al.}}]{LIGOScientific:2018mvr}%
  \BibitemOpen
  \bibfield  {author} {\bibinfo {author} {\bibfnamefont {B.~P.}\ \bibnamefont
  {Abbott}} \emph {et~al.} (\bibinfo {collaboration} {LIGO Scientific,
  Virgo}),\ }\bibfield  {title} {\enquote {\bibinfo {title} {{GWTC-1: A
  Gravitational-Wave Transient Catalog of Compact Binary Mergers Observed by
  LIGO and Virgo during the First and Second Observing Runs}},}\ }\href
  {\doibase 10.1103/PhysRevX.9.031040} {\bibfield  {journal} {\bibinfo
  {journal} {Phys. Rev.}\ }\textbf {\bibinfo {volume} {X9}},\ \bibinfo {pages}
  {031040} (\bibinfo {year} {2019}{\natexlab{a}})},\ \Eprint
  {http://arxiv.org/abs/1811.12907} {arXiv:1811.12907 [astro-ph.HE]}
  \BibitemShut {NoStop}%
\bibitem [{\citenamefont {Abbott}\ \emph
  {et~al.}(2020{\natexlab{a}})\citenamefont {Abbott} \emph
  {et~al.}}]{Abbott:2020niy}%
  \BibitemOpen
  \bibfield  {author} {\bibinfo {author} {\bibfnamefont {R.}~\bibnamefont
  {Abbott}} \emph {et~al.} (\bibinfo {collaboration} {LIGO Scientific,
  Virgo}),\ }\bibfield  {title} {\enquote {\bibinfo {title} {{GWTC-2: Compact
  Binary Coalescences Observed by LIGO and Virgo During the First Half of the
  Third Observing Run}},}\ }\href@noop {} {\  (\bibinfo {year}
  {2020}{\natexlab{a}})},\ \Eprint {http://arxiv.org/abs/2010.14527}
  {arXiv:2010.14527 [gr-qc]} \BibitemShut {NoStop}%
\bibitem [{\citenamefont {Aasi}\ \emph {et~al.}(2015)\citenamefont {Aasi} \emph
  {et~al.}}]{TheLIGOScientific:2014jea}%
  \BibitemOpen
  \bibfield  {author} {\bibinfo {author} {\bibfnamefont {J.}~\bibnamefont
  {Aasi}} \emph {et~al.} (\bibinfo {collaboration} {LIGO Scientific}),\
  }\bibfield  {title} {\enquote {\bibinfo {title} {{Advanced LIGO}},}\ }\href
  {\doibase 10.1088/0264-9381/32/7/074001} {\bibfield  {journal} {\bibinfo
  {journal} {Class. Quant. Grav.}\ }\textbf {\bibinfo {volume} {32}},\ \bibinfo
  {pages} {074001} (\bibinfo {year} {2015})},\ \Eprint
  {http://arxiv.org/abs/1411.4547} {arXiv:1411.4547 [gr-qc]} \BibitemShut
  {NoStop}%
\bibitem [{\citenamefont {Acernese}\ \emph {et~al.}(2015)\citenamefont
  {Acernese} \emph {et~al.}}]{TheVirgo:2014hva}%
  \BibitemOpen
  \bibfield  {author} {\bibinfo {author} {\bibfnamefont {F.}~\bibnamefont
  {Acernese}} \emph {et~al.} (\bibinfo {collaboration} {Virgo}),\ }\bibfield
  {title} {\enquote {\bibinfo {title} {{Advanced Virgo: a second-generation
  interferometric gravitational wave detector}},}\ }\href {\doibase
  10.1088/0264-9381/32/2/024001} {\bibfield  {journal} {\bibinfo  {journal}
  {Class. Quant. Grav.}\ }\textbf {\bibinfo {volume} {32}},\ \bibinfo {pages}
  {024001} (\bibinfo {year} {2015})},\ \Eprint {http://arxiv.org/abs/1408.3978}
  {arXiv:1408.3978 [gr-qc]} \BibitemShut {NoStop}%
\bibitem [{\citenamefont {{Field}}\ \emph {et~al.}(2014)\citenamefont
  {{Field}}, \citenamefont {{Galley}}, \citenamefont {{Hesthaven}},
  \citenamefont {{Kaye}},\ and\ \citenamefont {{Tiglio}}}]{Field:2013cfa}%
  \BibitemOpen
  \bibfield  {author} {\bibinfo {author} {\bibfnamefont {S.~E.}\ \bibnamefont
  {{Field}}}, \bibinfo {author} {\bibfnamefont {C.~R.}\ \bibnamefont
  {{Galley}}}, \bibinfo {author} {\bibfnamefont {J.~S.}\ \bibnamefont
  {{Hesthaven}}}, \bibinfo {author} {\bibfnamefont {J.}~\bibnamefont {{Kaye}}},
  \ and\ \bibinfo {author} {\bibfnamefont {M.}~\bibnamefont {{Tiglio}}},\
  }\bibfield  {title} {\enquote {\bibinfo {title} {{Fast Prediction and
  Evaluation of Gravitational Waveforms Using Surrogate Models}},}\ }\href
  {\doibase 10.1103/PhysRevX.4.031006} {\ \textbf {\bibinfo {volume} {4}},\
  \bibinfo {eid} {031006} (\bibinfo {year} {2014})},\ \Eprint
  {http://arxiv.org/abs/1308.3565} {arXiv:1308.3565 [gr-qc]} \BibitemShut
  {NoStop}%
\bibitem [{\citenamefont {P\"urrer}(2014)}]{Purrer:2014fza}%
  \BibitemOpen
  \bibfield  {author} {\bibinfo {author} {\bibfnamefont {Michael}\ \bibnamefont
  {P\"urrer}},\ }\bibfield  {title} {\enquote {\bibinfo {title} {{Frequency
  domain reduced order models for gravitational waves from aligned-spin compact
  binaries}},}\ }\href {\doibase 10.1088/0264-9381/31/19/195010} {\bibfield
  {journal} {\bibinfo  {journal} {Class. Quant. Grav.}\ }\textbf {\bibinfo
  {volume} {31}},\ \bibinfo {pages} {195010} (\bibinfo {year} {2014})},\
  \Eprint {http://arxiv.org/abs/1402.4146} {arXiv:1402.4146 [gr-qc]}
  \BibitemShut {NoStop}%
\bibitem [{\citenamefont {Blackman}\ \emph {et~al.}(2015)\citenamefont
  {Blackman}, \citenamefont {Field}, \citenamefont {Galley}, \citenamefont
  {Szilágyi}, \citenamefont {Scheel}, \citenamefont {Tiglio},\ and\
  \citenamefont {Hemberger}}]{Blackman:2015pia}%
  \BibitemOpen
  \bibfield  {author} {\bibinfo {author} {\bibfnamefont {Jonathan}\
  \bibnamefont {Blackman}}, \bibinfo {author} {\bibfnamefont {Scott~E.}\
  \bibnamefont {Field}}, \bibinfo {author} {\bibfnamefont {Chad~R.}\
  \bibnamefont {Galley}}, \bibinfo {author} {\bibfnamefont {Béla}\
  \bibnamefont {Szilágyi}}, \bibinfo {author} {\bibfnamefont {Mark~A.}\
  \bibnamefont {Scheel}}, \bibinfo {author} {\bibfnamefont {Manuel}\
  \bibnamefont {Tiglio}}, \ and\ \bibinfo {author} {\bibfnamefont {Daniel~A.}\
  \bibnamefont {Hemberger}},\ }\bibfield  {title} {\enquote {\bibinfo {title}
  {{Fast and Accurate Prediction of Numerical Relativity Waveforms from Binary
  Black Hole Coalescences Using Surrogate Models}},}\ }\href {\doibase
  10.1103/PhysRevLett.115.121102} {\bibfield  {journal} {\bibinfo  {journal}
  {Phys. Rev. Lett.}\ }\textbf {\bibinfo {volume} {115}},\ \bibinfo {pages}
  {121102} (\bibinfo {year} {2015})},\ \Eprint
  {http://arxiv.org/abs/1502.07758} {arXiv:1502.07758 [gr-qc]} \BibitemShut
  {NoStop}%
\bibitem [{\citenamefont {Blackman}\ \emph
  {et~al.}(2017{\natexlab{a}})\citenamefont {Blackman}, \citenamefont {Field},
  \citenamefont {Scheel}, \citenamefont {Galley}, \citenamefont {Ott},
  \citenamefont {Boyle}, \citenamefont {Kidder}, \citenamefont {Pfeiffer},\
  and\ \citenamefont {Szilágyi}}]{Blackman:2017pcm}%
  \BibitemOpen
  \bibfield  {author} {\bibinfo {author} {\bibfnamefont {Jonathan}\
  \bibnamefont {Blackman}}, \bibinfo {author} {\bibfnamefont {Scott~E.}\
  \bibnamefont {Field}}, \bibinfo {author} {\bibfnamefont {Mark~A.}\
  \bibnamefont {Scheel}}, \bibinfo {author} {\bibfnamefont {Chad~R.}\
  \bibnamefont {Galley}}, \bibinfo {author} {\bibfnamefont {Christian~D.}\
  \bibnamefont {Ott}}, \bibinfo {author} {\bibfnamefont {Michael}\ \bibnamefont
  {Boyle}}, \bibinfo {author} {\bibfnamefont {Lawrence~E.}\ \bibnamefont
  {Kidder}}, \bibinfo {author} {\bibfnamefont {Harald~P.}\ \bibnamefont
  {Pfeiffer}}, \ and\ \bibinfo {author} {\bibfnamefont {Béla}\ \bibnamefont
  {Szilágyi}},\ }\bibfield  {title} {\enquote {\bibinfo {title} {{Numerical
  relativity waveform surrogate model for generically precessing binary black
  hole mergers}},}\ }\href {\doibase 10.1103/PhysRevD.96.024058} {\bibfield
  {journal} {\bibinfo  {journal} {Phys. Rev.}\ }\textbf {\bibinfo {volume}
  {D96}},\ \bibinfo {pages} {024058} (\bibinfo {year} {2017}{\natexlab{a}})},\
  \Eprint {http://arxiv.org/abs/1705.07089} {arXiv:1705.07089 [gr-qc]}
  \BibitemShut {NoStop}%
\bibitem [{\citenamefont {Blackman}\ \emph
  {et~al.}(2017{\natexlab{b}})\citenamefont {Blackman}, \citenamefont {Field},
  \citenamefont {Scheel}, \citenamefont {Galley}, \citenamefont {Hemberger},
  \citenamefont {Schmidt},\ and\ \citenamefont {Smith}}]{Blackman:2017dfb}%
  \BibitemOpen
  \bibfield  {author} {\bibinfo {author} {\bibfnamefont {Jonathan}\
  \bibnamefont {Blackman}}, \bibinfo {author} {\bibfnamefont {Scott~E.}\
  \bibnamefont {Field}}, \bibinfo {author} {\bibfnamefont {Mark~A.}\
  \bibnamefont {Scheel}}, \bibinfo {author} {\bibfnamefont {Chad~R.}\
  \bibnamefont {Galley}}, \bibinfo {author} {\bibfnamefont {Daniel~A.}\
  \bibnamefont {Hemberger}}, \bibinfo {author} {\bibfnamefont {Patricia}\
  \bibnamefont {Schmidt}}, \ and\ \bibinfo {author} {\bibfnamefont {Rory}\
  \bibnamefont {Smith}},\ }\bibfield  {title} {\enquote {\bibinfo {title} {{A
  Surrogate Model of Gravitational Waveforms from Numerical Relativity
  Simulations of Precessing Binary Black Hole Mergers}},}\ }\href {\doibase
  10.1103/PhysRevD.95.104023} {\bibfield  {journal} {\bibinfo  {journal} {Phys.
  Rev.}\ }\textbf {\bibinfo {volume} {D95}},\ \bibinfo {pages} {104023}
  (\bibinfo {year} {2017}{\natexlab{b}})},\ \Eprint
  {http://arxiv.org/abs/1701.00550} {arXiv:1701.00550 [gr-qc]} \BibitemShut
  {NoStop}%
\bibitem [{\citenamefont {Varma}\ \emph
  {et~al.}(2019{\natexlab{a}})\citenamefont {Varma}, \citenamefont {Field},
  \citenamefont {Scheel}, \citenamefont {Blackman}, \citenamefont {Kidder},\
  and\ \citenamefont {Pfeiffer}}]{Varma:2018mmi}%
  \BibitemOpen
  \bibfield  {author} {\bibinfo {author} {\bibfnamefont {Vijay}\ \bibnamefont
  {Varma}}, \bibinfo {author} {\bibfnamefont {Scott~E.}\ \bibnamefont {Field}},
  \bibinfo {author} {\bibfnamefont {Mark~A.}\ \bibnamefont {Scheel}}, \bibinfo
  {author} {\bibfnamefont {Jonathan}\ \bibnamefont {Blackman}}, \bibinfo
  {author} {\bibfnamefont {Lawrence~E.}\ \bibnamefont {Kidder}}, \ and\
  \bibinfo {author} {\bibfnamefont {Harald~P.}\ \bibnamefont {Pfeiffer}},\
  }\bibfield  {title} {\enquote {\bibinfo {title} {{Surrogate model of
  hybridized numerical relativity binary black hole waveforms}},}\ }\href
  {\doibase 10.1103/PhysRevD.99.064045} {\bibfield  {journal} {\bibinfo
  {journal} {Phys. Rev.}\ }\textbf {\bibinfo {volume} {D99}},\ \bibinfo {pages}
  {064045} (\bibinfo {year} {2019}{\natexlab{a}})},\ \Eprint
  {http://arxiv.org/abs/1812.07865} {arXiv:1812.07865 [gr-qc]} \BibitemShut
  {NoStop}%
\bibitem [{\citenamefont {Chua}\ \emph {et~al.}(2019)\citenamefont {Chua},
  \citenamefont {Galley},\ and\ \citenamefont {Vallisneri}}]{Chua:2018woh}%
  \BibitemOpen
  \bibfield  {author} {\bibinfo {author} {\bibfnamefont {Alvin~J.K.}\
  \bibnamefont {Chua}}, \bibinfo {author} {\bibfnamefont {Chad~R.}\
  \bibnamefont {Galley}}, \ and\ \bibinfo {author} {\bibfnamefont {Michele}\
  \bibnamefont {Vallisneri}},\ }\bibfield  {title} {\enquote {\bibinfo {title}
  {{Reduced-order modeling with artificial neurons for gravitational-wave
  inference}},}\ }\href {\doibase 10.1103/PhysRevLett.122.211101} {\bibfield
  {journal} {\bibinfo  {journal} {Phys. Rev. Lett.}\ }\textbf {\bibinfo
  {volume} {122}},\ \bibinfo {pages} {211101} (\bibinfo {year} {2019})},\
  \Eprint {http://arxiv.org/abs/1811.05491} {arXiv:1811.05491 [astro-ph.IM]}
  \BibitemShut {NoStop}%
\bibitem [{\citenamefont {Lackey}\ \emph {et~al.}(2019)\citenamefont {Lackey},
  \citenamefont {P\"urrer}, \citenamefont {Taracchini},\ and\ \citenamefont
  {Marsat}}]{Lackey:2018zvw}%
  \BibitemOpen
  \bibfield  {author} {\bibinfo {author} {\bibfnamefont {Benjamin~D.}\
  \bibnamefont {Lackey}}, \bibinfo {author} {\bibfnamefont {Michael}\
  \bibnamefont {P\"urrer}}, \bibinfo {author} {\bibfnamefont {Andrea}\
  \bibnamefont {Taracchini}}, \ and\ \bibinfo {author} {\bibfnamefont
  {Sylvain}\ \bibnamefont {Marsat}},\ }\bibfield  {title} {\enquote {\bibinfo
  {title} {{Surrogate model for an aligned-spin effective one body waveform
  model of binary neutron star inspirals using Gaussian process regression}},}\
  }\href {\doibase 10.1103/PhysRevD.100.024002} {\bibfield  {journal} {\bibinfo
   {journal} {Phys. Rev. D}\ }\textbf {\bibinfo {volume} {100}},\ \bibinfo
  {pages} {024002} (\bibinfo {year} {2019})},\ \Eprint
  {http://arxiv.org/abs/1812.08643} {arXiv:1812.08643 [gr-qc]} \BibitemShut
  {NoStop}%
\bibitem [{\citenamefont {Varma}\ \emph
  {et~al.}(2019{\natexlab{b}})\citenamefont {Varma}, \citenamefont {Field},
  \citenamefont {Scheel}, \citenamefont {Blackman}, \citenamefont {Gerosa},
  \citenamefont {Stein}, \citenamefont {Kidder},\ and\ \citenamefont
  {Pfeiffer}}]{Varma:2019csw}%
  \BibitemOpen
  \bibfield  {author} {\bibinfo {author} {\bibfnamefont {Vijay}\ \bibnamefont
  {Varma}}, \bibinfo {author} {\bibfnamefont {Scott~E.}\ \bibnamefont {Field}},
  \bibinfo {author} {\bibfnamefont {Mark~A.}\ \bibnamefont {Scheel}}, \bibinfo
  {author} {\bibfnamefont {Jonathan}\ \bibnamefont {Blackman}}, \bibinfo
  {author} {\bibfnamefont {Davide}\ \bibnamefont {Gerosa}}, \bibinfo {author}
  {\bibfnamefont {Leo~C.}\ \bibnamefont {Stein}}, \bibinfo {author}
  {\bibfnamefont {Lawrence~E.}\ \bibnamefont {Kidder}}, \ and\ \bibinfo
  {author} {\bibfnamefont {Harald~P.}\ \bibnamefont {Pfeiffer}},\ }\bibfield
  {title} {\enquote {\bibinfo {title} {{Surrogate models for precessing binary
  black hole simulations with unequal masses}},}\ }\href {\doibase
  10.1103/PhysRevResearch.1.033015} {\bibfield  {journal} {\bibinfo  {journal}
  {Phys. Rev. Research.}\ }\textbf {\bibinfo {volume} {1}},\ \bibinfo {pages}
  {033015} (\bibinfo {year} {2019}{\natexlab{b}})},\ \Eprint
  {http://arxiv.org/abs/1905.09300} {arXiv:1905.09300 [gr-qc]} \BibitemShut
  {NoStop}%
\bibitem [{\citenamefont {Khan}\ and\ \citenamefont
  {Green}(2020)}]{Khan:2020fso}%
  \BibitemOpen
  \bibfield  {author} {\bibinfo {author} {\bibfnamefont {Sebastian}\
  \bibnamefont {Khan}}\ and\ \bibinfo {author} {\bibfnamefont {Rhys}\
  \bibnamefont {Green}},\ }\bibfield  {title} {\enquote {\bibinfo {title}
  {{Gravitational-wave surrogate models powered by artificial neural networks:
  The ANN-Sur for waveform generation}},}\ }\href@noop {} {\  (\bibinfo {year}
  {2020})},\ \Eprint {http://arxiv.org/abs/2008.12932} {arXiv:2008.12932
  [gr-qc]} \BibitemShut {NoStop}%
\bibitem [{\citenamefont {Williams}\ \emph {et~al.}(2020)\citenamefont
  {Williams}, \citenamefont {Heng}, \citenamefont {Gair}, \citenamefont
  {Clark},\ and\ \citenamefont {Khamesra}}]{Williams:2019vub}%
  \BibitemOpen
  \bibfield  {author} {\bibinfo {author} {\bibfnamefont {Daniel}\ \bibnamefont
  {Williams}}, \bibinfo {author} {\bibfnamefont {Ik~Siong}\ \bibnamefont
  {Heng}}, \bibinfo {author} {\bibfnamefont {Jonathan}\ \bibnamefont {Gair}},
  \bibinfo {author} {\bibfnamefont {James~A.}\ \bibnamefont {Clark}}, \ and\
  \bibinfo {author} {\bibfnamefont {Bhavesh}\ \bibnamefont {Khamesra}},\
  }\bibfield  {title} {\enquote {\bibinfo {title} {{Precessing numerical
  relativity waveform surrogate model for binary black holes: A Gaussian
  process regression approach}},}\ }\href {\doibase
  10.1103/PhysRevD.101.063011} {\bibfield  {journal} {\bibinfo  {journal}
  {Phys. Rev. D}\ }\textbf {\bibinfo {volume} {101}},\ \bibinfo {pages}
  {063011} (\bibinfo {year} {2020})},\ \Eprint
  {http://arxiv.org/abs/1903.09204} {arXiv:1903.09204 [gr-qc]} \BibitemShut
  {NoStop}%
\bibitem [{\citenamefont {Abbott}\ \emph
  {et~al.}(2019{\natexlab{b}})\citenamefont {Abbott} \emph
  {et~al.}}]{Salemi:2019owp}%
  \BibitemOpen
  \bibfield  {author} {\bibinfo {author} {\bibfnamefont {B.P.}\ \bibnamefont
  {Abbott}} \emph {et~al.} (\bibinfo {collaboration} {LIGO Scientific,
  Virgo}),\ }\bibfield  {title} {\enquote {\bibinfo {title} {{Search for
  Eccentric Binary Black Hole Mergers with Advanced LIGO and Advanced Virgo
  during their First and Second Observing Runs}},}\ }\href {\doibase
  10.3847/1538-4357/ab3c2d} {\bibfield  {journal} {\bibinfo  {journal}
  {Astrophys. J.}\ }\textbf {\bibinfo {volume} {883}},\ \bibinfo {pages} {149}
  (\bibinfo {year} {2019}{\natexlab{b}})},\ \Eprint
  {http://arxiv.org/abs/1907.09384} {arXiv:1907.09384 [astro-ph.HE]}
  \BibitemShut {NoStop}%
\bibitem [{\citenamefont {Romero-Shaw}\ \emph {et~al.}(2019)\citenamefont
  {Romero-Shaw}, \citenamefont {Lasky},\ and\ \citenamefont
  {Thrane}}]{Romero-Shaw:2019itr}%
  \BibitemOpen
  \bibfield  {author} {\bibinfo {author} {\bibfnamefont {Isobel~M.}\
  \bibnamefont {Romero-Shaw}}, \bibinfo {author} {\bibfnamefont {Paul~D.}\
  \bibnamefont {Lasky}}, \ and\ \bibinfo {author} {\bibfnamefont {Eric}\
  \bibnamefont {Thrane}},\ }\bibfield  {title} {\enquote {\bibinfo {title}
  {{Searching for Eccentricity: Signatures of Dynamical Formation in the First
  Gravitational-Wave Transient Catalogue of LIGO and Virgo}},}\ }\href
  {\doibase 10.1093/mnras/stz2996} {\bibfield  {journal} {\bibinfo  {journal}
  {Mon. Not. Roy. Astron. Soc.}\ }\textbf {\bibinfo {volume} {490}},\ \bibinfo
  {pages} {5210--5216} (\bibinfo {year} {2019})},\ \Eprint
  {http://arxiv.org/abs/1909.05466} {arXiv:1909.05466 [astro-ph.HE]}
  \BibitemShut {NoStop}%
\bibitem [{\citenamefont {Lenon}\ \emph {et~al.}(2020)\citenamefont {Lenon},
  \citenamefont {Nitz},\ and\ \citenamefont {Brown}}]{Lenon:2020oza}%
  \BibitemOpen
  \bibfield  {author} {\bibinfo {author} {\bibfnamefont {Amber~K.}\
  \bibnamefont {Lenon}}, \bibinfo {author} {\bibfnamefont {Alexander~H.}\
  \bibnamefont {Nitz}}, \ and\ \bibinfo {author} {\bibfnamefont {Duncan~A.}\
  \bibnamefont {Brown}},\ }\bibfield  {title} {\enquote {\bibinfo {title}
  {{Measuring the eccentricity of GW170817 and GW190425}},}\ }\href {\doibase
  10.1093/mnras/staa2120} {\bibfield  {journal} {\bibinfo  {journal} {Mon. Not.
  Roy. Astron. Soc.}\ }\textbf {\bibinfo {volume} {497}},\ \bibinfo {pages}
  {1966--1971} (\bibinfo {year} {2020})},\ \Eprint
  {http://arxiv.org/abs/2005.14146} {arXiv:2005.14146 [astro-ph.HE]}
  \BibitemShut {NoStop}%
\bibitem [{\citenamefont {Yun}\ \emph {et~al.}(2020)\citenamefont {Yun},
  \citenamefont {Han}, \citenamefont {Wang},\ and\ \citenamefont
  {Yang}}]{Yun:2020aow}%
  \BibitemOpen
  \bibfield  {author} {\bibinfo {author} {\bibfnamefont {Qian-Yun}\
  \bibnamefont {Yun}}, \bibinfo {author} {\bibfnamefont {Wen-Biao}\
  \bibnamefont {Han}}, \bibinfo {author} {\bibfnamefont {Gang}\ \bibnamefont
  {Wang}}, \ and\ \bibinfo {author} {\bibfnamefont {Shu-Cheng}\ \bibnamefont
  {Yang}},\ }\bibfield  {title} {\enquote {\bibinfo {title} {{Investigating
  eccentricities of the binary black hole signals from the LIGO-Virgo catalog
  GWTC-1}},}\ }\href@noop {} {\  (\bibinfo {year} {2020})},\ \Eprint
  {http://arxiv.org/abs/2002.08682} {arXiv:2002.08682 [gr-qc]} \BibitemShut
  {NoStop}%
\bibitem [{\citenamefont {Wu}\ \emph {et~al.}(2020)\citenamefont {Wu},
  \citenamefont {Cao},\ and\ \citenamefont {Zhu}}]{Wu:2020zwr}%
  \BibitemOpen
  \bibfield  {author} {\bibinfo {author} {\bibfnamefont {Shichao}\ \bibnamefont
  {Wu}}, \bibinfo {author} {\bibfnamefont {Zhoujian}\ \bibnamefont {Cao}}, \
  and\ \bibinfo {author} {\bibfnamefont {Zong-Hong}\ \bibnamefont {Zhu}},\
  }\bibfield  {title} {\enquote {\bibinfo {title} {{Measuring the eccentricity
  of binary black holes in GWTC-1 by using the inspiral-only waveform}},}\
  }\href {\doibase 10.1093/mnras/staa1176} {\bibfield  {journal} {\bibinfo
  {journal} {Mon. Not. Roy. Astron. Soc.}\ }\textbf {\bibinfo {volume} {495}},\
  \bibinfo {pages} {466--478} (\bibinfo {year} {2020})},\ \Eprint
  {http://arxiv.org/abs/2002.05528} {arXiv:2002.05528 [astro-ph.IM]}
  \BibitemShut {NoStop}%
\bibitem [{\citenamefont {Nitz}\ \emph {et~al.}(2019)\citenamefont {Nitz},
  \citenamefont {Lenon},\ and\ \citenamefont {Brown}}]{Nitz:2019spj}%
  \BibitemOpen
  \bibfield  {author} {\bibinfo {author} {\bibfnamefont {Alexander~H.}\
  \bibnamefont {Nitz}}, \bibinfo {author} {\bibfnamefont {Amber}\ \bibnamefont
  {Lenon}}, \ and\ \bibinfo {author} {\bibfnamefont {Duncan~A.}\ \bibnamefont
  {Brown}},\ }\bibfield  {title} {\enquote {\bibinfo {title} {{Search for
  Eccentric Binary Neutron Star Mergers in the first and second observing runs
  of Advanced LIGO}},}\ }\href {\doibase 10.3847/1538-4357/ab6611} {\bibfield
  {journal} {\bibinfo  {journal} {Astrophys. J.}\ }\textbf {\bibinfo {volume}
  {890}},\ \bibinfo {pages} {1} (\bibinfo {year} {2019})},\ \Eprint
  {http://arxiv.org/abs/1912.05464} {arXiv:1912.05464 [astro-ph.HE]}
  \BibitemShut {NoStop}%
\bibitem [{\citenamefont {Ramos-Buades}\ \emph
  {et~al.}(2020{\natexlab{a}})\citenamefont {Ramos-Buades}, \citenamefont
  {Tiwari}, \citenamefont {Haney},\ and\ \citenamefont
  {Husa}}]{Ramos-Buades:2020eju}%
  \BibitemOpen
  \bibfield  {author} {\bibinfo {author} {\bibfnamefont {Antoni}\ \bibnamefont
  {Ramos-Buades}}, \bibinfo {author} {\bibfnamefont {Shubhanshu}\ \bibnamefont
  {Tiwari}}, \bibinfo {author} {\bibfnamefont {Maria}\ \bibnamefont {Haney}}, \
  and\ \bibinfo {author} {\bibfnamefont {Sascha}\ \bibnamefont {Husa}},\
  }\bibfield  {title} {\enquote {\bibinfo {title} {{Impact of eccentricity on
  the gravitational wave searches for binary black holes: High mass case}},}\
  }\href {\doibase 10.1103/PhysRevD.102.043005} {\bibfield  {journal} {\bibinfo
   {journal} {Phys. Rev. D}\ }\textbf {\bibinfo {volume} {102}},\ \bibinfo
  {pages} {043005} (\bibinfo {year} {2020}{\natexlab{a}})},\ \Eprint
  {http://arxiv.org/abs/2005.14016} {arXiv:2005.14016 [gr-qc]} \BibitemShut
  {NoStop}%
\bibitem [{\citenamefont {Peters}(1964)}]{peters1964gravitational}%
  \BibitemOpen
  \bibfield  {author} {\bibinfo {author} {\bibfnamefont {P.~C.}\ \bibnamefont
  {Peters}},\ }\bibfield  {title} {\enquote {\bibinfo {title} {Gravitational
  radiation and the motion of two point masses},}\ }\href {\doibase
  10.1103/PhysRev.136.B1224} {\bibfield  {journal} {\bibinfo  {journal} {Phys.
  Rev.}\ }\textbf {\bibinfo {volume} {136}},\ \bibinfo {pages} {B1224--B1232}
  (\bibinfo {year} {1964})}\BibitemShut {NoStop}%
\bibitem [{\citenamefont {Giesler}\ \emph {et~al.}(2018)\citenamefont
  {Giesler}, \citenamefont {Clausen},\ and\ \citenamefont
  {Ott}}]{Giesler:2017uyu}%
  \BibitemOpen
  \bibfield  {author} {\bibinfo {author} {\bibfnamefont {Matthew}\ \bibnamefont
  {Giesler}}, \bibinfo {author} {\bibfnamefont {Drew}\ \bibnamefont {Clausen}},
  \ and\ \bibinfo {author} {\bibfnamefont {Christian~D.}\ \bibnamefont {Ott}},\
  }\bibfield  {title} {\enquote {\bibinfo {title} {{Low-mass X-ray binaries
  from black-hole retaining globular clusters}},}\ }\href {\doibase
  10.1093/mnras/sty659} {\bibfield  {journal} {\bibinfo  {journal} {Mon. Not.
  Roy. Astron. Soc.}\ }\textbf {\bibinfo {volume} {477}},\ \bibinfo {pages}
  {1853--1879} (\bibinfo {year} {2018})},\ \Eprint
  {http://arxiv.org/abs/1708.05915} {arXiv:1708.05915 [astro-ph.HE]}
  \BibitemShut {NoStop}%
\bibitem [{\citenamefont {Rodriguez}\ \emph
  {et~al.}(2018{\natexlab{a}})\citenamefont {Rodriguez} \emph
  {et~al.}}]{Rodriguez:2018pss}%
  \BibitemOpen
  \bibfield  {author} {\bibinfo {author} {\bibfnamefont {Carl~L.}\ \bibnamefont
  {Rodriguez}} \emph {et~al.},\ }\href {\doibase 10.1103/PhysRevD.98.123005}
  {\bibfield  {journal} {\bibinfo  {journal} {Phys. Rev.}\ }\textbf {\bibinfo
  {volume} {D98}},\ \bibinfo {pages} {123005} (\bibinfo {year}
  {2018}{\natexlab{a}})},\ \Eprint {http://arxiv.org/abs/1811.04926}
  {arXiv:1811.04926 [astro-ph.HE]} \BibitemShut {NoStop}%
\bibitem [{\citenamefont {O'Leary}\ \emph {et~al.}(2006)\citenamefont
  {O'Leary}, \citenamefont {Rasio}, \citenamefont {Fregeau}, \citenamefont
  {Ivanova},\ and\ \citenamefont {O'Shaughnessy}}]{OLeary:2005vqo}%
  \BibitemOpen
  \bibfield  {author} {\bibinfo {author} {\bibfnamefont {Ryan~M.}\ \bibnamefont
  {O'Leary}}, \bibinfo {author} {\bibfnamefont {Frederic~A.}\ \bibnamefont
  {Rasio}}, \bibinfo {author} {\bibfnamefont {John~M.}\ \bibnamefont
  {Fregeau}}, \bibinfo {author} {\bibfnamefont {Natalia}\ \bibnamefont
  {Ivanova}}, \ and\ \bibinfo {author} {\bibfnamefont {Richard~W.}\
  \bibnamefont {O'Shaughnessy}},\ }\bibfield  {title} {\enquote {\bibinfo
  {title} {{Binary mergers and growth of black holes in dense star
  clusters}},}\ }\href {\doibase 10.1086/498446} {\bibfield  {journal}
  {\bibinfo  {journal} {Astrophys. J.}\ }\textbf {\bibinfo {volume} {637}},\
  \bibinfo {pages} {937--951} (\bibinfo {year} {2006})},\ \Eprint
  {http://arxiv.org/abs/astro-ph/0508224} {arXiv:astro-ph/0508224} \BibitemShut
  {NoStop}%
\bibitem [{\citenamefont {Samsing}(2018)}]{Samsing:2017xmd}%
  \BibitemOpen
  \bibfield  {author} {\bibinfo {author} {\bibfnamefont {Johan}\ \bibnamefont
  {Samsing}},\ }\bibfield  {title} {\enquote {\bibinfo {title} {{Eccentric
  Black Hole Mergers Forming in Globular Clusters}},}\ }\href {\doibase
  10.1103/PhysRevD.97.103014} {\bibfield  {journal} {\bibinfo  {journal} {Phys.
  Rev. D}\ }\textbf {\bibinfo {volume} {97}},\ \bibinfo {pages} {103014}
  (\bibinfo {year} {2018})},\ \Eprint {http://arxiv.org/abs/1711.07452}
  {arXiv:1711.07452 [astro-ph.HE]} \BibitemShut {NoStop}%
\bibitem [{\citenamefont {Fragione}\ and\ \citenamefont
  {Kocsis}(2019)}]{Fragione:2019hqt}%
  \BibitemOpen
  \bibfield  {author} {\bibinfo {author} {\bibfnamefont {Giacomo}\ \bibnamefont
  {Fragione}}\ and\ \bibinfo {author} {\bibfnamefont {Bence}\ \bibnamefont
  {Kocsis}},\ }\bibfield  {title} {\enquote {\bibinfo {title} {{Black hole
  mergers from quadruples}},}\ }\href {\doibase 10.1093/mnras/stz1175}
  {\bibfield  {journal} {\bibinfo  {journal} {Mon. Not. Roy. Astron. Soc.}\
  }\textbf {\bibinfo {volume} {486}},\ \bibinfo {pages} {4781--4789} (\bibinfo
  {year} {2019})},\ \Eprint {http://arxiv.org/abs/1903.03112} {arXiv:1903.03112
  [astro-ph.GA]} \BibitemShut {NoStop}%
\bibitem [{\citenamefont {Kumamoto}\ \emph {et~al.}(2019)\citenamefont
  {Kumamoto}, \citenamefont {Fujii},\ and\ \citenamefont
  {Tanikawa}}]{Kumamoto:2018gdg}%
  \BibitemOpen
  \bibfield  {author} {\bibinfo {author} {\bibfnamefont {Jun}\ \bibnamefont
  {Kumamoto}}, \bibinfo {author} {\bibfnamefont {Michiko~S.}\ \bibnamefont
  {Fujii}}, \ and\ \bibinfo {author} {\bibfnamefont {Ataru}\ \bibnamefont
  {Tanikawa}},\ }\bibfield  {title} {\enquote {\bibinfo {title}
  {{Gravitational-Wave Emission from Binary Black Holes Formed in Open
  Clusters}},}\ }\href {\doibase 10.1093/mnras/stz1068} {\bibfield  {journal}
  {\bibinfo  {journal} {Mon. Not. Roy. Astron. Soc.}\ }\textbf {\bibinfo
  {volume} {486}},\ \bibinfo {pages} {3942--3950} (\bibinfo {year} {2019})},\
  \Eprint {http://arxiv.org/abs/1811.06726} {arXiv:1811.06726 [astro-ph.HE]}
  \BibitemShut {NoStop}%
\bibitem [{\citenamefont {O'Leary}\ \emph {et~al.}(2009)\citenamefont
  {O'Leary}, \citenamefont {Kocsis},\ and\ \citenamefont
  {Loeb}}]{OLeary:2008myb}%
  \BibitemOpen
  \bibfield  {author} {\bibinfo {author} {\bibfnamefont {Ryan~M.}\ \bibnamefont
  {O'Leary}}, \bibinfo {author} {\bibfnamefont {Bence}\ \bibnamefont {Kocsis}},
  \ and\ \bibinfo {author} {\bibfnamefont {Abraham}\ \bibnamefont {Loeb}},\
  }\bibfield  {title} {\enquote {\bibinfo {title} {{Gravitational waves from
  scattering of stellar-mass black holes in galactic nuclei}},}\ }\href
  {\doibase 10.1111/j.1365-2966.2009.14653.x} {\bibfield  {journal} {\bibinfo
  {journal} {Mon. Not. Roy. Astron. Soc.}\ }\textbf {\bibinfo {volume} {395}},\
  \bibinfo {pages} {2127--2146} (\bibinfo {year} {2009})},\ \Eprint
  {http://arxiv.org/abs/0807.2638} {arXiv:0807.2638 [astro-ph]} \BibitemShut
  {NoStop}%
\bibitem [{\citenamefont {Gondán}\ and\ \citenamefont
  {Kocsis}(2020)}]{Gondan:2020svr}%
  \BibitemOpen
  \bibfield  {author} {\bibinfo {author} {\bibfnamefont {László}\
  \bibnamefont {Gondán}}\ and\ \bibinfo {author} {\bibfnamefont {Bence}\
  \bibnamefont {Kocsis}},\ }\bibfield  {title} {\enquote {\bibinfo {title}
  {{High Eccentricities and High Masses Characterize Gravitational-wave
  Captures in Galactic Nuclei as Seen by Earth-based Detectors}},}\ }\href@noop
  {} {\  (\bibinfo {year} {2020})},\ \Eprint {http://arxiv.org/abs/2011.02507}
  {arXiv:2011.02507 [astro-ph.HE]} \BibitemShut {NoStop}%
\bibitem [{\citenamefont {Abbott}\ \emph
  {et~al.}(2020{\natexlab{b}})\citenamefont {Abbott} \emph
  {et~al.}}]{Abbott:2020tfl}%
  \BibitemOpen
  \bibfield  {author} {\bibinfo {author} {\bibfnamefont {R.}~\bibnamefont
  {Abbott}} \emph {et~al.} (\bibinfo {collaboration} {LIGO Scientific,
  Virgo}),\ }\bibfield  {title} {\enquote {\bibinfo {title} {{GW190521: A
  Binary Black Hole Merger with a Total Mass of $150 ~ M_{\odot}$}},}\ }\href
  {\doibase 10.1103/PhysRevLett.125.101102} {\bibfield  {journal} {\bibinfo
  {journal} {Phys. Rev. Lett.}\ }\textbf {\bibinfo {volume} {125}},\ \bibinfo
  {pages} {101102} (\bibinfo {year} {2020}{\natexlab{b}})},\ \Eprint
  {http://arxiv.org/abs/2009.01075} {arXiv:2009.01075 [gr-qc]} \BibitemShut
  {NoStop}%
\bibitem [{\citenamefont {Romero-Shaw}\ \emph {et~al.}(2020)\citenamefont
  {Romero-Shaw}, \citenamefont {Lasky}, \citenamefont {Thrane},\ and\
  \citenamefont {Bustillo}}]{Romero-Shaw:2020thy}%
  \BibitemOpen
  \bibfield  {author} {\bibinfo {author} {\bibfnamefont {Isobel~M.}\
  \bibnamefont {Romero-Shaw}}, \bibinfo {author} {\bibfnamefont {Paul~D.}\
  \bibnamefont {Lasky}}, \bibinfo {author} {\bibfnamefont {Eric}\ \bibnamefont
  {Thrane}}, \ and\ \bibinfo {author} {\bibfnamefont {Juan~Calderon}\
  \bibnamefont {Bustillo}},\ }\bibfield  {title} {\enquote {\bibinfo {title}
  {{GW190521: orbital eccentricity and signatures of dynamical formation in a
  binary black hole merger signal}},}\ }\href@noop {} {\  (\bibinfo {year}
  {2020})},\ \Eprint {http://arxiv.org/abs/2009.04771} {arXiv:2009.04771
  [astro-ph.HE]} \BibitemShut {NoStop}%
\bibitem [{\citenamefont {Gayathri}\ \emph {et~al.}(2020)\citenamefont
  {Gayathri}, \citenamefont {Healy}, \citenamefont {Lange}, \citenamefont
  {O'Brien}, \citenamefont {Szczepanczyk}, \citenamefont {Bartos},
  \citenamefont {Campanelli}, \citenamefont {Klimenko}, \citenamefont
  {Lousto},\ and\ \citenamefont {O'Shaughnessy}}]{Gayathri:2020coq}%
  \BibitemOpen
  \bibfield  {author} {\bibinfo {author} {\bibfnamefont {V.}~\bibnamefont
  {Gayathri}}, \bibinfo {author} {\bibfnamefont {J.}~\bibnamefont {Healy}},
  \bibinfo {author} {\bibfnamefont {J.}~\bibnamefont {Lange}}, \bibinfo
  {author} {\bibfnamefont {B.}~\bibnamefont {O'Brien}}, \bibinfo {author}
  {\bibfnamefont {M.}~\bibnamefont {Szczepanczyk}}, \bibinfo {author}
  {\bibfnamefont {I.}~\bibnamefont {Bartos}}, \bibinfo {author} {\bibfnamefont
  {M.}~\bibnamefont {Campanelli}}, \bibinfo {author} {\bibfnamefont
  {S.}~\bibnamefont {Klimenko}}, \bibinfo {author} {\bibfnamefont
  {C.}~\bibnamefont {Lousto}}, \ and\ \bibinfo {author} {\bibfnamefont
  {R.}~\bibnamefont {O'Shaughnessy}},\ }\bibfield  {title} {\enquote {\bibinfo
  {title} {{GW190521 as a Highly Eccentric Black Hole Merger}},}\ }\href@noop
  {} {\  (\bibinfo {year} {2020})},\ \Eprint {http://arxiv.org/abs/2009.05461}
  {arXiv:2009.05461 [astro-ph.HE]} \BibitemShut {NoStop}%
\bibitem [{\citenamefont {Calder\'on~Bustillo}\ \emph
  {et~al.}(2020{\natexlab{a}})\citenamefont {Calder\'on~Bustillo},
  \citenamefont {Sanchis-Gual}, \citenamefont {Torres-Forn\'e},\ and\
  \citenamefont {Font}}]{CalderonBustillo:2020odh}%
  \BibitemOpen
  \bibfield  {author} {\bibinfo {author} {\bibfnamefont {Juan}\ \bibnamefont
  {Calder\'on~Bustillo}}, \bibinfo {author} {\bibfnamefont {Nicolas}\
  \bibnamefont {Sanchis-Gual}}, \bibinfo {author} {\bibfnamefont {Alejandro}\
  \bibnamefont {Torres-Forn\'e}}, \ and\ \bibinfo {author} {\bibfnamefont
  {Jos\'e~A.}\ \bibnamefont {Font}},\ }\bibfield  {title} {\enquote {\bibinfo
  {title} {{Confusing head-on and precessing intermediate-mass binary black
  hole mergers}},}\ }\href@noop {} {\  (\bibinfo {year}
  {2020}{\natexlab{a}})},\ \Eprint {http://arxiv.org/abs/2009.01066}
  {arXiv:2009.01066 [gr-qc]} \BibitemShut {NoStop}%
\bibitem [{\citenamefont {Calder\'on~Bustillo}\ \emph
  {et~al.}(2020{\natexlab{b}})\citenamefont {Calder\'on~Bustillo},
  \citenamefont {Sanchis-Gual}, \citenamefont {Torres-Forn\'e}, \citenamefont
  {Font}, \citenamefont {Vajpeyi}, \citenamefont {Smith}, \citenamefont
  {Herdeiro}, \citenamefont {Radu},\ and\ \citenamefont
  {Leong}}]{CalderonBustillo:2020srq}%
  \BibitemOpen
  \bibfield  {author} {\bibinfo {author} {\bibfnamefont {Juan}\ \bibnamefont
  {Calder\'on~Bustillo}}, \bibinfo {author} {\bibfnamefont {Nicolas}\
  \bibnamefont {Sanchis-Gual}}, \bibinfo {author} {\bibfnamefont {Alejandro}\
  \bibnamefont {Torres-Forn\'e}}, \bibinfo {author} {\bibfnamefont {Jos\'e~A.}\
  \bibnamefont {Font}}, \bibinfo {author} {\bibfnamefont {Avi}\ \bibnamefont
  {Vajpeyi}}, \bibinfo {author} {\bibfnamefont {Rory}\ \bibnamefont {Smith}},
  \bibinfo {author} {\bibfnamefont {Carlos}\ \bibnamefont {Herdeiro}}, \bibinfo
  {author} {\bibfnamefont {Eugen}\ \bibnamefont {Radu}}, \ and\ \bibinfo
  {author} {\bibfnamefont {Samson~H.W.}\ \bibnamefont {Leong}},\ }\bibfield
  {title} {\enquote {\bibinfo {title} {{The (ultra) light in the dark: A
  potential vector boson of $8.7\times 10^{-13}$ eV from GW190521}},}\
  }\href@noop {} {\  (\bibinfo {year} {2020}{\natexlab{b}})},\ \Eprint
  {http://arxiv.org/abs/2009.05376} {arXiv:2009.05376 [gr-qc]} \BibitemShut
  {NoStop}%
\bibitem [{\citenamefont {Zevin}\ \emph {et~al.}(2019)\citenamefont {Zevin},
  \citenamefont {Samsing}, \citenamefont {Rodriguez}, \citenamefont {Haster},\
  and\ \citenamefont {Ramirez-Ruiz}}]{Zevin:2018kzq}%
  \BibitemOpen
  \bibfield  {author} {\bibinfo {author} {\bibfnamefont {Michael}\ \bibnamefont
  {Zevin}}, \bibinfo {author} {\bibfnamefont {Johan}\ \bibnamefont {Samsing}},
  \bibinfo {author} {\bibfnamefont {Carl}\ \bibnamefont {Rodriguez}}, \bibinfo
  {author} {\bibfnamefont {Carl-Johan}\ \bibnamefont {Haster}}, \ and\ \bibinfo
  {author} {\bibfnamefont {Enrico}\ \bibnamefont {Ramirez-Ruiz}},\ }\bibfield
  {title} {\enquote {\bibinfo {title} {{Eccentric Black Hole Mergers in Dense
  Star Clusters: The Role of Binary\textendash{}Binary Encounters}},}\ }\href
  {\doibase 10.3847/1538-4357/aaf6ec} {\bibfield  {journal} {\bibinfo
  {journal} {Astrophys. J.}\ }\textbf {\bibinfo {volume} {871}},\ \bibinfo
  {pages} {91} (\bibinfo {year} {2019})},\ \Eprint
  {http://arxiv.org/abs/1810.00901} {arXiv:1810.00901 [astro-ph.HE]}
  \BibitemShut {NoStop}%
\bibitem [{\citenamefont {Nishizawa}\ \emph {et~al.}(2017)\citenamefont
  {Nishizawa}, \citenamefont {Sesana}, \citenamefont {Berti},\ and\
  \citenamefont {Klein}}]{Nishizawa:2016eza}%
  \BibitemOpen
  \bibfield  {author} {\bibinfo {author} {\bibfnamefont {Atsushi}\ \bibnamefont
  {Nishizawa}}, \bibinfo {author} {\bibfnamefont {Alberto}\ \bibnamefont
  {Sesana}}, \bibinfo {author} {\bibfnamefont {Emanuele}\ \bibnamefont
  {Berti}}, \ and\ \bibinfo {author} {\bibfnamefont {Antoine}\ \bibnamefont
  {Klein}},\ }\bibfield  {title} {\enquote {\bibinfo {title} {{Constraining
  stellar binary black hole formation scenarios with eLISA eccentricity
  measurements}},}\ }\href {\doibase 10.1093/mnras/stw2993} {\bibfield
  {journal} {\bibinfo  {journal} {Mon. Not. Roy. Astron. Soc.}\ }\textbf
  {\bibinfo {volume} {465}},\ \bibinfo {pages} {4375--4380} (\bibinfo {year}
  {2017})},\ \Eprint {http://arxiv.org/abs/1606.09295} {arXiv:1606.09295
  [astro-ph.HE]} \BibitemShut {NoStop}%
\bibitem [{\citenamefont {Nishizawa}\ \emph {et~al.}(2016)\citenamefont
  {Nishizawa}, \citenamefont {Berti}, \citenamefont {Klein},\ and\
  \citenamefont {Sesana}}]{Nishizawa:2016jji}%
  \BibitemOpen
  \bibfield  {author} {\bibinfo {author} {\bibfnamefont {Atsushi}\ \bibnamefont
  {Nishizawa}}, \bibinfo {author} {\bibfnamefont {Emanuele}\ \bibnamefont
  {Berti}}, \bibinfo {author} {\bibfnamefont {Antoine}\ \bibnamefont {Klein}},
  \ and\ \bibinfo {author} {\bibfnamefont {Alberto}\ \bibnamefont {Sesana}},\
  }\bibfield  {title} {\enquote {\bibinfo {title} {{eLISA eccentricity
  measurements as tracers of binary black hole formation}},}\ }\href {\doibase
  10.1103/PhysRevD.94.064020} {\bibfield  {journal} {\bibinfo  {journal} {Phys.
  Rev.}\ }\textbf {\bibinfo {volume} {D94}},\ \bibinfo {pages} {064020}
  (\bibinfo {year} {2016})},\ \Eprint {http://arxiv.org/abs/1605.01341}
  {arXiv:1605.01341 [gr-qc]} \BibitemShut {NoStop}%
\bibitem [{\citenamefont {Breivik}\ \emph {et~al.}(2016)\citenamefont
  {Breivik}, \citenamefont {Rodriguez}, \citenamefont {Larson}, \citenamefont
  {Kalogera},\ and\ \citenamefont {Rasio}}]{Breivik:2016ddj}%
  \BibitemOpen
  \bibfield  {author} {\bibinfo {author} {\bibfnamefont {Katelyn}\ \bibnamefont
  {Breivik}}, \bibinfo {author} {\bibfnamefont {Carl~L.}\ \bibnamefont
  {Rodriguez}}, \bibinfo {author} {\bibfnamefont {Shane~L.}\ \bibnamefont
  {Larson}}, \bibinfo {author} {\bibfnamefont {Vassiliki}\ \bibnamefont
  {Kalogera}}, \ and\ \bibinfo {author} {\bibfnamefont {Frederic~A.}\
  \bibnamefont {Rasio}},\ }\bibfield  {title} {\enquote {\bibinfo {title}
  {{Distinguishing Between Formation Channels for Binary Black Holes with
  LISA}},}\ }\href {\doibase 10.3847/2041-8205/830/1/L18} {\bibfield  {journal}
  {\bibinfo  {journal} {Astrophys. J. Lett.}\ }\textbf {\bibinfo {volume}
  {830}},\ \bibinfo {pages} {L18} (\bibinfo {year} {2016})},\ \Eprint
  {http://arxiv.org/abs/1606.09558} {arXiv:1606.09558 [astro-ph.GA]}
  \BibitemShut {NoStop}%
\bibitem [{\citenamefont {Fang}\ \emph {et~al.}(2019)\citenamefont {Fang},
  \citenamefont {Thompson},\ and\ \citenamefont {Hirata}}]{Fang:2019dnh}%
  \BibitemOpen
  \bibfield  {author} {\bibinfo {author} {\bibfnamefont {Xiao}\ \bibnamefont
  {Fang}}, \bibinfo {author} {\bibfnamefont {Todd~A.}\ \bibnamefont
  {Thompson}}, \ and\ \bibinfo {author} {\bibfnamefont {Christopher~M.}\
  \bibnamefont {Hirata}},\ }\bibfield  {title} {\enquote {\bibinfo {title}
  {{The Population of Eccentric Binary Black Holes: Implications for mHz
  Gravitational Wave Experiments}},}\ }\href {\doibase
  10.3847/1538-4357/ab0e6a} {\bibfield  {journal} {\bibinfo  {journal}
  {Astrophys. J.}\ }\textbf {\bibinfo {volume} {875}},\ \bibinfo {pages} {75}
  (\bibinfo {year} {2019})},\ \Eprint {http://arxiv.org/abs/1901.05092}
  {arXiv:1901.05092 [astro-ph.HE]} \BibitemShut {NoStop}%
\bibitem [{\citenamefont {Rodriguez}\ \emph
  {et~al.}(2018{\natexlab{b}})\citenamefont {Rodriguez}, \citenamefont
  {Amaro-Seoane}, \citenamefont {Chatterjee},\ and\ \citenamefont
  {Rasio}}]{Rodriguez:2017pec}%
  \BibitemOpen
  \bibfield  {author} {\bibinfo {author} {\bibfnamefont {Carl~L.}\ \bibnamefont
  {Rodriguez}}, \bibinfo {author} {\bibfnamefont {Pau}\ \bibnamefont
  {Amaro-Seoane}}, \bibinfo {author} {\bibfnamefont {Sourav}\ \bibnamefont
  {Chatterjee}}, \ and\ \bibinfo {author} {\bibfnamefont {Frederic~A.}\
  \bibnamefont {Rasio}},\ }\bibfield  {title} {\enquote {\bibinfo {title}
  {{Post-Newtonian Dynamics in Dense Star Clusters: Highly-Eccentric,
  Highly-Spinning, and Repeated Binary Black Hole Mergers}},}\ }\href {\doibase
  10.1103/PhysRevLett.120.151101} {\bibfield  {journal} {\bibinfo  {journal}
  {Phys. Rev. Lett.}\ }\textbf {\bibinfo {volume} {120}},\ \bibinfo {pages}
  {151101} (\bibinfo {year} {2018}{\natexlab{b}})},\ \Eprint
  {http://arxiv.org/abs/1712.04937} {arXiv:1712.04937 [astro-ph.HE]}
  \BibitemShut {NoStop}%
\bibitem [{\citenamefont {Gond\'an}\ \emph {et~al.}(2018)\citenamefont
  {Gond\'an}, \citenamefont {Kocsis}, \citenamefont {Raffai},\ and\
  \citenamefont {Frei}}]{Gondan:2017wzd}%
  \BibitemOpen
  \bibfield  {author} {\bibinfo {author} {\bibfnamefont {L\'aszl\'o}\
  \bibnamefont {Gond\'an}}, \bibinfo {author} {\bibfnamefont {Bence}\
  \bibnamefont {Kocsis}}, \bibinfo {author} {\bibfnamefont {P\'eter}\
  \bibnamefont {Raffai}}, \ and\ \bibinfo {author} {\bibfnamefont {Zsolt}\
  \bibnamefont {Frei}},\ }\bibfield  {title} {\enquote {\bibinfo {title}
  {{Eccentric Black Hole Gravitational-Wave Capture Sources in Galactic Nuclei:
  Distribution of Binary Parameters}},}\ }\href {\doibase
  10.3847/1538-4357/aabfee} {\bibfield  {journal} {\bibinfo  {journal}
  {Astrophys. J.}\ }\textbf {\bibinfo {volume} {860}},\ \bibinfo {pages} {5}
  (\bibinfo {year} {2018})},\ \Eprint {http://arxiv.org/abs/1711.09989}
  {arXiv:1711.09989 [astro-ph.HE]} \BibitemShut {NoStop}%
\bibitem [{\citenamefont {Tagawa}\ \emph {et~al.}(2020)\citenamefont {Tagawa},
  \citenamefont {Kocsis}, \citenamefont {Haiman}, \citenamefont {Bartos},
  \citenamefont {Omukai},\ and\ \citenamefont {Samsing}}]{Tagawa:2020jnc}%
  \BibitemOpen
  \bibfield  {author} {\bibinfo {author} {\bibfnamefont {Hiromichi}\
  \bibnamefont {Tagawa}}, \bibinfo {author} {\bibfnamefont {Bence}\
  \bibnamefont {Kocsis}}, \bibinfo {author} {\bibfnamefont {Zoltan}\
  \bibnamefont {Haiman}}, \bibinfo {author} {\bibfnamefont {Imre}\ \bibnamefont
  {Bartos}}, \bibinfo {author} {\bibfnamefont {Kazuyuki}\ \bibnamefont
  {Omukai}}, \ and\ \bibinfo {author} {\bibfnamefont {Johan}\ \bibnamefont
  {Samsing}},\ }\bibfield  {title} {\enquote {\bibinfo {title} {{Eccentric
  Black Hole Mergers in Active Galactic Nuclei}},}\ }\href@noop {} {\
  (\bibinfo {year} {2020})},\ \Eprint {http://arxiv.org/abs/2010.10526}
  {arXiv:2010.10526 [astro-ph.HE]} \BibitemShut {NoStop}%
\bibitem [{\citenamefont {Ramos-Buades}\ \emph
  {et~al.}(2020{\natexlab{b}})\citenamefont {Ramos-Buades}, \citenamefont
  {Husa}, \citenamefont {Pratten}, \citenamefont {Estell\'es}, \citenamefont
  {Garc\'\i{}a-Quir\'os}, \citenamefont {Mateu-Lucena}, \citenamefont
  {Colleoni},\ and\ \citenamefont {Jaume}}]{Ramos-Buades:2019uvh}%
  \BibitemOpen
  \bibfield  {author} {\bibinfo {author} {\bibfnamefont {Antoni}\ \bibnamefont
  {Ramos-Buades}}, \bibinfo {author} {\bibfnamefont {Sascha}\ \bibnamefont
  {Husa}}, \bibinfo {author} {\bibfnamefont {Geraint}\ \bibnamefont {Pratten}},
  \bibinfo {author} {\bibfnamefont {H\'ector}\ \bibnamefont {Estell\'es}},
  \bibinfo {author} {\bibfnamefont {Cecilio}\ \bibnamefont
  {Garc\'\i{}a-Quir\'os}}, \bibinfo {author} {\bibfnamefont {Maite}\
  \bibnamefont {Mateu-Lucena}}, \bibinfo {author} {\bibfnamefont {Marta}\
  \bibnamefont {Colleoni}}, \ and\ \bibinfo {author} {\bibfnamefont {Rafel}\
  \bibnamefont {Jaume}},\ }\bibfield  {title} {\enquote {\bibinfo {title}
  {{First survey of spinning eccentric black hole mergers: Numerical relativity
  simulations, hybrid waveforms, and parameter estimation}},}\ }\href {\doibase
  10.1103/PhysRevD.101.083015} {\bibfield  {journal} {\bibinfo  {journal}
  {Phys. Rev. D}\ }\textbf {\bibinfo {volume} {101}},\ \bibinfo {pages}
  {083015} (\bibinfo {year} {2020}{\natexlab{b}})},\ \Eprint
  {http://arxiv.org/abs/1909.11011} {arXiv:1909.11011 [gr-qc]} \BibitemShut
  {NoStop}%
\bibitem [{\citenamefont {Klein}\ \emph {et~al.}(2018)\citenamefont {Klein},
  \citenamefont {Boetzel}, \citenamefont {Gopakumar}, \citenamefont {Jetzer},\
  and\ \citenamefont {de~Vittori}}]{Klein:2018ybm}%
  \BibitemOpen
  \bibfield  {author} {\bibinfo {author} {\bibfnamefont {Antoine}\ \bibnamefont
  {Klein}}, \bibinfo {author} {\bibfnamefont {Yannick}\ \bibnamefont
  {Boetzel}}, \bibinfo {author} {\bibfnamefont {Achamveedu}\ \bibnamefont
  {Gopakumar}}, \bibinfo {author} {\bibfnamefont {Philippe}\ \bibnamefont
  {Jetzer}}, \ and\ \bibinfo {author} {\bibfnamefont {Lorenzo}\ \bibnamefont
  {de~Vittori}},\ }\bibfield  {title} {\enquote {\bibinfo {title} {{Fourier
  domain gravitational waveforms for precessing eccentric binaries}},}\ }\href
  {\doibase 10.1103/PhysRevD.98.104043} {\bibfield  {journal} {\bibinfo
  {journal} {Phys. Rev.}\ }\textbf {\bibinfo {volume} {D98}},\ \bibinfo {pages}
  {104043} (\bibinfo {year} {2018})},\ \Eprint
  {http://arxiv.org/abs/1801.08542} {arXiv:1801.08542 [gr-qc]} \BibitemShut
  {NoStop}%
\bibitem [{\citenamefont {Tiwari}\ and\ \citenamefont
  {Gopakumar}(2020)}]{Tiwari:2020hsu}%
  \BibitemOpen
  \bibfield  {author} {\bibinfo {author} {\bibfnamefont {Srishti}\ \bibnamefont
  {Tiwari}}\ and\ \bibinfo {author} {\bibfnamefont {Achamveedu}\ \bibnamefont
  {Gopakumar}},\ }\bibfield  {title} {\enquote {\bibinfo {title} {{Combining
  post-circular and Padé approximations to compute Fourier domain templates
  for eccentric inspirals}},}\ }\href {\doibase 10.1103/PhysRevD.102.084042}
  {\bibfield  {journal} {\bibinfo  {journal} {Phys. Rev.}\ }\textbf {\bibinfo
  {volume} {D102}},\ \bibinfo {pages} {084042} (\bibinfo {year} {2020})},\
  \Eprint {http://arxiv.org/abs/2009.11333} {arXiv:2009.11333 [gr-qc]}
  \BibitemShut {NoStop}%
\bibitem [{\citenamefont {Moore}\ \emph {et~al.}(2018)\citenamefont {Moore},
  \citenamefont {Robson}, \citenamefont {Loutrel},\ and\ \citenamefont
  {Yunes}}]{Moore:2018kvz}%
  \BibitemOpen
  \bibfield  {author} {\bibinfo {author} {\bibfnamefont {Blake}\ \bibnamefont
  {Moore}}, \bibinfo {author} {\bibfnamefont {Travis}\ \bibnamefont {Robson}},
  \bibinfo {author} {\bibfnamefont {Nicholas}\ \bibnamefont {Loutrel}}, \ and\
  \bibinfo {author} {\bibfnamefont {Nicolas}\ \bibnamefont {Yunes}},\
  }\bibfield  {title} {\enquote {\bibinfo {title} {{Towards a Fourier domain
  waveform for non-spinning binaries with arbitrary eccentricity}},}\ }\href
  {\doibase 10.1088/1361-6382/aaea00} {\bibfield  {journal} {\bibinfo
  {journal} {Class. Quant. Grav.}\ }\textbf {\bibinfo {volume} {35}},\ \bibinfo
  {pages} {235006} (\bibinfo {year} {2018})},\ \Eprint
  {http://arxiv.org/abs/1807.07163} {arXiv:1807.07163 [gr-qc]} \BibitemShut
  {NoStop}%
\bibitem [{\citenamefont {Moore}\ and\ \citenamefont
  {Yunes}(2019)}]{Moore:2019xkm}%
  \BibitemOpen
  \bibfield  {author} {\bibinfo {author} {\bibfnamefont {Blake}\ \bibnamefont
  {Moore}}\ and\ \bibinfo {author} {\bibfnamefont {Nicolás}\ \bibnamefont
  {Yunes}},\ }\bibfield  {title} {\enquote {\bibinfo {title} {{A 3PN Fourier
  Domain Waveform for Non-Spinning Binaries with Moderate Eccentricity}},}\
  }\href {\doibase 10.1088/1361-6382/ab3778} {\bibfield  {journal} {\bibinfo
  {journal} {Class. Quant. Grav.}\ }\textbf {\bibinfo {volume} {36}},\ \bibinfo
  {pages} {185003} (\bibinfo {year} {2019})},\ \Eprint
  {http://arxiv.org/abs/1903.05203} {arXiv:1903.05203 [gr-qc]} \BibitemShut
  {NoStop}%
\bibitem [{\citenamefont {Liu}\ \emph {et~al.}(2020)\citenamefont {Liu},
  \citenamefont {Cao},\ and\ \citenamefont {Shao}}]{Liu:2019jpg}%
  \BibitemOpen
  \bibfield  {author} {\bibinfo {author} {\bibfnamefont {Xiaolin}\ \bibnamefont
  {Liu}}, \bibinfo {author} {\bibfnamefont {Zhoujian}\ \bibnamefont {Cao}}, \
  and\ \bibinfo {author} {\bibfnamefont {Lijing}\ \bibnamefont {Shao}},\
  }\bibfield  {title} {\enquote {\bibinfo {title} {{Validating the
  Effective-One-Body Numerical-Relativity Waveform Models for Spin-aligned
  Binary Black Holes along Eccentric Orbits}},}\ }\href {\doibase
  10.1103/PhysRevD.101.044049} {\bibfield  {journal} {\bibinfo  {journal}
  {Phys. Rev.}\ }\textbf {\bibinfo {volume} {D101}},\ \bibinfo {pages} {044049}
  (\bibinfo {year} {2020})},\ \Eprint {http://arxiv.org/abs/1910.00784}
  {arXiv:1910.00784 [gr-qc]} \BibitemShut {NoStop}%
\bibitem [{\citenamefont {Tanay}\ \emph {et~al.}(2019)\citenamefont {Tanay},
  \citenamefont {Klein}, \citenamefont {Berti},\ and\ \citenamefont
  {Nishizawa}}]{Tanay:2019knc}%
  \BibitemOpen
  \bibfield  {author} {\bibinfo {author} {\bibfnamefont {Sashwat}\ \bibnamefont
  {Tanay}}, \bibinfo {author} {\bibfnamefont {Antoine}\ \bibnamefont {Klein}},
  \bibinfo {author} {\bibfnamefont {Emanuele}\ \bibnamefont {Berti}}, \ and\
  \bibinfo {author} {\bibfnamefont {Atsushi}\ \bibnamefont {Nishizawa}},\
  }\bibfield  {title} {\enquote {\bibinfo {title} {{Convergence of
  Fourier-domain templates for inspiraling eccentric compact binaries}},}\
  }\href {\doibase 10.1103/PhysRevD.100.064006} {\bibfield  {journal} {\bibinfo
   {journal} {Phys. Rev.}\ }\textbf {\bibinfo {volume} {D100}},\ \bibinfo
  {pages} {064006} (\bibinfo {year} {2019})},\ \Eprint
  {http://arxiv.org/abs/1905.08811} {arXiv:1905.08811 [gr-qc]} \BibitemShut
  {NoStop}%
\bibitem [{\citenamefont {Hinderer}\ and\ \citenamefont
  {Babak}(2017)}]{Hinderer:2017jcs}%
  \BibitemOpen
  \bibfield  {author} {\bibinfo {author} {\bibfnamefont {Tanja}\ \bibnamefont
  {Hinderer}}\ and\ \bibinfo {author} {\bibfnamefont {Stanislav}\ \bibnamefont
  {Babak}},\ }\bibfield  {title} {\enquote {\bibinfo {title} {{Foundations of
  an effective-one-body model for coalescing binaries on eccentric orbits}},}\
  }\href {\doibase 10.1103/PhysRevD.96.104048} {\bibfield  {journal} {\bibinfo
  {journal} {Phys. Rev. D}\ }\textbf {\bibinfo {volume} {96}},\ \bibinfo
  {pages} {104048} (\bibinfo {year} {2017})},\ \Eprint
  {http://arxiv.org/abs/1707.08426} {arXiv:1707.08426 [gr-qc]} \BibitemShut
  {NoStop}%
\bibitem [{\citenamefont {Hinder}\ \emph {et~al.}(2018)\citenamefont {Hinder},
  \citenamefont {Kidder},\ and\ \citenamefont {Pfeiffer}}]{Hinder:2017sxy}%
  \BibitemOpen
  \bibfield  {author} {\bibinfo {author} {\bibfnamefont {Ian}\ \bibnamefont
  {Hinder}}, \bibinfo {author} {\bibfnamefont {Lawrence~E.}\ \bibnamefont
  {Kidder}}, \ and\ \bibinfo {author} {\bibfnamefont {Harald~P.}\ \bibnamefont
  {Pfeiffer}},\ }\bibfield  {title} {\enquote {\bibinfo {title} {{Eccentric
  binary black hole inspiral-merger-ringdown gravitational waveform model from
  numerical relativity and post-Newtonian theory}},}\ }\href {\doibase
  10.1103/PhysRevD.98.044015} {\bibfield  {journal} {\bibinfo  {journal} {Phys.
  Rev. D}\ }\textbf {\bibinfo {volume} {98}},\ \bibinfo {pages} {044015}
  (\bibinfo {year} {2018})},\ \Eprint {http://arxiv.org/abs/1709.02007}
  {arXiv:1709.02007 [gr-qc]} \BibitemShut {NoStop}%
\bibitem [{\citenamefont {Huerta}\ \emph {et~al.}(2018)\citenamefont {Huerta}
  \emph {et~al.}}]{Huerta:2017kez}%
  \BibitemOpen
  \bibfield  {author} {\bibinfo {author} {\bibfnamefont {E.A.}\ \bibnamefont
  {Huerta}} \emph {et~al.},\ }\bibfield  {title} {\enquote {\bibinfo {title}
  {{Eccentric, nonspinning, inspiral, Gaussian-process merger approximant for
  the detection and characterization of eccentric binary black hole
  mergers}},}\ }\href {\doibase 10.1103/PhysRevD.97.024031} {\bibfield
  {journal} {\bibinfo  {journal} {Phys. Rev. D}\ }\textbf {\bibinfo {volume}
  {97}},\ \bibinfo {pages} {024031} (\bibinfo {year} {2018})},\ \Eprint
  {http://arxiv.org/abs/1711.06276} {arXiv:1711.06276 [gr-qc]} \BibitemShut
  {NoStop}%
\bibitem [{\citenamefont {Chen}\ \emph {et~al.}(2020)\citenamefont {Chen},
  \citenamefont {Huerta}, \citenamefont {Adamo}, \citenamefont {Haas},
  \citenamefont {O'Shea}, \citenamefont {Kumar},\ and\ \citenamefont
  {Moore}}]{Chen:2020lzc}%
  \BibitemOpen
  \bibfield  {author} {\bibinfo {author} {\bibfnamefont {Zhuo}\ \bibnamefont
  {Chen}}, \bibinfo {author} {\bibfnamefont {E.A.}\ \bibnamefont {Huerta}},
  \bibinfo {author} {\bibfnamefont {Joseph}\ \bibnamefont {Adamo}}, \bibinfo
  {author} {\bibfnamefont {Roland}\ \bibnamefont {Haas}}, \bibinfo {author}
  {\bibfnamefont {Eamonn}\ \bibnamefont {O'Shea}}, \bibinfo {author}
  {\bibfnamefont {Prayush}\ \bibnamefont {Kumar}}, \ and\ \bibinfo {author}
  {\bibfnamefont {Chris}\ \bibnamefont {Moore}},\ }\bibfield  {title} {\enquote
  {\bibinfo {title} {{Observation of eccentric binary black hole mergers with
  second and third generation gravitational wave detector networks}},}\
  }\href@noop {} {\  (\bibinfo {year} {2020})},\ \Eprint
  {http://arxiv.org/abs/2008.03313} {arXiv:2008.03313 [gr-qc]} \BibitemShut
  {NoStop}%
\bibitem [{\citenamefont {Chiaramello}\ and\ \citenamefont
  {Nagar}(2020)}]{Chiaramello:2020ehz}%
  \BibitemOpen
  \bibfield  {author} {\bibinfo {author} {\bibfnamefont {Danilo}\ \bibnamefont
  {Chiaramello}}\ and\ \bibinfo {author} {\bibfnamefont {Alessandro}\
  \bibnamefont {Nagar}},\ }\bibfield  {title} {\enquote {\bibinfo {title}
  {{Faithful analytical effective-one-body waveform model for spin-aligned,
  moderately eccentric, coalescing black hole binaries}},}\ }\href {\doibase
  10.1103/PhysRevD.101.101501} {\bibfield  {journal} {\bibinfo  {journal}
  {Phys. Rev. D}\ }\textbf {\bibinfo {volume} {101}},\ \bibinfo {pages}
  {101501} (\bibinfo {year} {2020})},\ \Eprint
  {http://arxiv.org/abs/2001.11736} {arXiv:2001.11736 [gr-qc]} \BibitemShut
  {NoStop}%
\bibitem [{\citenamefont {Cao}\ and\ \citenamefont {Han}(2017)}]{Cao:2017ndf}%
  \BibitemOpen
  \bibfield  {author} {\bibinfo {author} {\bibfnamefont {Zhoujian}\
  \bibnamefont {Cao}}\ and\ \bibinfo {author} {\bibfnamefont {Wen-Biao}\
  \bibnamefont {Han}},\ }\bibfield  {title} {\enquote {\bibinfo {title}
  {{Waveform model for an eccentric binary black hole based on the
  effective-one-body-numerical-relativity formalism}},}\ }\href {\doibase
  10.1103/PhysRevD.96.044028} {\bibfield  {journal} {\bibinfo  {journal} {Phys.
  Rev. D}\ }\textbf {\bibinfo {volume} {96}},\ \bibinfo {pages} {044028}
  (\bibinfo {year} {2017})},\ \Eprint {http://arxiv.org/abs/1708.00166}
  {arXiv:1708.00166 [gr-qc]} \BibitemShut {NoStop}%
\bibitem [{\citenamefont {Taracchini}\ \emph {et~al.}(2012)\citenamefont
  {Taracchini}, \citenamefont {Pan}, \citenamefont {Buonanno}, \citenamefont
  {Barausse}, \citenamefont {Boyle}, \citenamefont {Chu}, \citenamefont
  {Lovelace}, \citenamefont {Pfeiffer},\ and\ \citenamefont
  {Scheel}}]{Taracchini:2012ig}%
  \BibitemOpen
  \bibfield  {author} {\bibinfo {author} {\bibfnamefont {Andrea}\ \bibnamefont
  {Taracchini}}, \bibinfo {author} {\bibfnamefont {Yi}~\bibnamefont {Pan}},
  \bibinfo {author} {\bibfnamefont {Alessandra}\ \bibnamefont {Buonanno}},
  \bibinfo {author} {\bibfnamefont {Enrico}\ \bibnamefont {Barausse}}, \bibinfo
  {author} {\bibfnamefont {Michael}\ \bibnamefont {Boyle}}, \bibinfo {author}
  {\bibfnamefont {Tony}\ \bibnamefont {Chu}}, \bibinfo {author} {\bibfnamefont
  {Geoffrey}\ \bibnamefont {Lovelace}}, \bibinfo {author} {\bibfnamefont
  {Harald~P.}\ \bibnamefont {Pfeiffer}}, \ and\ \bibinfo {author}
  {\bibfnamefont {Mark~A.}\ \bibnamefont {Scheel}},\ }\bibfield  {title}
  {\enquote {\bibinfo {title} {{Prototype effective-one-body model for
  nonprecessing spinning inspiral-merger-ringdown waveforms}},}\ }\href
  {\doibase 10.1103/PhysRevD.86.024011} {\bibfield  {journal} {\bibinfo
  {journal} {Phys. Rev. D}\ }\textbf {\bibinfo {volume} {86}},\ \bibinfo
  {pages} {024011} (\bibinfo {year} {2012})},\ \Eprint
  {http://arxiv.org/abs/1202.0790} {arXiv:1202.0790 [gr-qc]} \BibitemShut
  {NoStop}%
\bibitem [{\citenamefont {Nagar}\ \emph {et~al.}(2018)\citenamefont {Nagar}
  \emph {et~al.}}]{Nagar:2018zoe}%
  \BibitemOpen
  \bibfield  {author} {\bibinfo {author} {\bibfnamefont {Alessandro}\
  \bibnamefont {Nagar}} \emph {et~al.},\ }\bibfield  {title} {\enquote
  {\bibinfo {title} {{Time-domain effective-one-body gravitational waveforms
  for coalescing compact binaries with nonprecessing spins, tides and self-spin
  effects}},}\ }\href {\doibase 10.1103/PhysRevD.98.104052} {\bibfield
  {journal} {\bibinfo  {journal} {Phys. Rev. D}\ }\textbf {\bibinfo {volume}
  {98}},\ \bibinfo {pages} {104052} (\bibinfo {year} {2018})},\ \Eprint
  {http://arxiv.org/abs/1806.01772} {arXiv:1806.01772 [gr-qc]} \BibitemShut
  {NoStop}%
\bibitem [{\citenamefont {Nagar}\ \emph {et~al.}(2020)\citenamefont {Nagar},
  \citenamefont {Riemenschneider}, \citenamefont {Pratten}, \citenamefont
  {Rettegno},\ and\ \citenamefont {Messina}}]{Nagar:2020pcj}%
  \BibitemOpen
  \bibfield  {author} {\bibinfo {author} {\bibfnamefont {Alessandro}\
  \bibnamefont {Nagar}}, \bibinfo {author} {\bibfnamefont {Gunnar}\
  \bibnamefont {Riemenschneider}}, \bibinfo {author} {\bibfnamefont {Geraint}\
  \bibnamefont {Pratten}}, \bibinfo {author} {\bibfnamefont {Piero}\
  \bibnamefont {Rettegno}}, \ and\ \bibinfo {author} {\bibfnamefont
  {Francesco}\ \bibnamefont {Messina}},\ }\bibfield  {title} {\enquote
  {\bibinfo {title} {{Multipolar effective one body waveform model for
  spin-aligned black hole binaries}},}\ }\href {\doibase
  10.1103/PhysRevD.102.024077} {\bibfield  {journal} {\bibinfo  {journal}
  {Phys. Rev. D}\ }\textbf {\bibinfo {volume} {102}},\ \bibinfo {pages}
  {024077} (\bibinfo {year} {2020})},\ \Eprint
  {http://arxiv.org/abs/2001.09082} {arXiv:2001.09082 [gr-qc]} \BibitemShut
  {NoStop}%
\bibitem [{\citenamefont {Nagar}\ \emph {et~al.}(2021)\citenamefont {Nagar},
  \citenamefont {Bonino},\ and\ \citenamefont {Rettegno}}]{Nagar:2021gss}%
  \BibitemOpen
  \bibfield  {author} {\bibinfo {author} {\bibfnamefont {Alessandro}\
  \bibnamefont {Nagar}}, \bibinfo {author} {\bibfnamefont {Alice}\ \bibnamefont
  {Bonino}}, \ and\ \bibinfo {author} {\bibfnamefont {Piero}\ \bibnamefont
  {Rettegno}},\ }\bibfield  {title} {\enquote {\bibinfo {title} {{All in one:
  effective one body multipolar waveform model for spin-aligned,
  quasi-circular, eccentric, hyperbolic black hole binaries}},}\ }\href@noop {}
  {\  (\bibinfo {year} {2021})},\ \Eprint {http://arxiv.org/abs/2101.08624}
  {arXiv:2101.08624 [gr-qc]} \BibitemShut {NoStop}%
\bibitem [{\citenamefont {Setyawati}\ and\ \citenamefont
  {Ohme}(2021)}]{Setyawati:2021gom}%
  \BibitemOpen
  \bibfield  {author} {\bibinfo {author} {\bibfnamefont {Yoshinta}\
  \bibnamefont {Setyawati}}\ and\ \bibinfo {author} {\bibfnamefont {Frank}\
  \bibnamefont {Ohme}},\ }\bibfield  {title} {\enquote {\bibinfo {title}
  {{Adding eccentricity to quasi-circular binary-black-hole waveform
  models}},}\ }\href@noop {} {\  (\bibinfo {year} {2021})},\ \Eprint
  {http://arxiv.org/abs/2101.11033} {arXiv:2101.11033 [gr-qc]} \BibitemShut
  {NoStop}%
\bibitem [{\citenamefont {Habib}\ and\ \citenamefont
  {Huerta}(2019)}]{Habib:2019cui}%
  \BibitemOpen
  \bibfield  {author} {\bibinfo {author} {\bibfnamefont {Sarah}\ \bibnamefont
  {Habib}}\ and\ \bibinfo {author} {\bibfnamefont {E.A.}\ \bibnamefont
  {Huerta}},\ }\bibfield  {title} {\enquote {\bibinfo {title}
  {{Characterization of numerical relativity waveforms of eccentric binary
  black hole mergers}},}\ }\href {\doibase 10.1103/PhysRevD.100.044016}
  {\bibfield  {journal} {\bibinfo  {journal} {Phys. Rev. D}\ }\textbf {\bibinfo
  {volume} {100}},\ \bibinfo {pages} {044016} (\bibinfo {year} {2019})},\
  \Eprint {http://arxiv.org/abs/1904.09295} {arXiv:1904.09295 [gr-qc]}
  \BibitemShut {NoStop}%
\bibitem [{\citenamefont {Sperhake}\ \emph {et~al.}(2008)\citenamefont
  {Sperhake}, \citenamefont {Berti}, \citenamefont {Cardoso}, \citenamefont
  {Gonzalez}, \citenamefont {Bruegmann},\ and\ \citenamefont
  {Ansorg}}]{Sperhake:2007gu}%
  \BibitemOpen
  \bibfield  {author} {\bibinfo {author} {\bibfnamefont {Ulrich}\ \bibnamefont
  {Sperhake}}, \bibinfo {author} {\bibfnamefont {Emanuele}\ \bibnamefont
  {Berti}}, \bibinfo {author} {\bibfnamefont {Vitor}\ \bibnamefont {Cardoso}},
  \bibinfo {author} {\bibfnamefont {Jose~A.}\ \bibnamefont {Gonzalez}},
  \bibinfo {author} {\bibfnamefont {Bernd}\ \bibnamefont {Bruegmann}}, \ and\
  \bibinfo {author} {\bibfnamefont {Marcus}\ \bibnamefont {Ansorg}},\
  }\bibfield  {title} {\enquote {\bibinfo {title} {{Eccentric binary black-hole
  mergers: The Transition from inspiral to plunge in general relativity}},}\
  }\href {\doibase 10.1103/PhysRevD.78.064069} {\bibfield  {journal} {\bibinfo
  {journal} {Phys. Rev. D}\ }\textbf {\bibinfo {volume} {78}},\ \bibinfo
  {pages} {064069} (\bibinfo {year} {2008})},\ \Eprint
  {http://arxiv.org/abs/0710.3823} {arXiv:0710.3823 [gr-qc]} \BibitemShut
  {NoStop}%
\bibitem [{\citenamefont {Hinder}\ \emph {et~al.}(2008)\citenamefont {Hinder},
  \citenamefont {Vaishnav}, \citenamefont {Herrmann}, \citenamefont
  {Shoemaker},\ and\ \citenamefont {Laguna}}]{Hinder:2007qu}%
  \BibitemOpen
  \bibfield  {author} {\bibinfo {author} {\bibfnamefont {Ian}\ \bibnamefont
  {Hinder}}, \bibinfo {author} {\bibfnamefont {Birjoo}\ \bibnamefont
  {Vaishnav}}, \bibinfo {author} {\bibfnamefont {Frank}\ \bibnamefont
  {Herrmann}}, \bibinfo {author} {\bibfnamefont {Deirdre}\ \bibnamefont
  {Shoemaker}}, \ and\ \bibinfo {author} {\bibfnamefont {Pablo}\ \bibnamefont
  {Laguna}},\ }\bibfield  {title} {\enquote {\bibinfo {title} {{Universality
  and final spin in eccentric binary black hole inspirals}},}\ }\href {\doibase
  10.1103/PhysRevD.77.081502} {\bibfield  {journal} {\bibinfo  {journal} {Phys.
  Rev. D}\ }\textbf {\bibinfo {volume} {77}},\ \bibinfo {pages} {081502}
  (\bibinfo {year} {2008})},\ \Eprint {http://arxiv.org/abs/0710.5167}
  {arXiv:0710.5167 [gr-qc]} \BibitemShut {NoStop}%
\bibitem [{\citenamefont {Sopuerta}\ \emph {et~al.}(2007)\citenamefont
  {Sopuerta}, \citenamefont {Yunes},\ and\ \citenamefont
  {Laguna}}]{Sopuerta:2006et}%
  \BibitemOpen
  \bibfield  {author} {\bibinfo {author} {\bibfnamefont {Carlos~F.}\
  \bibnamefont {Sopuerta}}, \bibinfo {author} {\bibfnamefont {Nicolas}\
  \bibnamefont {Yunes}}, \ and\ \bibinfo {author} {\bibfnamefont {Pablo}\
  \bibnamefont {Laguna}},\ }\bibfield  {title} {\enquote {\bibinfo {title}
  {{Gravitational recoil velocities from eccentric binary black hole
  mergers}},}\ }\href {\doibase 10.1086/512067} {\bibfield  {journal} {\bibinfo
   {journal} {Astrophys. J. Lett.}\ }\textbf {\bibinfo {volume} {656}},\
  \bibinfo {pages} {L9--L12} (\bibinfo {year} {2007})},\ \Eprint
  {http://arxiv.org/abs/astro-ph/0611110} {arXiv:astro-ph/0611110 [astro-ph]}
  \BibitemShut {NoStop}%
\bibitem [{\citenamefont {Sperhake}\ \emph {et~al.}(2020)\citenamefont
  {Sperhake}, \citenamefont {Rosca-Mead}, \citenamefont {Gerosa},\ and\
  \citenamefont {Berti}}]{Sperhake:2019wwo}%
  \BibitemOpen
  \bibfield  {author} {\bibinfo {author} {\bibfnamefont {U.}~\bibnamefont
  {Sperhake}}, \bibinfo {author} {\bibfnamefont {R.}~\bibnamefont
  {Rosca-Mead}}, \bibinfo {author} {\bibfnamefont {D.}~\bibnamefont {Gerosa}},
  \ and\ \bibinfo {author} {\bibfnamefont {E.}~\bibnamefont {Berti}},\
  }\bibfield  {title} {\enquote {\bibinfo {title} {{Amplification of superkicks
  in black-hole binaries through orbital eccentricity}},}\ }\href {\doibase
  10.1103/PhysRevD.101.024044} {\bibfield  {journal} {\bibinfo  {journal}
  {Phys. Rev.}\ }\textbf {\bibinfo {volume} {D101}},\ \bibinfo {pages} {024044}
  (\bibinfo {year} {2020})},\ \Eprint {http://arxiv.org/abs/1910.01598}
  {arXiv:1910.01598 [gr-qc]} \BibitemShut {NoStop}%
\bibitem [{\citenamefont {Haegel}\ and\ \citenamefont
  {Husa}(2020)}]{Haegel:2019uop}%
  \BibitemOpen
  \bibfield  {author} {\bibinfo {author} {\bibfnamefont {Le\"\i{}la}\
  \bibnamefont {Haegel}}\ and\ \bibinfo {author} {\bibfnamefont {Sascha}\
  \bibnamefont {Husa}},\ }\bibfield  {title} {\enquote {\bibinfo {title}
  {{Predicting the properties of black-hole merger remnants with deep neural
  networks}},}\ }\href {\doibase 10.1088/1361-6382/ab905c} {\bibfield
  {journal} {\bibinfo  {journal} {Class. Quant. Grav.}\ }\textbf {\bibinfo
  {volume} {37}},\ \bibinfo {pages} {135005} (\bibinfo {year} {2020})},\
  \Eprint {http://arxiv.org/abs/1911.01496} {arXiv:1911.01496 [gr-qc]}
  \BibitemShut {NoStop}%
\bibitem [{SpE()}]{SpECwebsite}%
  \BibitemOpen
  \href@noop {} {\enquote {\bibinfo {title} {The {S}pectral {E}instein
  {C}ode},}\ }\bibinfo {note}
  {\url{http://www.black-holes.org/SpEC.html}}\BibitemShut {NoStop}%
\bibitem [{SXS()}]{SXSWebsite}%
  \BibitemOpen
  \href@noop {} {\enquote {\bibinfo {title} {Simulating e{X}treme
  {S}pacetimes},}\ }\bibinfo {note}
  {\url{http://www.black-holes.org/}}\BibitemShut {NoStop}%
\bibitem [{\citenamefont {Chatziioannou}\ \emph {et~al.}(2021)\citenamefont
  {Chatziioannou}, \citenamefont {Pfeiffer} \emph
  {et~al.}}]{Chatziioannou:ecccontrol_inprep}%
  \BibitemOpen
  \bibfield  {author} {\bibinfo {author} {\bibfnamefont {Katerina}\
  \bibnamefont {Chatziioannou}}, \bibinfo {author} {\bibfnamefont {Harald~P.}\
  \bibnamefont {Pfeiffer}},  \emph {et~al.},\ }\href@noop {} {\  (\bibinfo
  {year} {2021})},\ \bibinfo {note} {in preparation}\BibitemShut {NoStop}%
\bibitem [{\citenamefont {York}(1999)}]{York:1998hy}%
  \BibitemOpen
  \bibfield  {author} {\bibinfo {author} {\bibfnamefont {James~W.}\
  \bibnamefont {York}, \bibfnamefont {Jr.}},\ }\bibfield  {title} {\enquote
  {\bibinfo {title} {{Conformal 'thin sandwich' data for the initial-value
  problem}},}\ }\href {\doibase 10.1103/PhysRevLett.82.1350} {\bibfield
  {journal} {\bibinfo  {journal} {Phys. Rev. Lett.}\ }\textbf {\bibinfo
  {volume} {82}},\ \bibinfo {pages} {1350--1353} (\bibinfo {year} {1999})},\
  \Eprint {http://arxiv.org/abs/gr-qc/9810051} {arXiv:gr-qc/9810051 [gr-qc]}
  \BibitemShut {NoStop}%
\bibitem [{\citenamefont {Pfeiffer}\ and\ \citenamefont
  {York}(2003)}]{Pfeiffer:2002iy}%
  \BibitemOpen
  \bibfield  {author} {\bibinfo {author} {\bibfnamefont {Harald~P.}\
  \bibnamefont {Pfeiffer}}\ and\ \bibinfo {author} {\bibfnamefont {James~W.}\
  \bibnamefont {York}, \bibfnamefont {Jr.}},\ }\bibfield  {title} {\enquote
  {\bibinfo {title} {{Extrinsic curvature and the Einstein constraints}},}\
  }\href {\doibase 10.1103/PhysRevD.67.044022} {\bibfield  {journal} {\bibinfo
  {journal} {Phys. Rev.}\ }\textbf {\bibinfo {volume} {D67}},\ \bibinfo {pages}
  {044022} (\bibinfo {year} {2003})},\ \Eprint
  {http://arxiv.org/abs/gr-qc/0207095} {arXiv:gr-qc/0207095 [gr-qc]}
  \BibitemShut {NoStop}%
\bibitem [{\citenamefont {Varma}\ \emph {et~al.}(2018)\citenamefont {Varma},
  \citenamefont {Scheel},\ and\ \citenamefont {Pfeiffer}}]{Varma:2018sqd}%
  \BibitemOpen
  \bibfield  {author} {\bibinfo {author} {\bibfnamefont {Vijay}\ \bibnamefont
  {Varma}}, \bibinfo {author} {\bibfnamefont {Mark~A.}\ \bibnamefont {Scheel}},
  \ and\ \bibinfo {author} {\bibfnamefont {Harald~P.}\ \bibnamefont
  {Pfeiffer}},\ }\bibfield  {title} {\enquote {\bibinfo {title} {{Comparison of
  binary black hole initial data sets}},}\ }\href {\doibase
  10.1103/PhysRevD.98.104011} {\bibfield  {journal} {\bibinfo  {journal} {Phys.
  Rev.}\ }\textbf {\bibinfo {volume} {D98}},\ \bibinfo {pages} {104011}
  (\bibinfo {year} {2018})},\ \Eprint {http://arxiv.org/abs/1808.08228}
  {arXiv:1808.08228 [gr-qc]} \BibitemShut {NoStop}%
\bibitem [{\citenamefont {Lindblom}\ \emph {et~al.}(2006)\citenamefont
  {Lindblom}, \citenamefont {Scheel}, \citenamefont {Kidder}, \citenamefont
  {Owen},\ and\ \citenamefont {Rinne}}]{Lindblom:2005qh}%
  \BibitemOpen
  \bibfield  {author} {\bibinfo {author} {\bibfnamefont {Lee}\ \bibnamefont
  {Lindblom}}, \bibinfo {author} {\bibfnamefont {Mark~A.}\ \bibnamefont
  {Scheel}}, \bibinfo {author} {\bibfnamefont {Lawrence~E.}\ \bibnamefont
  {Kidder}}, \bibinfo {author} {\bibfnamefont {Robert}\ \bibnamefont {Owen}}, \
  and\ \bibinfo {author} {\bibfnamefont {Oliver}\ \bibnamefont {Rinne}},\
  }\bibfield  {title} {\enquote {\bibinfo {title} {{A New generalized harmonic
  evolution system}},}\ }\href {\doibase 10.1088/0264-9381/23/16/S09}
  {\bibfield  {journal} {\bibinfo  {journal} {Class. Quant. Grav.}\ }\textbf
  {\bibinfo {volume} {23}},\ \bibinfo {pages} {S447--S462} (\bibinfo {year}
  {2006})},\ \Eprint {http://arxiv.org/abs/gr-qc/0512093} {arXiv:gr-qc/0512093
  [gr-qc]} \BibitemShut {NoStop}%
\bibitem [{\citenamefont {Rinne}\ \emph {et~al.}(2009)\citenamefont {Rinne},
  \citenamefont {Buchman}, \citenamefont {Scheel},\ and\ \citenamefont
  {Pfeiffer}}]{Rinne:2008vn}%
  \BibitemOpen
  \bibfield  {author} {\bibinfo {author} {\bibfnamefont {Oliver}\ \bibnamefont
  {Rinne}}, \bibinfo {author} {\bibfnamefont {Luisa~T.}\ \bibnamefont
  {Buchman}}, \bibinfo {author} {\bibfnamefont {Mark~A.}\ \bibnamefont
  {Scheel}}, \ and\ \bibinfo {author} {\bibfnamefont {Harald~P.}\ \bibnamefont
  {Pfeiffer}},\ }\bibfield  {title} {\enquote {\bibinfo {title}
  {{Implementation of higher-order absorbing boundary conditions for the
  Einstein equations}},}\ }\href {\doibase 10.1088/0264-9381/26/7/075009}
  {\bibfield  {journal} {\bibinfo  {journal} {Class. Quant. Grav.}\ }\textbf
  {\bibinfo {volume} {26}},\ \bibinfo {pages} {075009} (\bibinfo {year}
  {2009})},\ \Eprint {http://arxiv.org/abs/0811.3593} {arXiv:0811.3593 [gr-qc]}
  \BibitemShut {NoStop}%
\bibitem [{\citenamefont {Boyle}\ \emph {et~al.}(2019)\citenamefont {Boyle}
  \emph {et~al.}}]{Boyle:2019kee}%
  \BibitemOpen
  \bibfield  {author} {\bibinfo {author} {\bibfnamefont {Michael}\ \bibnamefont
  {Boyle}} \emph {et~al.},\ }\bibfield  {title} {\enquote {\bibinfo {title}
  {{The SXS Collaboration catalog of binary black hole simulations}},}\ }\href
  {\doibase 10.1088/1361-6382/ab34e2} {\bibfield  {journal} {\bibinfo
  {journal} {Class. Quant. Grav.}\ }\textbf {\bibinfo {volume} {36}},\ \bibinfo
  {pages} {195006} (\bibinfo {year} {2019})},\ \Eprint
  {http://arxiv.org/abs/1904.04831} {arXiv:1904.04831 [gr-qc]} \BibitemShut
  {NoStop}%
\bibitem [{\citenamefont {{SXS Collaboration}}()}]{SXSCatalog}%
  \BibitemOpen
  \bibfield  {author} {\bibinfo {author} {\bibnamefont {{SXS Collaboration}}},\
  }\href@noop {} {\enquote {\bibinfo {title} {The {SXS} collaboration catalog
  of gravitational waveforms},}\ }\bibinfo {note}
  {\url{http://www.black-holes.org/waveforms}}\BibitemShut {NoStop}%
\bibitem [{\citenamefont {Boyle}\ and\ \citenamefont
  {Mroue}(2009)}]{Boyle:2009vi}%
  \BibitemOpen
  \bibfield  {author} {\bibinfo {author} {\bibfnamefont {Michael}\ \bibnamefont
  {Boyle}}\ and\ \bibinfo {author} {\bibfnamefont {Abdul~H.}\ \bibnamefont
  {Mroue}},\ }\bibfield  {title} {\enquote {\bibinfo {title} {{Extrapolating
  gravitational-wave data from numerical simulations}},}\ }\href {\doibase
  10.1103/PhysRevD.80.124045} {\bibfield  {journal} {\bibinfo  {journal} {Phys.
  Rev.}\ }\textbf {\bibinfo {volume} {D80}},\ \bibinfo {pages} {124045}
  (\bibinfo {year} {2009})},\ \Eprint {http://arxiv.org/abs/0905.3177}
  {arXiv:0905.3177 [gr-qc]} \BibitemShut {NoStop}%
\bibitem [{\citenamefont {Boyle}(2016)}]{Boyle:2015nqa}%
  \BibitemOpen
  \bibfield  {author} {\bibinfo {author} {\bibfnamefont {Michael}\ \bibnamefont
  {Boyle}},\ }\bibfield  {title} {\enquote {\bibinfo {title} {{Transformations
  of asymptotic gravitational-wave data}},}\ }\href {\doibase
  10.1103/PhysRevD.93.084031} {\bibfield  {journal} {\bibinfo  {journal} {Phys.
  Rev.}\ }\textbf {\bibinfo {volume} {D93}},\ \bibinfo {pages} {084031}
  (\bibinfo {year} {2016})},\ \Eprint {http://arxiv.org/abs/1509.00862}
  {arXiv:1509.00862 [gr-qc]} \BibitemShut {NoStop}%
\bibitem [{\citenamefont {Boyle}()}]{scri}%
  \BibitemOpen
  \bibfield  {author} {\bibinfo {author} {\bibfnamefont {Michael}\ \bibnamefont
  {Boyle}},\ }\href@noop {} {\enquote {\bibinfo {title} {Scri},}\ }\bibinfo
  {note} {\url{https://github.com/moble/scri}}\BibitemShut {NoStop}%
\bibitem [{\citenamefont {Favata}(2009)}]{Favata:2008yd}%
  \BibitemOpen
  \bibfield  {author} {\bibinfo {author} {\bibfnamefont {Marc}\ \bibnamefont
  {Favata}},\ }\bibfield  {title} {\enquote {\bibinfo {title} {{Post-Newtonian
  corrections to the gravitational-wave memory for quasi-circular, inspiralling
  compact binaries}},}\ }\href {\doibase 10.1103/PhysRevD.80.024002} {\bibfield
   {journal} {\bibinfo  {journal} {Phys. Rev. D}\ }\textbf {\bibinfo {volume}
  {80}},\ \bibinfo {pages} {024002} (\bibinfo {year} {2009})},\ \Eprint
  {http://arxiv.org/abs/0812.0069} {arXiv:0812.0069 [gr-qc]} \BibitemShut
  {NoStop}%
\bibitem [{\citenamefont {Mitman}\ \emph
  {et~al.}(2020{\natexlab{a}})\citenamefont {Mitman}, \citenamefont {Moxon},
  \citenamefont {Scheel}, \citenamefont {Teukolsky}, \citenamefont {Boyle},
  \citenamefont {Deppe}, \citenamefont {Kidder},\ and\ \citenamefont
  {Throwe}}]{Mitman:2020pbt}%
  \BibitemOpen
  \bibfield  {author} {\bibinfo {author} {\bibfnamefont {Keefe}\ \bibnamefont
  {Mitman}}, \bibinfo {author} {\bibfnamefont {Jordan}\ \bibnamefont {Moxon}},
  \bibinfo {author} {\bibfnamefont {Mark~A.}\ \bibnamefont {Scheel}}, \bibinfo
  {author} {\bibfnamefont {Saul~A.}\ \bibnamefont {Teukolsky}}, \bibinfo
  {author} {\bibfnamefont {Michael}\ \bibnamefont {Boyle}}, \bibinfo {author}
  {\bibfnamefont {Nils}\ \bibnamefont {Deppe}}, \bibinfo {author}
  {\bibfnamefont {Lawrence~E.}\ \bibnamefont {Kidder}}, \ and\ \bibinfo
  {author} {\bibfnamefont {William}\ \bibnamefont {Throwe}},\ }\bibfield
  {title} {\enquote {\bibinfo {title} {{Computation of displacement and spin
  gravitational memory in numerical relativity}},}\ }\href {\doibase
  10.1103/PhysRevD.102.104007} {\bibfield  {journal} {\bibinfo  {journal}
  {Phys. Rev. D}\ }\textbf {\bibinfo {volume} {102}},\ \bibinfo {pages}
  {104007} (\bibinfo {year} {2020}{\natexlab{a}})},\ \Eprint
  {http://arxiv.org/abs/2007.11562} {arXiv:2007.11562 [gr-qc]} \BibitemShut
  {NoStop}%
\bibitem [{\citenamefont {Barkett}\ \emph {et~al.}(2020)\citenamefont
  {Barkett}, \citenamefont {Moxon}, \citenamefont {Scheel},\ and\ \citenamefont
  {Szil\'agyi}}]{Barkett:2019uae}%
  \BibitemOpen
  \bibfield  {author} {\bibinfo {author} {\bibfnamefont {Kevin}\ \bibnamefont
  {Barkett}}, \bibinfo {author} {\bibfnamefont {Jordan}\ \bibnamefont {Moxon}},
  \bibinfo {author} {\bibfnamefont {Mark~A.}\ \bibnamefont {Scheel}}, \ and\
  \bibinfo {author} {\bibfnamefont {B\'ela}\ \bibnamefont {Szil\'agyi}},\
  }\bibfield  {title} {\enquote {\bibinfo {title} {{Spectral
  Cauchy-Characteristic Extraction of the Gravitational Wave News Function}},}\
  }\href {\doibase 10.1103/PhysRevD.102.024004} {\bibfield  {journal} {\bibinfo
   {journal} {Phys. Rev. D}\ }\textbf {\bibinfo {volume} {102}},\ \bibinfo
  {pages} {024004} (\bibinfo {year} {2020})},\ \Eprint
  {http://arxiv.org/abs/1910.09677} {arXiv:1910.09677 [gr-qc]} \BibitemShut
  {NoStop}%
\bibitem [{\citenamefont {Moxon}\ \emph {et~al.}(2020)\citenamefont {Moxon},
  \citenamefont {Scheel},\ and\ \citenamefont {Teukolsky}}]{Moxon:2020gha}%
  \BibitemOpen
  \bibfield  {author} {\bibinfo {author} {\bibfnamefont {Jordan}\ \bibnamefont
  {Moxon}}, \bibinfo {author} {\bibfnamefont {Mark~A.}\ \bibnamefont {Scheel}},
  \ and\ \bibinfo {author} {\bibfnamefont {Saul~A.}\ \bibnamefont
  {Teukolsky}},\ }\bibfield  {title} {\enquote {\bibinfo {title} {{Improved
  Cauchy-characteristic evolution system for high-precision numerical
  relativity waveforms}},}\ }\href {\doibase 10.1103/PhysRevD.102.044052}
  {\bibfield  {journal} {\bibinfo  {journal} {Phys. Rev. D}\ }\textbf {\bibinfo
  {volume} {102}},\ \bibinfo {pages} {044052} (\bibinfo {year} {2020})},\
  \Eprint {http://arxiv.org/abs/2007.01339} {arXiv:2007.01339 [gr-qc]}
  \BibitemShut {NoStop}%
\bibitem [{\citenamefont {Mitman}\ \emph
  {et~al.}(2020{\natexlab{b}})\citenamefont {Mitman} \emph
  {et~al.}}]{Mitman:2020bjf}%
  \BibitemOpen
  \bibfield  {author} {\bibinfo {author} {\bibfnamefont {Keefe}\ \bibnamefont
  {Mitman}} \emph {et~al.},\ }\bibfield  {title} {\enquote {\bibinfo {title}
  {{Adding Gravitational Memory to Waveform Catalogs using BMS Balance
  Laws}},}\ }\href@noop {} {\  (\bibinfo {year} {2020}{\natexlab{b}})},\
  \Eprint {http://arxiv.org/abs/2011.01309} {arXiv:2011.01309 [gr-qc]}
  \BibitemShut {NoStop}%
\bibitem [{\citenamefont {Varma}\ \emph
  {et~al.}(2019{\natexlab{c}})\citenamefont {Varma}, \citenamefont {Gerosa},
  \citenamefont {Stein}, \citenamefont {Hébert},\ and\ \citenamefont
  {Zhang}}]{Varma:2018aht}%
  \BibitemOpen
  \bibfield  {author} {\bibinfo {author} {\bibfnamefont {Vijay}\ \bibnamefont
  {Varma}}, \bibinfo {author} {\bibfnamefont {Davide}\ \bibnamefont {Gerosa}},
  \bibinfo {author} {\bibfnamefont {Leo~C.}\ \bibnamefont {Stein}}, \bibinfo
  {author} {\bibfnamefont {François}\ \bibnamefont {Hébert}}, \ and\ \bibinfo
  {author} {\bibfnamefont {Hao}\ \bibnamefont {Zhang}},\ }\bibfield  {title}
  {\enquote {\bibinfo {title} {{High-accuracy mass, spin, and recoil
  predictions of generic black-hole merger remnants}},}\ }\href {\doibase
  10.1103/PhysRevLett.122.011101} {\bibfield  {journal} {\bibinfo  {journal}
  {Phys. Rev. Lett.}\ }\textbf {\bibinfo {volume} {122}},\ \bibinfo {pages}
  {011101} (\bibinfo {year} {2019}{\natexlab{c}})},\ \Eprint
  {http://arxiv.org/abs/1809.09125} {arXiv:1809.09125 [gr-qc]} \BibitemShut
  {NoStop}%
\bibitem [{\citenamefont {Varma}\ \emph {et~al.}(2020)\citenamefont {Varma},
  \citenamefont {Isi},\ and\ \citenamefont {Biscoveanu}}]{Varma:2020nbm}%
  \BibitemOpen
  \bibfield  {author} {\bibinfo {author} {\bibfnamefont {Vijay}\ \bibnamefont
  {Varma}}, \bibinfo {author} {\bibfnamefont {Maximiliano}\ \bibnamefont
  {Isi}}, \ and\ \bibinfo {author} {\bibfnamefont {Sylvia}\ \bibnamefont
  {Biscoveanu}},\ }\bibfield  {title} {\enquote {\bibinfo {title} {{Extracting
  the Gravitational Recoil from Black Hole Merger Signals}},}\ }\href {\doibase
  10.1103/PhysRevLett.124.101104} {\bibfield  {journal} {\bibinfo  {journal}
  {Phys. Rev. Lett.}\ }\textbf {\bibinfo {volume} {124}},\ \bibinfo {pages}
  {101104} (\bibinfo {year} {2020})},\ \Eprint
  {http://arxiv.org/abs/2002.00296} {arXiv:2002.00296 [gr-qc]} \BibitemShut
  {NoStop}%
\bibitem [{\citenamefont {Chandrasekhar}(1998)}]{Chandrasekhar:1983mtbh}%
  \BibitemOpen
  \bibfield  {author} {\bibinfo {author} {\bibfnamefont {S.}~\bibnamefont
  {Chandrasekhar}},\ }\href@noop {} {\emph {\bibinfo {title} {The Mathematical
  Theory of Black Holes}}}\ (\bibinfo  {publisher} {Clarendon Press},\ \bibinfo
  {year} {1998})\BibitemShut {NoStop}%
\bibitem [{\citenamefont {Healy}\ \emph {et~al.}(2017)\citenamefont {Healy},
  \citenamefont {Lousto}, \citenamefont {Nakano},\ and\ \citenamefont
  {Zlochower}}]{Healy:2017zqj}%
  \BibitemOpen
  \bibfield  {author} {\bibinfo {author} {\bibfnamefont {James}\ \bibnamefont
  {Healy}}, \bibinfo {author} {\bibfnamefont {Carlos~O.}\ \bibnamefont
  {Lousto}}, \bibinfo {author} {\bibfnamefont {Hiroyuki}\ \bibnamefont
  {Nakano}}, \ and\ \bibinfo {author} {\bibfnamefont {Yosef}\ \bibnamefont
  {Zlochower}},\ }\bibfield  {title} {\enquote {\bibinfo {title}
  {{Post-Newtonian Quasicircular Initial Orbits for Numerical Relativity}},}\
  }\href {\doibase 10.1088/1361-6382/aa7929} {\bibfield  {journal} {\bibinfo
  {journal} {Class. Quant. Grav.}\ }\textbf {\bibinfo {volume} {34}},\ \bibinfo
  {pages} {145011} (\bibinfo {year} {2017})},\ \Eprint
  {http://arxiv.org/abs/1702.00872} {arXiv:1702.00872 [gr-qc]} \BibitemShut
  {NoStop}%
\bibitem [{\citenamefont {Mroue}\ \emph {et~al.}(2010)\citenamefont {Mroue},
  \citenamefont {Pfeiffer}, \citenamefont {Kidder},\ and\ \citenamefont
  {Teukolsky}}]{Mroue:2010re}%
  \BibitemOpen
  \bibfield  {author} {\bibinfo {author} {\bibfnamefont {Abdul~H.}\
  \bibnamefont {Mroue}}, \bibinfo {author} {\bibfnamefont {Harald~P.}\
  \bibnamefont {Pfeiffer}}, \bibinfo {author} {\bibfnamefont {Lawrence~E.}\
  \bibnamefont {Kidder}}, \ and\ \bibinfo {author} {\bibfnamefont {Saul~A.}\
  \bibnamefont {Teukolsky}},\ }\bibfield  {title} {\enquote {\bibinfo {title}
  {{Measuring orbital eccentricity and periastron advance in quasi-circular
  black hole simulations}},}\ }\href {\doibase 10.1103/PhysRevD.82.124016}
  {\bibfield  {journal} {\bibinfo  {journal} {Phys. Rev.}\ }\textbf {\bibinfo
  {volume} {D82}},\ \bibinfo {pages} {124016} (\bibinfo {year} {2010})},\
  \Eprint {http://arxiv.org/abs/1004.4697} {arXiv:1004.4697 [gr-qc]}
  \BibitemShut {NoStop}%
\bibitem [{\citenamefont {Purrer}\ \emph {et~al.}(2012)\citenamefont {Purrer},
  \citenamefont {Husa},\ and\ \citenamefont {Hannam}}]{Purrer:2012wy}%
  \BibitemOpen
  \bibfield  {author} {\bibinfo {author} {\bibfnamefont {Michael}\ \bibnamefont
  {Purrer}}, \bibinfo {author} {\bibfnamefont {Sascha}\ \bibnamefont {Husa}}, \
  and\ \bibinfo {author} {\bibfnamefont {Mark}\ \bibnamefont {Hannam}},\
  }\bibfield  {title} {\enquote {\bibinfo {title} {{An Efficient iterative
  method to reduce eccentricity in numerical-relativity simulations of compact
  binary inspiral}},}\ }\href {\doibase 10.1103/PhysRevD.85.124051} {\bibfield
  {journal} {\bibinfo  {journal} {Phys. Rev. D}\ }\textbf {\bibinfo {volume}
  {85}},\ \bibinfo {pages} {124051} (\bibinfo {year} {2012})},\ \Eprint
  {http://arxiv.org/abs/1203.4258} {arXiv:1203.4258 [gr-qc]} \BibitemShut
  {NoStop}%
\bibitem [{\citenamefont {Mora}\ and\ \citenamefont
  {Will}(2002)}]{Mora:2002gf}%
  \BibitemOpen
  \bibfield  {author} {\bibinfo {author} {\bibfnamefont {Thierry}\ \bibnamefont
  {Mora}}\ and\ \bibinfo {author} {\bibfnamefont {Clifford~M.}\ \bibnamefont
  {Will}},\ }\bibfield  {title} {\enquote {\bibinfo {title} {{Numerically
  generated quasiequilibrium orbits of black holes: Circular or eccentric?}}}\
  }\href {\doibase 10.1103/PhysRevD.66.101501} {\bibfield  {journal} {\bibinfo
  {journal} {Phys. Rev. D}\ }\textbf {\bibinfo {volume} {66}},\ \bibinfo
  {pages} {101501} (\bibinfo {year} {2002})},\ \Eprint
  {http://arxiv.org/abs/gr-qc/0208089} {arXiv:gr-qc/0208089} \BibitemShut
  {NoStop}%
\bibitem [{\citenamefont {Field}\ \emph {et~al.}(2011)\citenamefont {Field},
  \citenamefont {Galley}, \citenamefont {Herrmann}, \citenamefont {Hesthaven},
  \citenamefont {Ochsner},\ and\ \citenamefont {Tiglio}}]{Field:2011mf}%
  \BibitemOpen
  \bibfield  {author} {\bibinfo {author} {\bibfnamefont {Scott~E.}\
  \bibnamefont {Field}}, \bibinfo {author} {\bibfnamefont {Chad~R.}\
  \bibnamefont {Galley}}, \bibinfo {author} {\bibfnamefont {Frank}\
  \bibnamefont {Herrmann}}, \bibinfo {author} {\bibfnamefont {Jan~S.}\
  \bibnamefont {Hesthaven}}, \bibinfo {author} {\bibfnamefont {Evan}\
  \bibnamefont {Ochsner}}, \ and\ \bibinfo {author} {\bibfnamefont {Manuel}\
  \bibnamefont {Tiglio}},\ }\bibfield  {title} {\enquote {\bibinfo {title}
  {{Reduced basis catalogs for gravitational wave templates}},}\ }\href
  {\doibase 10.1103/PhysRevLett.106.221102} {\bibfield  {journal} {\bibinfo
  {journal} {Phys. Rev. Lett.}\ }\textbf {\bibinfo {volume} {106}},\ \bibinfo
  {pages} {221102} (\bibinfo {year} {2011})},\ \Eprint
  {http://arxiv.org/abs/1101.3765} {arXiv:1101.3765 [gr-qc]} \BibitemShut
  {NoStop}%
\bibitem [{\citenamefont {Maday}\ \emph {et~al.}(2009)\citenamefont {Maday},
  \citenamefont {Nguyen}, \citenamefont {Patera},\ and\ \citenamefont
  {Pau}}]{Maday:2009}%
  \BibitemOpen
  \bibfield  {author} {\bibinfo {author} {\bibfnamefont {Y.}~\bibnamefont
  {Maday}}, \bibinfo {author} {\bibfnamefont {N.~.C}\ \bibnamefont {Nguyen}},
  \bibinfo {author} {\bibfnamefont {A.~T.}\ \bibnamefont {Patera}}, \ and\
  \bibinfo {author} {\bibfnamefont {S.~H.}\ \bibnamefont {Pau}},\ }\bibfield
  {title} {\enquote {\bibinfo {title} {A general multipurpose interpolation
  procedure: the magic points},}\ }\href {\doibase 10.3934/cpaa.2009.8.383}
  {\bibfield  {journal} {\bibinfo  {journal} {Communications on Pure and
  Applied Analysis}\ }\textbf {\bibinfo {volume} {8}},\ \bibinfo {pages}
  {383--404} (\bibinfo {year} {2009})}\BibitemShut {NoStop}%
\bibitem [{\citenamefont {Chaturantabut}\ and\ \citenamefont
  {Sorensen}(2010)}]{chaturantabut2010nonlinear}%
  \BibitemOpen
  \bibfield  {author} {\bibinfo {author} {\bibfnamefont {Saifon}\ \bibnamefont
  {Chaturantabut}}\ and\ \bibinfo {author} {\bibfnamefont {Danny~C}\
  \bibnamefont {Sorensen}},\ }\bibfield  {title} {\enquote {\bibinfo {title}
  {Nonlinear model reduction via discrete empirical interpolation},}\
  }\href@noop {} {\bibfield  {journal} {\bibinfo  {journal} {SIAM Journal on
  Scientific Computing}\ }\textbf {\bibinfo {volume} {32}},\ \bibinfo {pages}
  {2737--2764} (\bibinfo {year} {2010})}\BibitemShut {NoStop}%
\bibitem [{\citenamefont {Canizares}\ \emph {et~al.}(2015)\citenamefont
  {Canizares}, \citenamefont {Field}, \citenamefont {Gair}, \citenamefont
  {Raymond}, \citenamefont {Smith},\ and\ \citenamefont
  {Tiglio}}]{Canizares:2014fya}%
  \BibitemOpen
  \bibfield  {author} {\bibinfo {author} {\bibfnamefont {Priscilla}\
  \bibnamefont {Canizares}}, \bibinfo {author} {\bibfnamefont {Scott~E.}\
  \bibnamefont {Field}}, \bibinfo {author} {\bibfnamefont {Jonathan}\
  \bibnamefont {Gair}}, \bibinfo {author} {\bibfnamefont {Vivien}\ \bibnamefont
  {Raymond}}, \bibinfo {author} {\bibfnamefont {Rory}\ \bibnamefont {Smith}}, \
  and\ \bibinfo {author} {\bibfnamefont {Manuel}\ \bibnamefont {Tiglio}},\
  }\bibfield  {title} {\enquote {\bibinfo {title} {{Accelerated
  gravitational-wave parameter estimation with reduced order modeling}},}\
  }\href {\doibase 10.1103/PhysRevLett.114.071104} {\bibfield  {journal}
  {\bibinfo  {journal} {Phys. Rev. Lett.}\ }\textbf {\bibinfo {volume} {114}},\
  \bibinfo {pages} {071104} (\bibinfo {year} {2015})},\ \Eprint
  {http://arxiv.org/abs/1404.6284} {arXiv:1404.6284 [gr-qc]} \BibitemShut
  {NoStop}%
\bibitem [{\citenamefont {Taylor}\ and\ \citenamefont
  {Varma}(2020)}]{Taylor:2020bmj}%
  \BibitemOpen
  \bibfield  {author} {\bibinfo {author} {\bibfnamefont {Afura}\ \bibnamefont
  {Taylor}}\ and\ \bibinfo {author} {\bibfnamefont {Vijay}\ \bibnamefont
  {Varma}},\ }\bibfield  {title} {\enquote {\bibinfo {title} {{Gravitational
  wave peak luminosity model for precessing binary black holes}},}\ }\href
  {\doibase 10.1103/PhysRevD.102.104047} {\  (\bibinfo {year} {2020}),\
  10.1103/PhysRevD.102.104047},\ \Eprint {http://arxiv.org/abs/2010.00120}
  {arXiv:2010.00120 [gr-qc]} \BibitemShut {NoStop}%
\bibitem [{\citenamefont {Varma}\ and\ \citenamefont
  {Ajith}(2017)}]{Varma:2016dnf}%
  \BibitemOpen
  \bibfield  {author} {\bibinfo {author} {\bibfnamefont {Vijay}\ \bibnamefont
  {Varma}}\ and\ \bibinfo {author} {\bibfnamefont {Parameswaran}\ \bibnamefont
  {Ajith}},\ }\bibfield  {title} {\enquote {\bibinfo {title} {{Effects of
  nonquadrupole modes in the detection and parameter estimation of black hole
  binaries with nonprecessing spins}},}\ }\href {\doibase
  10.1103/PhysRevD.96.124024} {\bibfield  {journal} {\bibinfo  {journal} {Phys.
  Rev.}\ }\textbf {\bibinfo {volume} {D96}},\ \bibinfo {pages} {124024}
  (\bibinfo {year} {2017})},\ \Eprint {http://arxiv.org/abs/1612.05608}
  {arXiv:1612.05608 [gr-qc]} \BibitemShut {NoStop}%
\bibitem [{\citenamefont {McKechan}\ \emph {et~al.}(2010)\citenamefont
  {McKechan}, \citenamefont {Robinson},\ and\ \citenamefont
  {Sathyaprakash}}]{McKechan:2010kp}%
  \BibitemOpen
  \bibfield  {author} {\bibinfo {author} {\bibfnamefont {D.~J.~A.}\
  \bibnamefont {McKechan}}, \bibinfo {author} {\bibfnamefont {C.}~\bibnamefont
  {Robinson}}, \ and\ \bibinfo {author} {\bibfnamefont {B.~S.}\ \bibnamefont
  {Sathyaprakash}},\ }\bibfield  {title} {\enquote {\bibinfo {title} {{A
  tapering window for time-domain templates and simulated signals in the
  detection of gravitational waves from coalescing compact binaries}},}\
  }\bibfield  {booktitle} {\emph {\bibinfo {booktitle} {{Gravitational waves.
  Proceedings, 8th Edoardo Amaldi Conference, Amaldi 8, New York, USA, June
  22-26, 2009}}},\ }\href {\doibase 10.1088/0264-9381/27/8/084020} {\bibfield
  {journal} {\bibinfo  {journal} {Class. Quant. Grav.}\ }\textbf {\bibinfo
  {volume} {27}},\ \bibinfo {pages} {084020} (\bibinfo {year} {2010})},\
  \Eprint {http://arxiv.org/abs/1003.2939} {arXiv:1003.2939 [gr-qc]}
  \BibitemShut {NoStop}%
\bibitem [{\citenamefont {{LIGO Scientific
  Collaboration}}(2018)}]{aLIGODesignNoiseCurve}%
  \BibitemOpen
  \bibfield  {author} {\bibinfo {author} {\bibnamefont {{LIGO Scientific
  Collaboration}}},\ }\href@noop {} {\emph {\bibinfo {title} {Updated Advanced
  LIGO sensitivity design curve}}},\ \bibinfo {type} {Tech. Rep.}\ (\bibinfo
  {year} {2018})\ \bibinfo {note}
  {\url{https://dcc.ligo.org/LIGO-T1800044/public}}\BibitemShut {NoStop}%
\bibitem [{\citenamefont {Newman}\ and\ \citenamefont
  {Penrose}(1966)}]{Newman:1966ub}%
  \BibitemOpen
  \bibfield  {author} {\bibinfo {author} {\bibfnamefont {E.T.}\ \bibnamefont
  {Newman}}\ and\ \bibinfo {author} {\bibfnamefont {R.}~\bibnamefont
  {Penrose}},\ }\bibfield  {title} {\enquote {\bibinfo {title} {{Note on the
  Bondi-Metzner-Sachs group}},}\ }\href {\doibase 10.1063/1.1931221} {\bibfield
   {journal} {\bibinfo  {journal} {J. Math. Phys.}\ }\textbf {\bibinfo {volume}
  {7}},\ \bibinfo {pages} {863--870} (\bibinfo {year} {1966})}\BibitemShut
  {NoStop}%
\bibitem [{\citenamefont {Goldberg}\ \emph {et~al.}(1967)\citenamefont
  {Goldberg}, \citenamefont {MacFarlane}, \citenamefont {Newman}, \citenamefont
  {Rohrlich},\ and\ \citenamefont {Sudarshan}}]{Goldberg:1966uu}%
  \BibitemOpen
  \bibfield  {author} {\bibinfo {author} {\bibfnamefont {J.N.}\ \bibnamefont
  {Goldberg}}, \bibinfo {author} {\bibfnamefont {A.J.}\ \bibnamefont
  {MacFarlane}}, \bibinfo {author} {\bibfnamefont {E.T.}\ \bibnamefont
  {Newman}}, \bibinfo {author} {\bibfnamefont {F.}~\bibnamefont {Rohrlich}}, \
  and\ \bibinfo {author} {\bibfnamefont {E.C.G.}\ \bibnamefont {Sudarshan}},\
  }\bibfield  {title} {\enquote {\bibinfo {title} {{Spin s spherical harmonics
  and edth}},}\ }\href {\doibase 10.1063/1.1705135} {\bibfield  {journal}
  {\bibinfo  {journal} {J. Math. Phys.}\ }\textbf {\bibinfo {volume} {8}},\
  \bibinfo {pages} {2155} (\bibinfo {year} {1967})}\BibitemShut {NoStop}%
\bibitem [{\citenamefont {Teukolsky}(1973)}]{Teukolsky:1973ha}%
  \BibitemOpen
  \bibfield  {author} {\bibinfo {author} {\bibfnamefont {Saul~A.}\ \bibnamefont
  {Teukolsky}},\ }\bibfield  {title} {\enquote {\bibinfo {title}
  {{Perturbations of a rotating black hole. 1. Fundamental equations for
  gravitational electromagnetic and neutrino field perturbations}},}\ }\href
  {\doibase 10.1086/152444} {\bibfield  {journal} {\bibinfo  {journal}
  {Astrophys. J.}\ }\textbf {\bibinfo {volume} {185}},\ \bibinfo {pages}
  {635--647} (\bibinfo {year} {1973})}\BibitemShut {NoStop}%
\bibitem [{\citenamefont {Teukolsky}(1972)}]{Teukolsky:1972my}%
  \BibitemOpen
  \bibfield  {author} {\bibinfo {author} {\bibfnamefont {S.A.}\ \bibnamefont
  {Teukolsky}},\ }\bibfield  {title} {\enquote {\bibinfo {title} {{Rotating
  black holes - separable wave equations for gravitational and electromagnetic
  perturbations}},}\ }\href {\doibase 10.1103/PhysRevLett.29.1114} {\bibfield
  {journal} {\bibinfo  {journal} {Phys. Rev. Lett.}\ }\textbf {\bibinfo
  {volume} {29}},\ \bibinfo {pages} {1114--1118} (\bibinfo {year}
  {1972})}\BibitemShut {NoStop}%
\bibitem [{\citenamefont {Berti}\ and\ \citenamefont
  {Klein}(2014)}]{Berti:2014fga}%
  \BibitemOpen
  \bibfield  {author} {\bibinfo {author} {\bibfnamefont {Emanuele}\
  \bibnamefont {Berti}}\ and\ \bibinfo {author} {\bibfnamefont {Antoine}\
  \bibnamefont {Klein}},\ }\bibfield  {title} {\enquote {\bibinfo {title}
  {{Mixing of spherical and spheroidal modes in perturbed Kerr black holes}},}\
  }\href {\doibase 10.1103/PhysRevD.90.064012} {\bibfield  {journal} {\bibinfo
  {journal} {Phys. Rev. D}\ }\textbf {\bibinfo {volume} {90}},\ \bibinfo
  {pages} {064012} (\bibinfo {year} {2014})},\ \Eprint
  {http://arxiv.org/abs/1408.1860} {arXiv:1408.1860 [gr-qc]} \BibitemShut
  {NoStop}%
\bibitem [{\citenamefont {Szilagyi}\ \emph {et~al.}(2009)\citenamefont
  {Szilagyi}, \citenamefont {Lindblom},\ and\ \citenamefont
  {Scheel}}]{Szilagyi:2009qz}%
  \BibitemOpen
  \bibfield  {author} {\bibinfo {author} {\bibfnamefont {Bela}\ \bibnamefont
  {Szilagyi}}, \bibinfo {author} {\bibfnamefont {Lee}\ \bibnamefont
  {Lindblom}}, \ and\ \bibinfo {author} {\bibfnamefont {Mark~A.}\ \bibnamefont
  {Scheel}},\ }\bibfield  {title} {\enquote {\bibinfo {title} {{Simulations of
  Binary Black Hole Mergers Using Spectral Methods}},}\ }\href {\doibase
  10.1103/PhysRevD.80.124010} {\bibfield  {journal} {\bibinfo  {journal} {Phys.
  Rev.}\ }\textbf {\bibinfo {volume} {D80}},\ \bibinfo {pages} {124010}
  (\bibinfo {year} {2009})},\ \Eprint {http://arxiv.org/abs/0909.3557}
  {arXiv:0909.3557 [gr-qc]} \BibitemShut {NoStop}%
\end{thebibliography}%

\newpage

\end{document}